\documentclass[journal,letter,twoside]{IEEEtran}

\usepackage{cite}
\usepackage{amsmath,amssymb,amsfonts}
\usepackage{bm}
\usepackage{algorithmic}
\usepackage{textcomp}
\usepackage{comment}
\usepackage{verbatim}

\usepackage{booktabs} 

\usepackage{floatrow}
\usepackage[caption=false,font=footnotesize]{subfig}

\usepackage[table, dvipsnames]{xcolor}  
\usepackage{fancyhdr}
\usepackage{graphicx}
\usepackage{url}
\usepackage{verbatim}
\usepackage{multirow}
\usepackage{hhline}
\usepackage[us]{datetime}
\usepackage{paralist}
\usepackage{color}
\usepackage{soul}
\usepackage[implicit=false,hidelinks]{hyperref}
\usepackage{stfloats}
\usepackage[ruled, vlined, norelsize]{algorithm2e} 
\SetAlFnt{\sf} 

\graphicspath{{figures/}}

\usepackage{soul}
\newcommand{\drop}[1]{\textcolor{red}{#1}}
\renewcommand{\drop}[1]{}

\newdimen\arrayruleHwidth
\makeatletter
\def\Hline{\noalign{\ifnum0=`}\fi\hrule \@height \arrayruleHwidth
\futurelet \@tempa\@xhline}
\makeatother

\newcolumntype{P}[1]{>{\centering\arraybackslash}p{#1}}

\makeatletter
\def\blfootnote{\xdef\@thefnmark{}\@footnotetext}
\makeatother

\shortdate
\settimeformat{ampmtime}

\ifCLASSINFOpdf
\else
\fi
\begin{document}

\newcommand\thetitle{\textit{UNSAIL}: Thwarting Oracle-Less Machine Learning Attacks on Logic Locking}

\twocolumn
\setcounter{page}{1}

\title{\thetitle}

\author{Lilas~Alrahis,
Satwik~Patnaik,~\IEEEmembership{Member,~IEEE,}
Johann~Knechtel,~\IEEEmembership{Member,~IEEE,}
Hani~Saleh,~\IEEEmembership{Senior~Member,~IEEE,}
Baker~Mohammad,~\IEEEmembership{Senior~Member,~IEEE,}
Mahmoud~Al-Qutayri,~\IEEEmembership{Senior~Member,~IEEE,} and 
Ozgur~Sinanoglu,~\IEEEmembership{Senior~Member,~IEEE}

\thanks{Manuscript received June 12, 2020; revised December 8, 2020; accepted January 21, 2021.
This article was recommended by Associate Editor T.\ Güneysu. (\textit{Corresponding authors: Lilas Alrahis and Satwik Patnaik.})
}
\thanks{This work was supported in part by Khalifa University under Award No. [RC2-2018-020] and by the Center for Cyber Security at NYU New York/Abu Dhabi (NYU/NYUAD). The work of S. Patnaik was supported by the Global Ph.D. Fellowship at NYU/NYUAD.}
\IEEEcompsocitemizethanks{\IEEEcompsocthanksitem Lilas~Alrahis, Hani~Saleh, Baker~Mohammad, and Mahmoud~Al-Qutayri are with the System on Chip Center (SoCC), Khalifa University, Abu Dhabi 127788, UAE (email: lilasrahis@gmail.com).\protect
\IEEEcompsocthanksitem Satwik~Patnaik was with the Department of Electrical and Computer Engineering, Tandon School of Engineering, New York University, Brooklyn, NY 11201, USA.
He is currently with the Department of Electrical and Computer Engineering, Texas A\&M University, College Station, TX 77843 USA (e-mail: satwik.patnaik@tamu.edu).\protect
\IEEEcompsocthanksitem Johann~Knechtel and Ozgur~Sinanoglu are with the Division of Engineering, New York University Abu Dhabi, Abu Dhabi 129188, UAE (e-mail: johann@nyu.edu, ozgursin@nyu.edu).
}
\thanks{Digital Object Identifier 10.1109/TIFS.2021.3057576}
}

\markboth{IEEE Transactions on Information Forensics and Security}
{Alrahis \MakeLowercase{\textit{et al.\ }}: \textit{UNSAIL}: Thwarting Oracle-Less Machine Learning Attacks on Logic Locking}

\IEEEtitleabstractindextext{

\begin{abstract}
Logic locking aims to protect the intellectual property (IP) of integrated circuit (IC) designs throughout the globalized supply chain.
The SAIL attack, based on tailored machine learning (ML) models, circumvents combinational logic locking with high accuracy and is amongst the most potent attacks as it does not require a functional IC acting as an oracle.  
In this work, we propose \textit{UNSAIL}, a logic locking technique that inserts key-gate structures with the specific aim to confuse ML models
like those used in SAIL. More specifically, \textit{UNSAIL} serves to prevent attacks seeking to resolve the structural transformations of synthesis-induced obfuscation, which
is an essential step for logic locking.
Our approach is generic; it can protect any local structure of key-gates against such ML-based attacks in an \textit{oracle-less} setting. 
We develop a reference implementation for the SAIL attack and launch it on both traditionally locked and \textit{UNSAIL}-locked designs. 
For SAIL, two ML models have been proposed (which we implement accordingly), namely a change-prediction model and a reconstruction model;
the change-prediction model is used to determine which key-gate structures to restore using the reconstruction model. 
Our study on benchmarks ranging from the ISCAS-85 and ITC-99 suites to the
OpenRISC Reference Platform System-on-Chip (ORPSoC) confirms that \textit{UNSAIL} degrades the accuracy of the change-prediction model and the reconstruction model by an average of \texttt{20.13} and \texttt{17} percentage points (\texttt{pp}), respectively. 
When the aforementioned models are combined, which is the most powerful scenario for SAIL, \textit{UNSAIL} reduces the attack accuracy of SAIL by an average of \texttt{11pp}. 
We further demonstrate that \textit{UNSAIL} thwarts other \textit{oracle-less} attacks, i.e., SWEEP and the redundancy attack, indicating the generic nature and strength of our approach.
Detailed layout-level evaluations illustrate that \textit{UNSAIL} incurs minimal area and power overheads of \texttt{0.26\%} and
\texttt{0.61\%}, respectively, on the million-gate ORPSoC design.
\end{abstract}

}

\maketitle

\renewcommand{\headrulewidth}{0.0pt}
\thispagestyle{fancy}
\lhead{}
\rhead{}
\chead{\copyright~2021 IEEE.
This is the author's version of the work. It is posted here for personal use.
Not for redistribution.	The definitive Version of Record is published in
IEEE TIFS, DOI 10.1109/TIFS.2021.3057576}
\cfoot{}

\IEEEdisplaynontitleabstractindextext
\IEEEpeerreviewmaketitle

\section{Introduction}
\label{sec:introduction}

\IEEEPARstart{T}{he} substantial and continuously increasing manufacturing costs have led most of the semiconductor industry to adopt a fabless business model. 
Leading semiconductor design houses such as Apple\textsuperscript{\textregistered} and Qualcomm\textsuperscript{\textregistered}
outsource their fabrication to off-shore foundries, which may be potentially \textit{untrustworthy}. 
Attackers present in the integrated circuit (IC) supply chain can compromise the security of the underlying hardware during fabrication, testing, assembly, and packaging.
Several hardware-focused attacks can be launched by attackers, which include (but are not limited to) reverse engineering, illegal overproduction, intellectual property (IP) piracy, and implantation of malicious circuits known as hardware Trojans~\cite{rostami2014primer}. 
Several design-for-security techniques seek to prevent IP piracy during the untrusted manufacturing stage, such as state-space obfuscation~\cite{chakraborty2009harpoon}, split manufacturing~\cite{patnaik2018concerted,patnaik2018raise}, hardware metering~\cite{alkabani2007active}, and logic locking~\cite{epic_journal,chakraborty2019keynote}. 
The prime focus of this paper is to address the shortcomings of traditional logic locking techniques in offering protection during the untrusted manufacturing stage.
The important technical terms used in the paper are defined in Table~\ref{tab:common_terms} to ease readability.

\begin{table*}[tb]
\centering
\caption{Definition of Common Terms}
\label{tab:common_terms}
\resizebox{\textwidth}{!}{
\begin{tabular}{ll}
\hline
\textbf{Term} & \textbf{Description} 
\\ \hline
Key-gate & 
\begin{tabular}[c]{@{}l@{}}A key-gate is a newly added gate and more precisely interposed into the design, driven by an original wire from the netlist and a newly introduced \textit{key-input \texttt{k}}.\\
Only when the correct \textit{key-bit value} for \texttt{k} is assigned, that key-gate would
maintain/restore the functionality of the design; otherwise, it would remain locked,\\
i.e., non-functional.
\end{tabular} 
\\ \hline
Locked netlist & A netlist where selected nets are locked using key-gates driven by key-inputs (connected to a tamper-proof memory).
\\ \hline
\textit{Oracle-less} attack & \begin{tabular}[c]{@{}l@{}}An attacker with access only to the GDSII representation of a locked design performs reverse engineering to obtain the locked netlist.\\
Therefore, the attacker needs to get around the logic locking scheme and all its key-gates focusing on structural analysis.\\
This is in contrast to the majority of attacks being \textit{oracle-guided}, where the attacker holds a working chip, essential for functional verification.
\end{tabular} 
\\ \hline
Subgraph & Locality around a key-gate; different sizes of subgraphs serve to capture different fan-in and fan-out cones of key-gates.
\\ \hline
Pre-/Post-subgraph & Pre-/Post-synthesis key-gate subgraph. These are essential to describe the obfuscation of key-gates induced by re-synthesis.
\\ \hline
\end{tabular}
}
\end{table*}

\begin{figure}[tb]
\centering
\includegraphics[width=\textwidth]{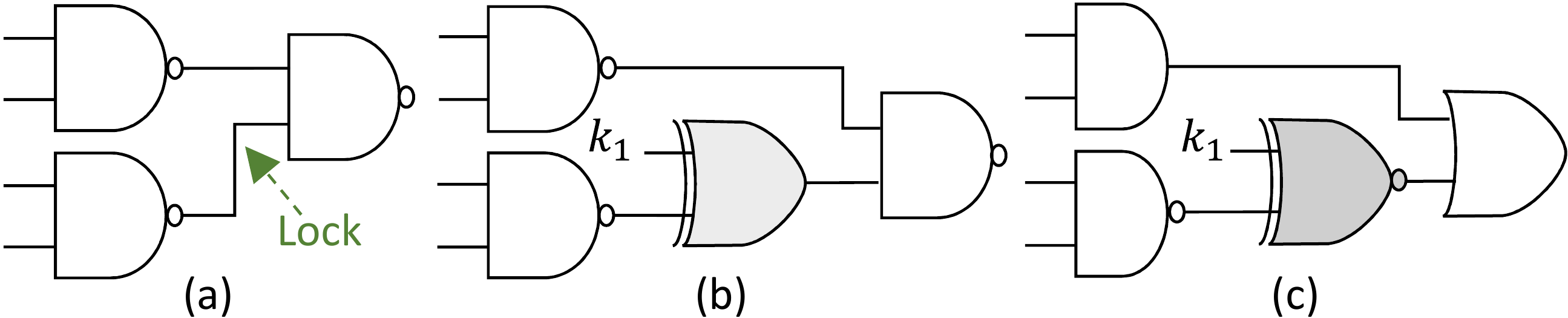}
\caption{Illustration of logic locking using X(N)OR logic gate. 
(a) Original design. 
(b) Locked design using an XOR key-gate (light gray) controlled by key-input \texttt{k1}; the correct key is \texttt{0}.
(c) Re-synthesized locked design. 
The key-gate is transformed to an XNOR key-gate (dark gray), along with localized transformations of other gates, but the correct key remains \texttt{0}.}
\label{fig:logic_locking_example}
\end{figure}

\subsection{Logic Locking}

In logic locking, additional key-gates are inserted in the original design to obfuscate its underlying functionality. 
These key-gates are controlled by key-inputs $\vec{\texttt{k}}$, driven by an on-chip tamper-proof memory. 
The locked design functions properly only after programming the correct key.
An example of logic locking is illustrated in Fig.~\ref{fig:logic_locking_example}. 
The original design is shown in Fig.~\ref{fig:logic_locking_example}(a) where a suitable place for key-gate insertion is marked. 
An XOR key-gate is inserted, which is driven by an original signal from the netlist and the newly introduced key-input \texttt{k1} as depicted in Fig.~\ref{fig:logic_locking_example}(b). 
The correct key-bit for an XOR key-gate must be \texttt{0} to maintain the original functionality of the design. 
However, as one can observe, it would be trivial for an attacker to identify the correct key owing to the one-to-one mapping between the type of key-gate and the corresponding key value.

To decorrelate this kind of inference/information leakage, logic locking schemes incur \textit{obfuscation} through iterative rounds of \textit{synthesis}.\footnote{Synthesis is a design stage which ``compiles'' an algorithmic/behavioral description to an optimized hardware implementation consisting of logic gates.} 
We refer to such an obfuscation procedure as ``re-synthesis'' throughout the paper. 
The locked and re-synthesized design is shown in Fig.~\ref{fig:logic_locking_example}(c).
Still, without any further efforts, the scale and persuasiveness of such obfuscations are subject alone to the synthesis tool, whose objectives and metrics are \emph{not} focused toward security.
These and other risks incurred by design tools during the implementation of secure schemes have been acknowledged, e.g., see~\cite{yang2019stripped,knechtel2020towards}.

\begin{figure*}[tb]
\centering
\includegraphics[width=0.93\textwidth]{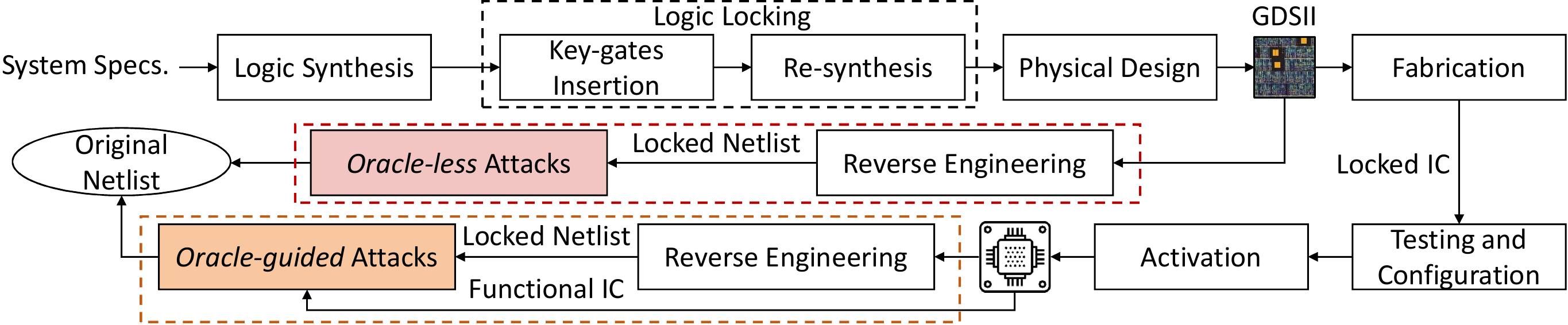}
\caption{Integration of logic locking in the IC supply chain. 
An attacker in the fab can launch an \textit{oracle-less} attack on a locked netlist, obtaining the original design. 
Such a threat is considered more dangerous than that of an \textit{oracle-guided} attack, which requires a working chip, as it can be easily
launched by further adversaries at an earlier stage in the supply chain.
The flow of an \textit{oracle-less} attack is highlighted in the red dotted rectangle. 
In contrast, the flow of an \textit{oracle-guided} attack is illustrated in the orange dotted rectangle.
In this work, we focus on \textit{oracle-less} attacks only.}
\label{fig:threat_models}
\end{figure*}

Fig.~\ref{fig:threat_models} illustrates the integration of logic locking in the IC supply chain.
Logic locking is commonly implemented on a synthesized design where the key is the designer's secret.
Post-testing, the correct key is loaded into the chips either by the design house or another trustworthy entity.

\subsection{Threat Models and Assumptions}
\label{sec:threat_models}

Most of the existing research on logic locking assumes an \textit{oracle-guided} threat model (orange dotted box in Fig.~\ref{fig:threat_models}).
In such a scenario, an attacker has access to (i)~a locked netlist and (ii)~a functional IC holding the correct key (in a tamper- and access-proof memory), useful for functional verification of any attack inference. 
An attacker can obtain the locked netlist by reverse-engineering the layout of a chip, and another chip can be used as an oracle for functional verification.
However, these chips must be obtained from the open market; they become available only sometime after fabrication. 

Another threat arises from fab-based adversaries who have access to all the structural information of an IC during manufacturing but do not possess an activated, working IC required for functional verification. 
That is, if adversaries can devise attacks using only the locked netlist, such \textit{oracle-less} attacks could compromise the security offered by logic locking very early in the supply chain, which represents a more potent threat model.
The red dotted box in Fig.~\ref{fig:threat_models} indicates the flow of an \textit{oracle-less} attack.

\subsection{Scope of This Work}
\label{sec:contributions}

This work is motivated by the recent emergence of \textit{oracle-less}, machine learning (ML)-based attacks on logic locking.
Our objective is to develop an effective technique to delineate the learning of an ML-based framework, leading to low accuracy of such otherwise powerful attacks.
We propose \textit{UNSAIL}, a defense mechanism that can be integrated with any traditional logic locking technique, to protect
locking against learning-based \textit{oracle-less} attacks, mainly the SAIL~\cite{chakraborty2018sail} and SWEEP~\cite{alaql2019sweep} attacks.

Again, the main focus of this work on \textit{UNSAIL} is to thwart \textit{oracle-less} attacks and protect the design during the untrusted manufacturing stage.
Nevertheless, \textit{UNSAIL} can be readily integrated with some SAT-attack resilient locking technique, to achieve a two-layer defense protecting against both \textit{oracle-guided} and \textit{oracle-less} attacks.

The primary contributions of this work are as follows: 

\begin{enumerate}

\item We implement a framework for the \textit{UNSAIL} defense which can be easily integrated with any combinational logic locking technique and any design-tool suite (e.g., \textit{Synopsys Design Compiler}).

\item We develop a reference framework of SAIL and implement the algorithms as outlined in~\cite{chakraborty2018sail}.

\item We perform a thorough and detailed analysis of our proposed \textit{UNSAIL} technique.
To that end, we have studied the effect of different types of key-gates, different key-sizes, and other key-gate insertion algorithms.
We also study the effect of randomizing the selection of key-gates when securing against different \textit{oracle-less} attacks. The related \textit{UNSAIL}-locked benchmarks are released in~\cite{ours}.

\end{enumerate}

Through our elaborate experimental study, the effectiveness of \textit{UNSAIL} for protecting combinational logic locking techniques against the \textit{oracle-less} ML-based SAIL~\cite{chakraborty2018sail} and SWEEP~\cite{alaql2019sweep} attacks is showcased. 
To the best of our knowledge, no other defense mechanisms have been proposed to mitigate these potent attacks. 
Additionally, our experiments also demonstrate that \textit{UNSAIL} can thwart non-ML-based \textit{oracle-less} attacks such as the redundancy attack~\cite{li2019piercing}.

More specifically, five sets of comprehensive experiments are performed to validate the effectiveness of \textit{UNSAIL}:
(i)~evaluating the change-prediction model of SAIL;
(ii)~evaluating the reconstruction model of SAIL;
(iii)~evaluating the full SAIL attack, where both models are used in conjunction;
(iv) evaluating the SWEEP attack;
and (v) evaluating the redundancy attack.
Throughout our experiments using benchmarks from the ISCAS-85 and ITC-99 suite and the OpenRISC Reference Platform System-on-Chip (ORPSoC), the
proposed defense is shown to effectively reduce the accuracy of the different ML models of SAIL by an average of \texttt{20.13}, \texttt{17}, and \texttt{11} percentage points (\texttt{pp}), respectively.
In addition, the accuracy for the SWEEP attack reduces by an average of \texttt{15pp},
showcasing that our defense is resilient against another powerful \textit{oracle-less}, learning-based attack.
The average percentage of \textit{UNSAIL}'s key-bits recovered by the redundancy attack is \texttt{38\%}, demonstrating that \textit{UNSAIL} also protects against this non-ML-based \textit{oracle-less} attack.
Layout-level evaluations show that our defense incurs minimal area and power overheads of \texttt{0.26\%} and \texttt{0.61\%}, respectively, on the million-gate ORPSoC design.

The remainder of this paper is organized as follows.   
The landscape of attack and defense strategies for logic locking is reviewed in Sec.~\ref{sec:background_logic_locking}.
Details about the relevant prior art of \textit{oracle-less} ML-based attacks on logic locking are provided in Sec.~\ref{sec:background}.
Section~\ref{sec:proposed_defense} presents the concept of our proposed \textit{UNSAIL} scheme, whereas the implementation details of both the SAIL attack and the \textit{UNSAIL} defense are given in Sec.~\ref{sec:methodology}. 
The experimental setup is described in Sec.~\ref{sec:experiments}, and our detailed experimental study is presented in Sec.~\ref{sec:security_analysis}.
Section~\ref{sec:discussion} provides a discussion, and we conclude in Sec.~\ref{sec:conclusion}.
\section{Logic Locking Techniques and Related Attacks}
\label{sec:background_logic_locking}

\subsection{Brief Overview of Logic Locking}

\textbf{Traditional Logic Locking.} Early research in logic locking focused on finding suitable places for the insertion of key-gates.
Researchers proposed several key-gate insertion algorithms such as random logic locking (RLL)~\cite{epic_journal}, fault analysis-based logic locking (FLL)~\cite{JV-Tcomp-2013}, and strong/secure logic locking (SLL)~\cite{yasin_TCAD_2016}. 
Researchers also investigated the use of different logic gates for obfuscation, e.g., X(N)OR gates~\cite{epic_journal,JV-Tcomp-2013}, AND/OR
gates~\cite{dupuis2014novel}, multiplexers (MUXes)~\cite{JV-Tcomp-2013}, etc. One advantage of such schemes is the high output corruption induced upon the application of incorrect keys.
 
\textbf{SAT-Attack Resilient Logic Locking.} An adversary with access to a functional IC can launch several attacks on the aforementioned logic locking techniques. 
The most potent \textit{oracle-guided} attack is the SAT-based attack, which compromised all existing logic locking schemes at that
time~\cite{Subramanyan_host_2015}.
In response to this powerful attack, researchers began developing SAT-attack resilient solutions such as
Anti-SAT~\cite{CHES2016YANGXIE}, stripped functionality logic locking (SFLL)~\cite{yasin_CCS_2017}, and SFLL-fault~\cite{sengupta2018atpg}. With each newly introduced SAT-attack resilient scheme, tailored \textit{oracle-guided} attacks emerged, such as \textit{Double-DIP}~\cite{shen2017double}, \textit{AppSAT}~\cite{shamsi2017appsat}, etc.

\subsection{\textit{Oracle-Less} Attacks on Logic Locking}
\label{sec:oracle_less_attacks}

The initial research was primarily focused on protecting logic locking from the SAT-based~\cite{Subramanyan_host_2015} and other derivative attacks~\cite{scansat,scansat2}, which require an
oracle.
Recently, various \textit{oracle-less} attacks have been proposed that rely only on structural properties of the locked design. 
They can be classified as follows.

\begin{figure*}[tb]
\centering
\includegraphics[width=0.9\textwidth]{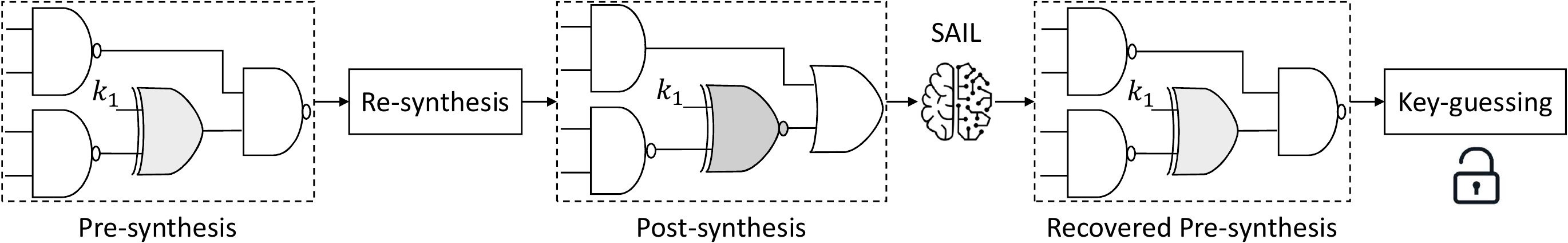}
\caption{High-level explanation for SAIL~\cite{chakraborty2018sail}. 
SAIL seeks to learn the structural changes induced by re-synthesis.
Such a trained SAIL model can then revert the changes and recover the pre-synthesis locked design, where the correct key-bit values can be obtained by analyzing the types of key-gates.}
\label{fig:SAIL_idea_final}
\end{figure*}

\textbf{\textit{Oracle-Less} Attacks on SAT-Attack Resilient Logic Locking.} Such attacks focus on structural properties of SAT-attack resilient techniques and attempt to circumvent their security promise by identifying and removing the added protection logic, thereby isolating the original circuit cone. 
Examples include the SFLL-hd--unlocked attack~\cite{yang2019stripped}, the functional reverse engineering-based attack~\cite{alrahis2019functional}, and the functional analysis attacks~\cite{sirone2020functional}. Most recently, GNNUnlock~\cite{alrahis2020gnnunlock} was proposed as a holistic attack on SAT-attack resilient logic locking, breaking several schemes under different parameters and synthesis settings. One of the drawbacks of SAT-attack resilient locking techniques~\cite{CHES2016YANGXIE,yasin_CCS_2017,sengupta2018atpg} is that they thwart the SAT-based attack~\cite{Subramanyan_host_2015} by inducing a low output corruption for incorrect key-assignments. 
As a result, SAT-attack resilient locking techniques are usually integrated with a high-corruptibility locking technique such as RLL or FLL~\cite{yasin_CCS_2017}, to achieve security against removal attacks. 
Such an integration is commonly referred to as \textit{two-layer locking}.
In order to recover the original design, both layers must be broken.

\textbf{\textit{Oracle-Less} Attacks on Traditional Logic Locking.} This category includes the de-synthesis attack~\cite{massad2017logic}, 
the redundancy attack~\cite{li2019piercing}, and the ML-based SAIL~\cite{chakraborty2018sail} and SWEEP~\cite{alaql2019sweep} attacks, all
of which target traditional logic locking.
For example, with the redundancy attack by Li \textit{et al.}~\cite{li2019piercing}, the authors observed that incorrect assignment(s) of the key-bit value(s) result in more redundancies in the netlist, enabling them to prune out incorrect key-assignments.

Most relevant for this work, SAIL~\cite{chakraborty2018sail} leverages ML to learn localized structural changes induced by re-synthesis for obfuscation of logic locking (see Fig.~\ref{fig:SAIL_idea_final}).
SWEEP~\cite{alaql2019sweep} utilizes a feature weighting algorithm to learn and perform a mapping between design features and the correct key.
Both these attacks are discussed in more detail next.
\section{Prior Work on Oracle-Less ML-Based Attacks on Traditional Logic Locking}
\label{sec:background}
\label{sec:SAIL_background}

\begin{figure}[tb]
\centering
\includegraphics[width=\textwidth]{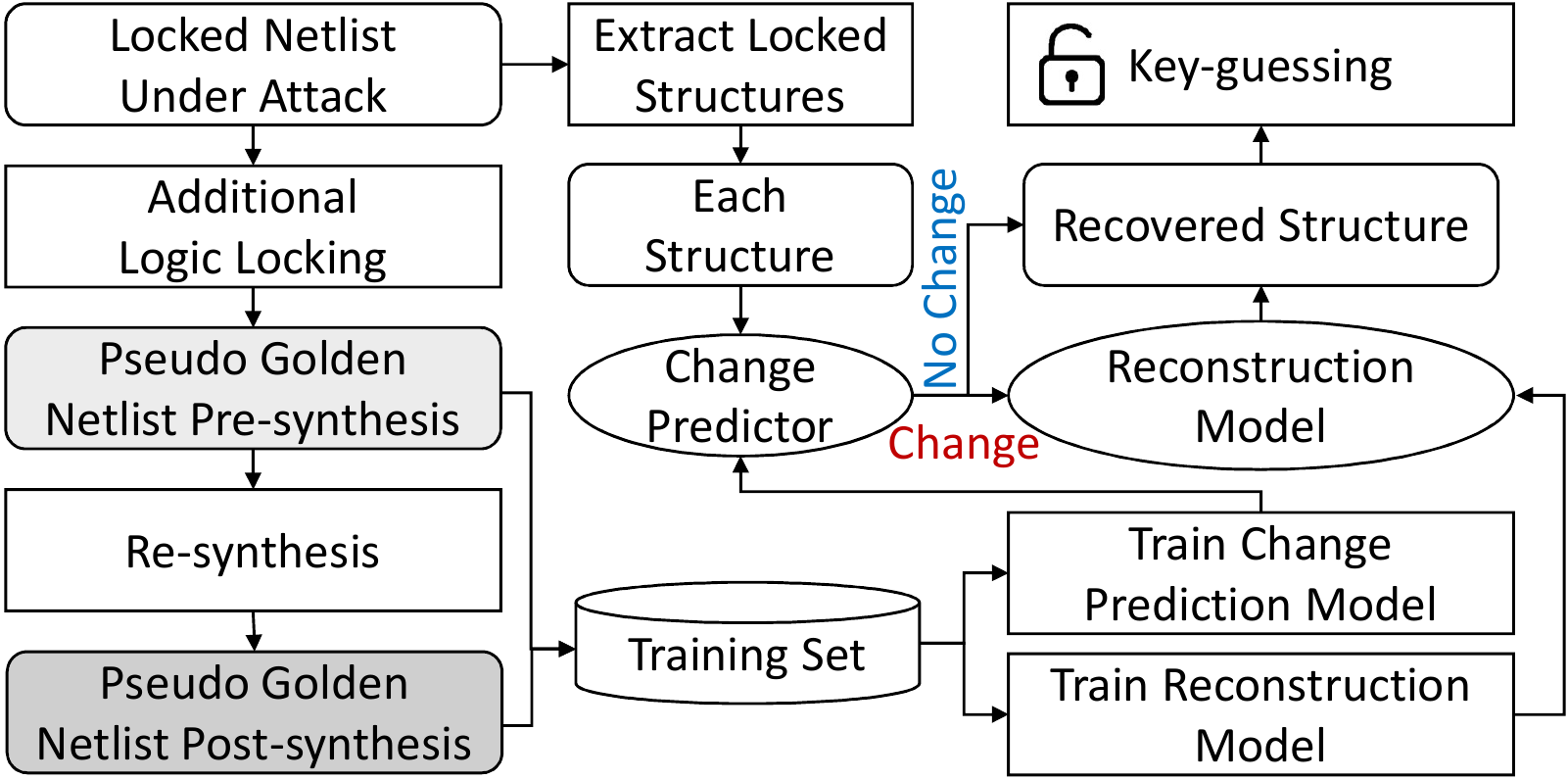}
\caption{The SAIL attack flow~\cite{chakraborty2018sail}.}
\label{fig:SAIL_attack_flow}
\end{figure}

\subsection{SAIL~\cite{chakraborty2018sail}}

In SAIL~\cite{chakraborty2018sail}, the authors have shown that ML models can learn and revert the structural changes induced by re-synthesis, for traditional logic locking techniques (e.g., RLL, FLL, and SLL).
Often there is a direct correspondence between the type of key-gate and the
key-bit value, as illustrated earlier in Fig.~\ref{fig:logic_locking_example}.
Hence, logic locking techniques seek to apply structural changes for all the key-gates using iterative re-synthesis, thereby obfuscating the correspondence between the type of key-gate and the key-bit value.

Utilizing the locked design $\texttt{C'}(\vec{\texttt{x}},\vec{\texttt{k}})$, SAIL seeks to recover the original netlist
$\texttt{C}(\vec{\texttt{x}})$, where $\vec{\texttt{x}}$ is the vector of regular inputs and $\vec{\texttt{k}}$ is the vector of additional key-inputs. 
The notion of SAIL is to learn the structural changes introduced on key-gates by deterministic design tools, logic synthesis in particular.
The authors of~\cite{chakraborty2018sail} have shown that SAIL succeeds in (i)~retrieving the locked design before synthesis, and (ii)~obtaining the key values by analyzing the type of retrieved key-gates. 
A conceptual example is illustrated in Fig.~\ref{fig:SAIL_idea_final}.

The flow of SAIL is shown in Fig.~\ref{fig:SAIL_attack_flow}. 
In addition to the locked netlist $\texttt{C'}$, the attacker also requires knowledge of the underlying logic locking technique $\texttt{C'}=\texttt{lock}(\texttt{C},\vec{\texttt{k}})$ and the synthesis setup. 
To generate training data, another round of locking is first implemented on top of the locked design, providing $\texttt{C''}(\vec{\texttt{x}},\vec{\texttt{k}},\vec{\texttt{k}_{\texttt{1}}})=\texttt{lock}(\texttt{C'},\vec{\texttt{k}_{\texttt{1}}})$. 
Local structures around key-gates, which are identified by their connection to key-inputs $\vec{\texttt{k}_{\texttt{1}}}$, are extracted as \textit{pre-synthesis subgraphs} or \textit{pre-subgraphs} for short, denoted as set $\texttt{S}$. 
Next, the design $\texttt{C''}$ is synthesized and, similarly, structures around key-gates connected to $\vec{\texttt{k}_{\texttt{1}}}$ are extracted as \textit{post-synthesis subgraphs} or \textit{post-subgraphs} for short, denoted as set $\texttt{S'}$. 
This procedure is repeated multiple times to generate an extensive data set of pre-subgraphs and post-subgraphs of various degrees.\footnote{%
Some examples of subgraphs can be found in Fig.~\ref{fig:UNSAIL} and Fig.~\ref{fig:unsail_nopatch}.}
This data set is then used to train two ML models: $\texttt{ML}_{\texttt{1}}$, a change-prediction model and $\texttt{ML}_{\texttt{2}}$, a reconstruction model. 
Given a key-gate and some surrounding structure from the locked design under attack, $\texttt{ML}_{\texttt{1}}$ shall predict whether the related subgraph went through a structural change due to re-synthesis and, if any change is predicted, $\texttt{ML}_{\texttt{2}}$ shall revert the change, providing the original, un-obfuscated key-gate.
More details for the two ML models are discussed next.

\textbf{Change-Prediction Model $\texttt{ML}_{\texttt{1}}$.} This model is built using the notion of Random Forest (RF), an ensemble-based model scheme used for classification or regression problems.
In general, an RF is composed of multiple decision trees trained independently, and the model outputs the most voted class by individual trees.
For SAIL, the sets \texttt{S} and \texttt{S'} are to be compared, providing a Boolean change indicator. 
The set [\texttt{S'}, change indicator] is then provided to the RF classifier in which each decision tree is trained using separately
bootstrapped samples from the data set~\cite{efron1992bootstrap}.
Moreover, a subset of attributes is randomly chosen from the available attributes to split each tree at each node, as is common practice for RF~\cite{breiman2001random}.
However, further details, such as the number of decision trees employed, have not been provided in~\cite{chakraborty2018sail}.

\textbf{Reconstruction Model $\texttt{ML}_{\texttt{2}}$.} This model consists of a multi-layer, multi-channel neural network.
The model is trained using the set [\texttt{S'},\texttt{S}]; however, the details of the network structure, training algorithm, or weight initialization have not been provided in~\cite{chakraborty2018sail}.
Generating the training data and conducting the training procedure is to be conducted again for each newly introduced design and each newly
introduced logic locking scheme, while the form of training data remains the same for both $\texttt{ML}_{\texttt{1}}$ and $\texttt{ML}_{\texttt{2}}$.

\subsection{SWEEP~\cite{alaql2019sweep}}
\label{sec:SWEEP_background}

SWEEP~\cite{alaql2019sweep} is a constant-propagation attack developed to circumvent MUX-based logic locking~\cite{JV-Tcomp-2013}.
In MUX-based logic locking, the key-gates are commonly \texttt{2:1} MUXes, where the ``true input'' is connected with the intended signal of the original design, while the ``false input'' is connected to another signal of the original design~\cite{JV-Tcomp-2013}. 
As the true input can be easily made either to be the first or the second input pin of the MUX,
the key-bit provided at the \textit{select} pin of the MUX can be accordingly either \texttt{0} or \texttt{1}---there is no inherent information leakage as there is no fixed correspondence of correct key-bit values and MUX key-gates.

Still, SWEEP succeeded in learning the synthesis-induced structural changes as follows.
The attack is made aware of (i)~the obfuscation algorithm leveraged for selecting signals for true and false inputs, and the (ii)~locked design.
The attack performs a training stage in which the obfuscation algorithm is utilized to generate some locked designs with known correct key-bit values. 
Next, an iterative procedure is followed. 
Each key-input is visited twice, setting the correct/incorrect key value as a constant and synthesizing the locked design for both cases.
A set of features is extracted from the synthesis reports obtained for both cases (correct versus incorrect key-bit assignment). 
A feature weighting algorithm is utilized to learn the correlation between the extracted features and the correct key-bit values. 
Once training is completed, the same constant-propagation technique used in the training stage is utilized to extract the features from the design under attack. 
Finally, the correct key-bit values are identified using the weighted function generated from the training data.
\section{Proposed Defense}
\label{sec:proposed_defense}

\begin{figure}[tb]
\centering
\includegraphics[width=0.9\textwidth]{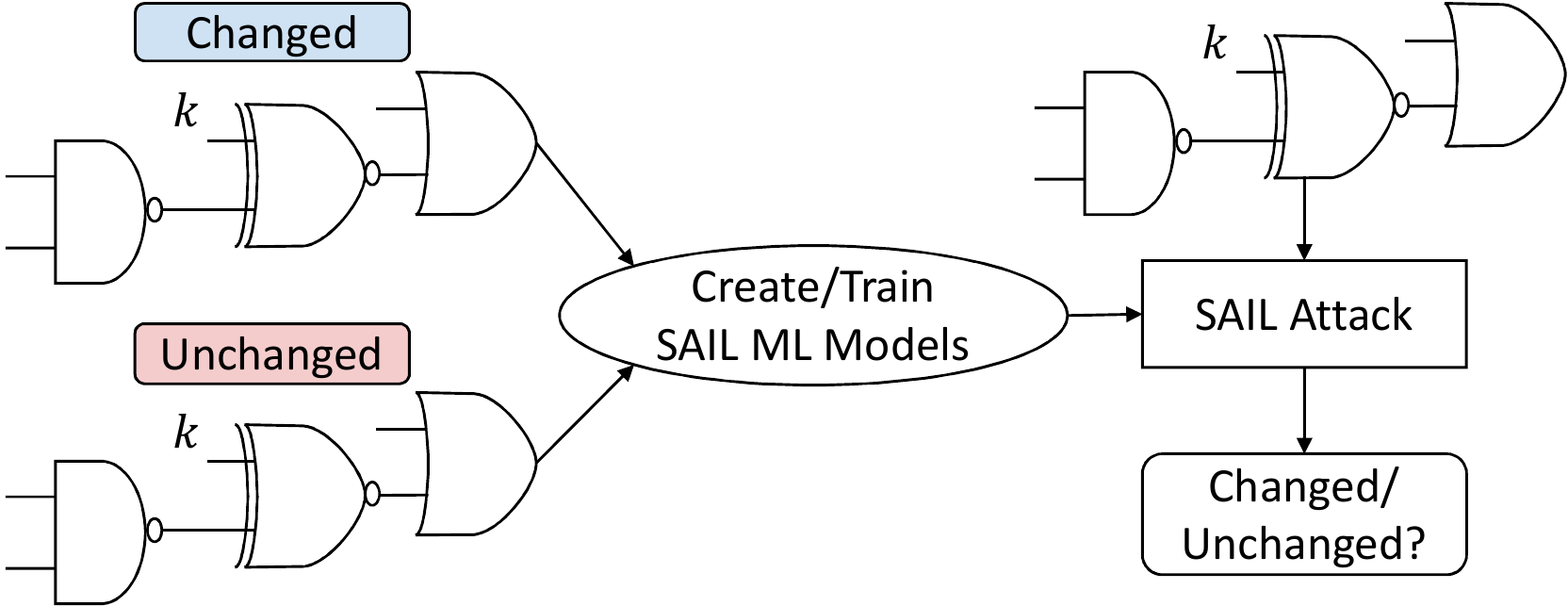}
\caption{High-level concept of \textit{UNSAIL}.
Left: training SAIL models after incorporating \textit{UNSAIL}. 
The same subgraph has two distinct labels, fundamentally undermining training efforts for SAIL.
Right: accordingly, SAIL cannot determine whether the subgraph went through change or not, which is an essential step for the attack's efficacy.}
\label{fig:UNSAIL}
\end{figure}

In this work, we propose \textit{UNSAIL}, a defense scheme that aims to confuse \textit{oracle-less} ML-based attacks on logic locking,
   particularly but not exclusively the ML models used in SAIL.

In general, the quality of training data determines the accuracy and performance of any ML system. 
Thus, a specific goal for \textit{UNSAIL} is to
inject ``bad data'' during learning.
As shown in Fig.~\ref{fig:UNSAIL}, our key idea is to replicate data that evokes different attack responses.
Specifically, we introduce identical subgraphs in the final locked netlist that could readily be classified as either ``\textit{Changed}'' or
``\textit{Unchanged},'' thereby confusing both $\texttt{ML}_{\texttt{1}}$ and $\texttt{ML}_{\texttt{2}}$ of SAIL.
In other words, the ``bad data'' injected by \textit{UNSAIL} propagates through SAIL, feeding into its ML models and inducing flawed inferences.

\begin{figure*}[tb]
\centering
\includegraphics[width=0.9\textwidth]{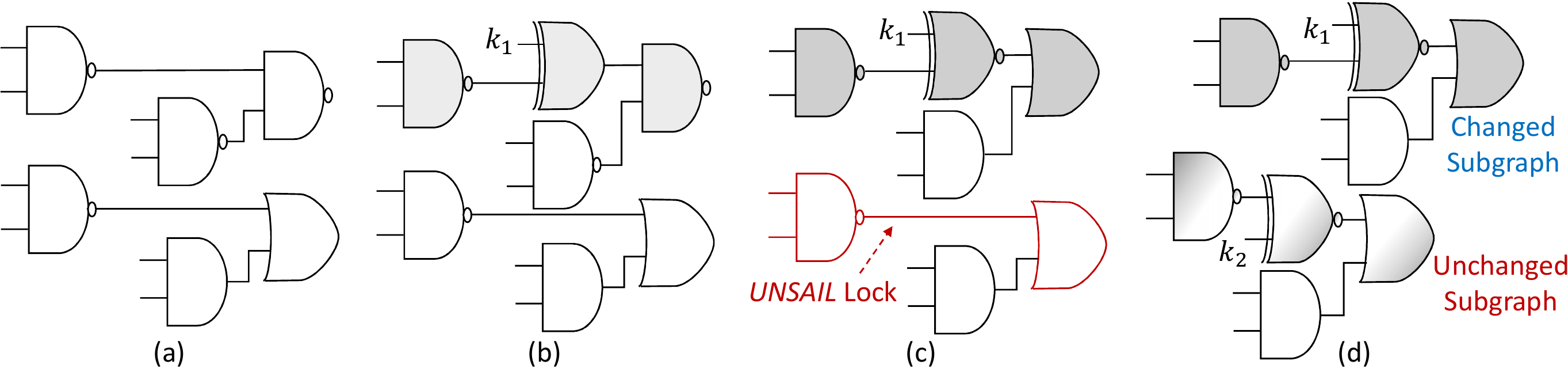}
\caption{Example of \textit{UNSAIL} integrated with X(N)OR logic locking scheme. 
(a) Original design. 
(b) Locked design using one XOR key-gate; the correct key value is \texttt{k1=0}. 
The gates highlighted in light gray represent a pre-subgraph of size \texttt{sub=3}. 
(c) Post-synthesized design. 
Based on the observed subgraph, \textit{UNSAIL} searches for a \textit{NAND-OR} structure suitable to insert an additional, corresponding key-gate, namely an XNOR.
(d) Final locked design with two identical subgraphs, one generated by synthesis and one added by \textit{UNSAIL}. 
The correct keys are \texttt{k1=0, k2=1}.}
\label{fig:unsail_nopatch}
\end{figure*}

Next, we outline the working principle of \textit{UNSAIL}.
A motivational example is shown in Fig.~\ref{fig:unsail_nopatch}. Note that more implementation details for \textit{UNSAIL} are also given in Sec.~\ref{sec:UNSAIL_implementation}.

First, a design is locked traditionally, using any combinational logic locking technique of choice, with only a subset of all the desired key-gates being inserted at this stage (say, \texttt{K/2}). 
This locked design is then passed through synthesis, which transforms some of the key-gates and surrounding structures (i.e., the subgraphs).
For the remaining key-gates (say, the other \texttt{K/2}), \textit{UNSAIL} then carefully ``injects'' identical subgraphs in that post-synthesized design.
Essentially, there are two parts to this.
First, we tackle the subgraphs, which remained unchanged during synthesis.
For that, we want to revisit the synthesis stage and add transformable \textit{UNSAIL} key-gate structures matching to the specific sets of gates which remained unchanged so far. 
In other words, we want synthesis to work on \textit{UNSAIL} structures, which will then undergo changes.
Second, we can achieve a similar effect for the structures that already went through change due to the earlier synthesis step by adding the same structures in the post-synthesized design. 
The newly added structures are not generated by the synthesis tool and did not undergo any structural change. 
Since the resulting \textit{UNSAIL}-locked design now contains genuinely transformed subgraphs and ``injected'' ones in sufficiently large numbers, SAIL will fail to identify structural changes with confidence.
\section{Methodology}
\label{sec:methodology}

\subsection{Implementation of SAIL} The concept of SAIL was presented in~\cite{chakraborty2018sail}; we have also reviewed it in Sec.~\ref{sec:SAIL_background}. 
However, the precise setup details were not provided in~\cite{chakraborty2018sail}.
Thus, in this work, we implement SAIL according to the best of our understanding.
As it is forming an essential part as a baseline for our work, we discuss our SAIL implementation in some detail.

We encode the subgraphs as vectors that describe the order of gates in the subgraph structures.
We consider the following sizes of subgraphs, or \textit{sub-sizes} for short: \texttt{sub=3}, \texttt{sub=5}, and \texttt{sub=6}.
A subgraph is extracted and encoded as an individual vector for each key-input in the locked design for all sub-sizes. 
For example, if the design is locked using \texttt{K=64} and sub-sizes of \texttt{sub=3}, \texttt{sub=5}, and \texttt{sub=6} are considered, then \texttt{64}
subgraphs/vectors are extracted from the locked design for each sub-size, or \texttt{192} subgraphs/vectors in total.
We note that in SAIL, the authors considered sub-sizes from \texttt{sub=3} up to \texttt{sub=10}, and they observed that (a)~the accuracy of the SAIL classifier increased with
the sub-size but also (b)~the average accuracy saturated for sub-size of \texttt{sub=5} to \texttt{sub=6}~\cite{chakraborty2018sail}.

After extracting all vectors, we apply a one-hot encoding on them, and we feed the encoded vectors to a classifier model $\texttt{ML}_{\texttt{1}}$ that
we built as described in~\cite{chakraborty2018sail} and reviewed in Sec.~\ref{sec:SAIL_background}. Concerning the reconstruction model
$\texttt{ML}_{\texttt{2}}$, the authors in~\cite{chakraborty2018sail} described it merely as a multi-input multi-channel network; however, no details were given regarding the type of the network used nor regarding its dimensions. 
In order to reproduce the results of SAIL, we had to implement and test several network types for $\texttt{ML}_{\texttt{2}}$, such as feedforward neural network and
recurrent neural network.
The model that showed the best results was a sequence-to-sequence (Seq2Seq) encoder-decoder model with attention.\footnote{In general, such models consist
of an encoder processing the input and a decoder processing the output, while the attention mechanism serves for the
decoder to focus on the relevant parts of the encoded input when generating the translation.
Such models are commonly used in automated translation and, for our work,
the goal is to translate the post-subgraphs to pre-subgraphs.}
We implemented the encoder using an embedding layer and two long short-term memory (LSTM) layers.
The embedding converts the textual encoded vectors (representing subgraphs)
into vectors of real numbers. The decoder model is implemented using two LSTMs with attention. $\texttt{ML}_{\texttt{2}}$ is an ensemble model;
hence, we have trained three such models for the different sub-sizes considered (\texttt{sub=3}, \texttt{sub=5}, and
\texttt{sub=6}). The models were combined using a cumulative voting scheme.
More implementation and setup details are also provided in Sec.~\ref{sec:experiments}.

\subsection{Implementation of \textit{UNSAIL}}
\label{sec:UNSAIL_implementation}

\begin{figure}[tb]
\centering
\includegraphics[width=0.95\textwidth]{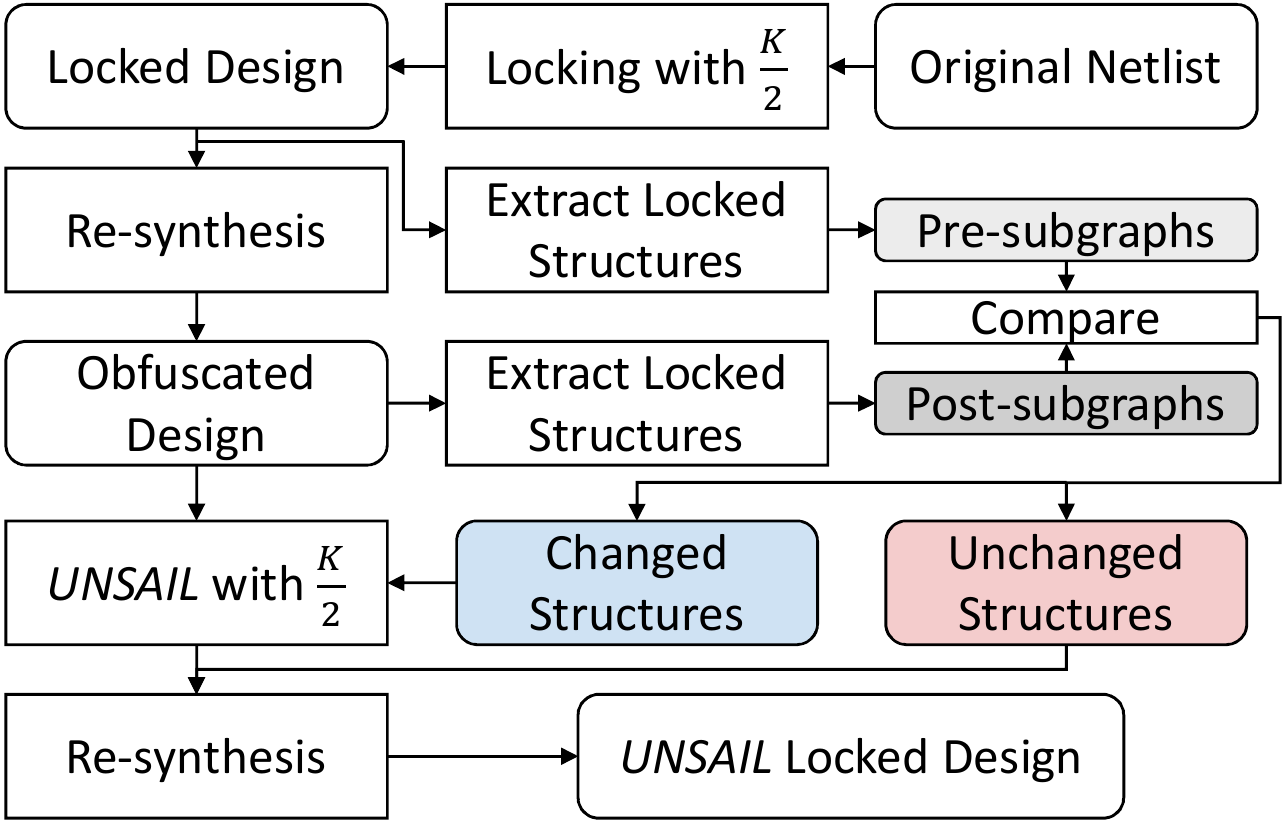}
\caption{Integration of \textit{UNSAIL} with combinational logic locking.}
\label{fig:unsail_flow}
\end{figure}

Fig.~\ref{fig:unsail_flow} illustrates the \textit{UNSAIL} flow, which can be integrated with any combinational logic locking technique.
To protect the original design \texttt{C} using a key-size of \texttt{K}, \texttt{C} is first locked using $\texttt{K}/\texttt{2}$ key-bits
following the logic locking technique of the designer's choice.
Next, the locked design \texttt{C'} is passed through synthesis, to obtain an obfuscated design $\texttt{C'}_{\texttt{o}}$. 
The local structures around all key-gates (i.e., subgraphs) are obtained from both \texttt{C'} and $\texttt{C'}_{\texttt{o}}$ and compared to
detect any changes induced by re-synthesis. 
The subgraphs
that went through synthesis-induced changes are selected and stored in a dictionary data structure.
Note that subgraphs which did not go through changes during synthesis are also stored for later use, in a separate set \texttt{U}.

Next, the dictionary is used to guide the insertion of the remaining $\texttt{K}/\texttt{2}$ key-gates.
The search is carried out until all the remaining key-bits are assigned with key-gates.
Note that, if some particular entry cannot be found in the locked netlist, we can easily ``fill-up'' remaining key-bits by leveraging some more
instances of other entries previously found.
Some of the \textit{UNSAIL}-locked structures are synthesized to the specific set of gates in \texttt{U} to confuse the learning of the ML-models further.

\begin{figure*}[tb]
\centering
\includegraphics[width=.945\textwidth]{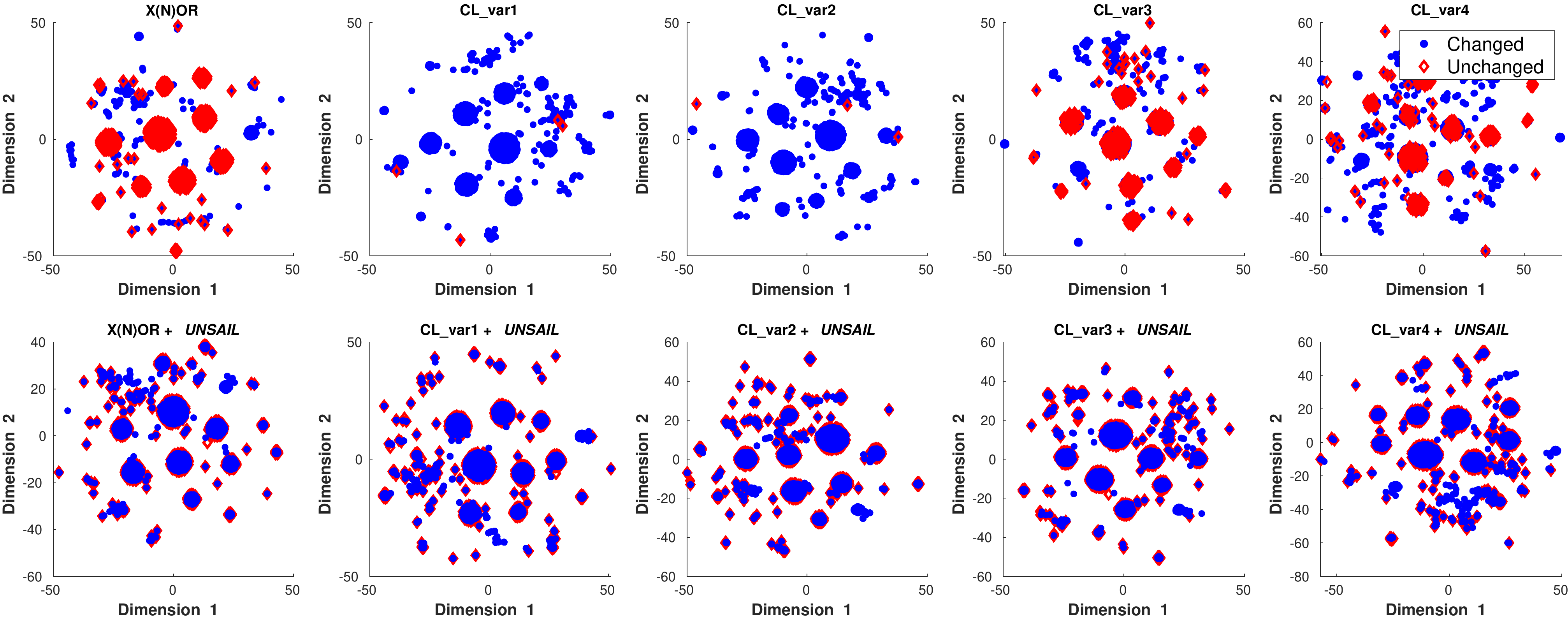}
\caption{Visualizing the pre-subgraph and post-subgraph data using t-distributed stochastic neighbor embedding~\cite{maaten2008visualizing}.
(Top) Aggressive synthesis optimization is performed on the RLL benchmarks and, thus, in most cases, the ``\textit{Changed}'' and
``\textit{Unchanged} post-synthesis subgraphs can be clearly differentiated, which is the main requirement for SAIL.
(Bottom) Once \textit{UNSAIL} is incorporated, these classes are largely overlapping, and the more substantial the overlap, the larger the
complexity for the classification problem in general, i.e., for any classifier model, including the one used for SAIL.
}
\label{fig:tsne}
\end{figure*}

In this work, the subgraph extraction, \textit{UNSAIL} key-gates insertion stages, and RLL are implemented using \textit{Perl} scripts that operate directly on \textit{Verilog} netlists.

\subsection{Scope and Effect of \textit{UNSAIL}}
\label{sec:effect_of_UNSAIL}

We emphasize that for SAIL~\cite{chakraborty2018sail}, only X(N)OR-based logic locking is evaluated. 
In our work, however, we thoroughly investigate the impact of various types of key-gates (Table~\ref{tab:variations}).

The motivation for considering different structures is as follows.
X(N)OR key-gates, which are commonly applied for many logic locking schemes, can be replaced by MUXes, which are more resilient against most attacks when compared to simple gates (as also indicated in Sec.~\ref{sec:SWEEP_background}).
For such replacement, the locked net and its negated signal would be connected to the MUX inputs in one of the two possible orders, and the resulting key-bit is connected to the MUX select line (see also Fig.~\ref{fig:MUX_locking_example}(a) on page~\pageref{fig:MUX_locking_example}).
However, given that the negated signal might be easy to identify from the netlist structure, such an otherwise resilient MUX could be tackled by SAIL and similar attacks. 
Thus, we advocate to vary and mix different types of key-gates, and we study the effect of such \textit{compound locking} (CL).
Note that, when locking the designs with the CL variations (Table~\ref{tab:variations}), the type of the key-gate is chosen randomly from the set of key-gates. 
We perform such random selection to break the \textit{deterministic} nature of the mapping problem targeted at by the ML models of SAIL.

\begin{table}[tb]
\centering
\scriptsize
\caption{Variations of Key-Gates Used in This Work}
\label{tab:variations}
\begin{tabular}{@{}cc@{}}
\hline
\textbf{Locking Variations} & \textbf{Types of Key-Gates} 
\\ \hline
X(N)OR & X(N)OR key-gates 
\\ \hline
CL\_v1 & \begin{tabular}[c]{@{}c@{}}Multiplexers key-gates constructed using AND, OR gates\\ 
\& multiplexers key-gates constructed using NAND gates
\end{tabular} 
\\ \hline
CL\_v2 & \begin{tabular}[c]{@{}c@{}}
Multiplexers key-gates constructed using NOR gates\\
\& CL\_v1
\end{tabular} 
\\ \hline
CL\_v3 & \begin{tabular}[c]{@{}c@{}}
X(N)OR key-gates\\
\& CL\_v2
\end{tabular} 
\\ \hline
CL\_v4 & \begin{tabular}[c]{@{}c@{}}
AND/OR key-gates\\
\& CL\_v3
\end{tabular} 
\\ \hline
\end{tabular}\\[1em]
CL is short for \textit{compound locking} where a mix of various key-gates are used.
\end{table}

Next, we conduct an exploratory experiment on two cases for the ITC-99 benchmark \texttt{b17\_C} to understand the impact of \textit{UNSAIL} on the final structure of locked designs.
For case a), we lock the benchmark with \texttt{K=512}, using only RLL, but considering all the different key-gate structures listed in Table~\ref{tab:variations}.
For each structure considered, \texttt{20} instances of RLL designs are generated. 
For case b), we lock the benchmark using both RLL and \textit{UNSAIL}; each technique is employed to realize \texttt{256} key-bits, resulting in \texttt{K=512}. 
As in a), we consider all the different structures, and \texttt{20} locked instances are generated for each of the structures.

For both cases, first, the post-subgraphs and pre-subgraphs are extracted.
The post-subgraphs that went through change due to synthesis are labeled as ``\textit{Changed}'' and those unaltered as ``\textit{Unchanged}.''
The related data is then projected non-linearly to 2D using t-distributed stochastic neighbor embedding (t-SNE)~\cite{maaten2008visualizing}.
Fig.~\ref{fig:tsne}(top) represents the data set for RLL traditionally employed---for most key-gate structures; one cluster is dominant, namely that for ``\textit{Changed}'' post-subgraphs.
Applying \textit{UNSAIL}, however, can render such classification significantly more difficult---clusters are primarily overlapping, as shown in Fig.~\ref{fig:tsne}(bottom). 
\ul{In short, enhancing logic locking through \textit{UNSAIL}, we can expect a large overlap between classes, essentially ensuring ``bad data,'' thereby hindering appropriate training of SAIL (or, for that matter, any ML-based attack on structural properties of locked netlists).}

To further quantify the difficulty of separating/learning the classes, we study the classification accuracy in detail.
Although we investigate various classification models, as discussed in more detail in Sec.~\ref{sec:accuracy}, the robustness of such insights
depends on the classifier choices/parameters.
Thus, our first goal is to support our claim that \textit{UNSAIL} can incur a difficult classification problem for any classifier type.

The related notion of meta-analysis of supervised ML models is a research area that aims to correlate the inherent complexity of a dataset with the
performance of the classifiers~\cite{kalousis2004data}.
Several metrics, including the maximum \textit{Fisher's discriminant} ratio \texttt{F1}~\cite{ho2001data}, have been proposed to characterize the
classification complexity inherent to datasets~\cite{Complexity_measures,lorena2019complex}. \texttt{F1} has
been shown to be effective for quantifying the difficulty in separating the data into corresponding classes and, hence, in portraying the complexity of the respective classification problem~\cite{Complexity_measures,lorena2019complex,example_f1}. 
In general, \textit{Fisher's discriminant} ratio \texttt{f} measures how strongly two classes differ along with a specific feature and is defined as follows:
\begin{equation*}
 \texttt{f}=\dfrac{(\mu_{\texttt{1}}\texttt{-}\mu_{\texttt{2}})^{\texttt{2}}}{(\sigma_{\texttt{1}})^{\texttt{2}}+(\sigma_{\texttt{2}})^{\texttt{2}}}
\end{equation*}
where \texttt{$\mu_{x}$} represent the mean of the feature values for class \texttt{x} and \texttt{$\sigma_{x}$} represents the standard deviation of the feature values. 
The range of the ratio is $[\texttt{0}\longrightarrow\infty]$. 
A small ratio indicates a substantial overlap between the classes.
The larger the ratio, the easier is the separation of the two classes using that feature/attribute. 
Hence, to measure the overlap between two classes in general, \texttt{f} is calculated for all of the considered features. 
Then the maximum ratio \texttt{F1} is selected to judge the separability of the classes.

It is expected that \texttt{F1} will be lower for \textit{UNSAIL} when compared to RLL.
We quantify the \texttt{F1} ratio for both \textit{UNSAIL} and RLL on selected ITC-99 benchmarks for \texttt{K=512} and \texttt{sub=3} (Table~\ref{tab:F1_512}).
Indeed, the results support our claim: \ul{On average, \textit{UNSAIL} achieves a \texttt{55.72\%} reduction in the \texttt{F1} ratio, which translates to complex classification problems in general.}
It is expected that the effect of \textit{UNSAIL} on the change-prediction classification should be more prominent for those cases where a significant reduction can be observed, i.e., for particular flavors of compound locking. 
This is verified in Sec.~\ref{sec:accuracy}, as the results obtained there show that compound-based \textit{UNSAIL} locking affects the classification stage more than X(N)OR-based \textit{UNSAIL} locking.

\begin{table}[tb]
\centering
\scriptsize
\setlength{\tabcolsep}{0.25mm}
\caption{Maximum Fisher's Discriminant for \texttt{K=512} and \texttt{sub=3} on Selected ITC-99 Benchmarks}
\label{tab:F1_512}
\begin{tabular}{*{12}{c}}
\hline
\textbf{Key-gates}
& \multicolumn{2}{c}{\textbf{X(N)OR}} 
& \multicolumn{2}{c}{\textbf{CL\_v1}} 
& \multicolumn{2}{c}{\textbf{CL\_v2}} 
& \multicolumn{2}{c}{\textbf{CL\_v3}} 
& \multicolumn{2}{c}{\textbf{CL\_v4}} 
\\ \hline
\textbf{Insertion}
& \textbf{RLL} & \textbf{\textit{UNSAIL}}
& \textbf{RLL} & \textbf{\textit{UNSAIL}} 
& \textbf{RLL} & \textbf{\textit{UNSAIL}} 
& \textbf{RLL} & \textbf{\textit{UNSAIL}} 
& \textbf{RLL} & \textbf{\textit{UNSAIL}} 
\\ \hline
b14\_C & 
2.15 & 1.03 & 
 
2.13 & 0.57 & 
 
3.75 & 0.66 & 
 
6.37 & 0.71 & 
 
1.41 & 0.90 
\\ \hline
b15\_C & 
1.30 & 0.85 & 
 
2.06 & 0.61 & 
 
3.73 & 0.75 & 
 
5.47 & 0.80 & 
 
1.04 & 0.65 
\\ \hline
b20\_C & 
1.60 &0.90 & 
 
1.98 & 0.60 & 
 
4.75 & 0.82 & 
 
2.64 & 0.84 & 
 
1.22 & 0.98 
\\ \hline
b21\_C & 
1.56 &2.00 & 
 
2.00 & 0.65 & 
 
4.75 & 0.63 & 
 
3.09 & 2.00 & 
 
2.00 & 2.00 
\\ \hline
b22\_C & 
1.76 & 0.98 & 
 
1.94 & 0.62 & 
 
3.01 & 0.86 & 
 
5.61 & 0.62 & 
 
1.21 & 1.00 
\\ \hline
b17\_C & 
0.97 & 0.83 & 
 
2.04 & 0.69 & 
 
3.95 & 0.45 & 
 
9.60 & 0.87 & 
 
2.00 & 2.00 
\\ \hline
\textbf{Average} 
& \textbf{1.56} & \textbf{1.10} & 
 
\textbf{2.03} & \textbf{0.62} & 
 
\textbf{3.99} & \textbf{0.70} & 
 
\textbf{5.46} & \textbf{0.97} & 
 
\textbf{1.48} & \textbf{1.25} 
\\ \hline
\end{tabular}
\end{table}
\section{Experimental Setup}
\label{sec:experiments}

\begin{figure}[tb]
\centering
\includegraphics[width=0.9\textwidth]{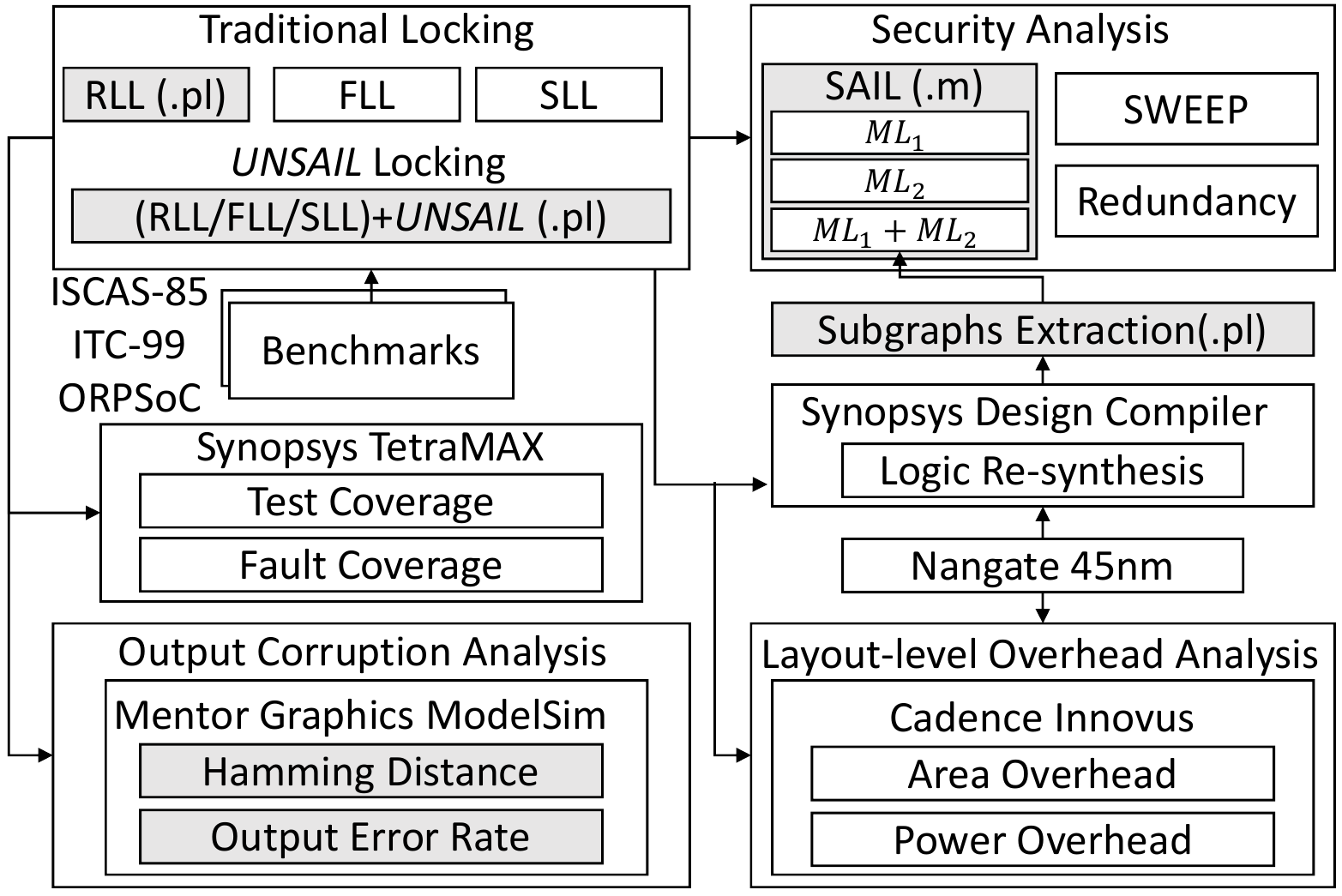}
\caption{The various components employed in our study. Gray colored boxes represent scripts developed/implemented in-house.}
\label{fig:Setup}
\end{figure}

Here, we provide the details regarding the experimental setup followed in our work. 
See also Fig.~\ref{fig:Setup} for an overview.

\subsection{Test Cases}

We study the effectiveness of \textit{UNSAIL} on eight combinational benchmarks from the ISCAS-85 suite and six combinational benchmarks from the ITC-99 suite.
Similar to the exploratory experiment in Sec.~\ref{sec:effect_of_UNSAIL}, we consider two cases: a)~RLL, and b)~\textit{UNSAIL} based on RLL. 
For both cases, each benchmark is locked using all the different key-gate structures in Table~\ref{tab:variations}.
For case a), ISCAS-85 benchmarks are locked with \texttt{K=64} and \texttt{K=128}, respectively, while ITC-99 benchmarks are locked with \texttt{K=256} and \texttt{K=512}, respectively.
Moreover, each benchmark is locked \texttt{20} times for each structure, resulting in a total of \texttt{2,800} RLL instances. 
For each structure, key-size, and benchmark, one of these \texttt{20} locked instances is considered as circuit under attack and excluded from the training set, resulting in a total of \texttt{140} attacked RLL instances.
For case b), same benchmarks are locked, considering the same parameters as above, while following the \textit{UNSAIL} procedure.

Besides leveraging RLL, we also integrate \textit{UNSAIL} with FLL~\cite{JV-Tcomp-2013} and SLL~\cite{yasin_TCAD_2016}, respectively. Since the aforementioned logic locking techniques are essentially X(N)OR-based logic locking, we train the model using RLL-based logic locking and launch the attack on the FLL-based and SLL-based locked instances with \texttt{K=128}.
Independently, we also investigate \textit{UNSAIL} when locking the GPS module in the million-gate ORPSoC design~\cite{CEP_github}.

\subsection{Setup for Security Evaluation}

The SAIL model $\texttt{ML}_{\texttt{1}}$ is implemented as a RF model with \texttt{50} decision trees and also as a support vector machine (SVM) using a radial basis function Gaussian kernel. 
The hyper-parameters for the SVM classifier were re-evaluated for each trained model and, thus, the kernel parameters such as the scale vary depending on the circuit under attack and its corresponding training set. 
The SAIL model $\texttt{ML}_{\texttt{2}}$ is implemented as an ensemble of Seq2Seq encoder-decoder models with attention, as described in Sec.~\ref{sec:methodology}. 
We use an embedding dimension of \texttt{256} for both the encoder and decoder. 
All of the LSTM layers consist of \texttt{200} hidden units. 
In order to mitigate over-fitting during training, a random-dropout probability of \texttt{0.05} was set. 
Each Seq2Seq model is trained for \texttt{60} epochs with a mini-batch size of \texttt{20}. In each epoch, the subgraphs are shuffled, and then mini-batches are constructed. The network parameters are updated using Adam optimizer after each mini-batch. A learning rate of \texttt{0.002}, a gradient decay factor of \texttt{0.9}, and a squared gradient decay factor of \texttt{0.999} are set.
Both SAIL models are implemented in \textit{MATLAB}. 

We also launch SWEEP on RLL and \textit{UNSAIL}-locked instances using the open-source attack tool~\cite{alaql2019sweep}.
The default margin value of \texttt{0} was used. The higher the margin is, the lower the chance of performing a wild guess by the attack, and the lower the reported accuracy is. We also evaluate the resilience of \textit{UNSAIL} against another \textit{oracle-less} attack on logic locking, the redundancy
attack~\cite{li2019piercing}.

The output corruption enforced by \textit{UNSAIL} is measured by the Hamming distance (HD) between the outputs of the original design and the outputs of the locked design, under the application of random incorrect keys. 
The output error rate (OER) is calculated, as well.
Ideal values for HD and OER would be \texttt{50\%} and \texttt{100\%}, respectively.
The simulations for HD and OER are performed using \textit{Mentor Graphics ModelSim}.

\subsection{Setup for Synthesis, Testing, and Layout Evaluation}

Synthesis is performed using \textit{Synopsys Design Compiler} for the slow process corner with particular focus on area minimization and iso-performance timing closure.
\textit{Synopsys Tetramax} was used to generate a minimal set of test patterns for the locked benchmarks.
Test coverage and fault coverage are also calculated using the same tool.
For layout-level assessment, we employ the public \textit{Nangate 45nm Open Cell Library} with ten metal layers and use \textit{Cadence Innovus}.
Layout overheads are calculated at \texttt{0.95V}, \texttt{125}$^\circ$\texttt{C}, with the slow process corner and input switching activity of \texttt{0.20}.
\begin{table*}[tb]
\centering
\caption{Accuracy for SAIL Random Forest (RF) Classifier for Selected ISCAS-85 and ITC-99 Benchmarks Using \texttt{sub=3}}
\label{tab:RF_Classifier_sub3}
\resizebox{\textwidth}{!}{%
\begin{tabular}{ccccccccccccccccccccccccccccccc}
\hline
& 
\multicolumn{14}{c}{\textbf{K=64}} & 
& 
& 
\multicolumn{14}{c}{\textbf{K=128}} 
\\ \hline
\textbf{Key-gates} & \multicolumn{2}{c}{\textbf{X(N)OR}} & 
& 
\multicolumn{2}{c}{\textbf{CL\_v1}} & 
& 
\multicolumn{2}{c}{\textbf{CL\_v2}} & 
& 
\multicolumn{2}{c}{\textbf{CL\_v3}} & 
& 
\multicolumn{2}{c}{\textbf{CL\_v4}} & 
& 
& 
\multicolumn{2}{c}{\textbf{X(N)OR}} & 
& 
\multicolumn{2}{c}{\textbf{CL\_v1}} & 
& 
\multicolumn{2}{c}{\textbf{CL\_v2}} & 
& 
\multicolumn{2}{c}{\textbf{CL\_v3}} & 
& 
\multicolumn{2}{c}{\textbf{CL\_v4}} 
\\ \hline
\textbf{Insertion} & 
\textbf{RLL} & \textit{\textbf{UNSAIL}} & 
& 
\textbf{RLL} & \textit{\textbf{UNSAIL}} & 
& 
\textbf{RLL} & \textit{\textbf{UNSAIL}} & 
& 
\textbf{RLL} & \textit{\textbf{UNSAIL}} & 
& 
\textbf{RLL} & \textit{\textbf{UNSAIL}} & 
& 
& 
\textbf{RLL} & \textit{\textbf{UNSAIL}} & 
& 
\textbf{RLL} & \textit{\textbf{UNSAIL}} & 
& 
\textbf{RLL} & \textit{\textbf{UNSAIL}} & 
& 
\textbf{RLL} & \textit{\textbf{UNSAIL}} & 
& 
\textbf{RLL} & \textit{\textbf{UNSAIL}} 
\\ \hline
c880 & 
0.72 & 0.58 & 
& 
{1.00}& 0.64 & 
& 
{1.00}& 0.59 & 
& 
0.97 & 0.56 & 
& 
0.92 & 0.62 & 
& 
& 
0.70 & 0.57 & 
& 
{1.00}& 0.58 & 
& 
{1.00}& 0.63 & 
& 
0.93 & 0.60 & 
& 
0.95 & 0.62 
\\ \hline
c1355 & 
0.92 & 0.65 & 
& 
{1.00}& 0.74 & 
& 
{1.00}& 0.74 & 
& 
0.97 & 0.72 & 
& 
0.98 & 0.72 & 
& 
& 
0.94 & 0.75 & 
& 
{1.00}& 0.71 & 
& 
{1.00}& 0.71 & 
& 
0.98 & 0.68 & 
& 
0.98 & 0.78 
\\ \hline
c1908 & 
0.83 & 0.75 & 
& 
{1.00}& 0.69 & 
& 
{1.00}& 0.70 & 
& 
0.97 & 0.62 & 
& 
0.97 & 0.74 & 
& 
& 
0.91 & 0.67 & 
& 
{1.00}& 0.69 & 
& 
{1.00}& 0.62 & 
& 
0.96 & 0.64 & 
& 
0.95 & 0.70 
\\ \hline
c2670 & 
0.68 & 0.59 & 
& 
{1.00}& 0.59 & 
& 
{1.00}& 0.56 & 
& 
0.94 & 0.59 & 
& 
0.95 & 0.62 & 
& 
& 
0.71 & 0.60 & 
& 
{1.00}& 0.54 & 
& 
{1.00}& 0.60 & 
& 
0.91 & 0.58 & 
& 
0.91 & 0.59 
\\ \hline
c3540 & 
0.80 & 0.56 & 
& 
{1.00}& 0.67 & 
& 
{1.00}& 0.68 & 
& 
0.89 & 0.58 & 
& 
0.95 & 0.61 & 
& 
& 
0.78 & 0.56 & 
& 
{1.00}& 0.55 & 
& 
{1.00}& 0.57 & 
& 
0.93 & 0.59 & 
& 
0.93 & 0.68 
\\ \hline
c5315 & 
0.85 & 0.62 & 
& 
{1.00}& 0.61 & 
& 
{1.00}& 0.62 & 
& 
0.97 & 0.56 & 
& 
0.89 & 0.66 & 
& 
& 
0.78 & 0.58 & 
& 
{1.00}& 0.59 & 
& 
{1.00}& 0.59 & 
& 
0.89 & 0.56 & 
& 
0.95 & 0.61 
\\ \hline
c6288 & 
0.88 & 0.64 & 
& 
{1.00}& 0.73 & 
& 
0.98 & 0.64 & 
& 
0.98 & 0.71 & 
& 
0.97 & 0.70 & 
& 
& 
0.87 & 0.65 & 
& 
0.99 & 0.66 & 
& 
{1.00}& 0.66 & 
& 
0.98 & 0.67 & 
& 
0.95 & 0.72 
\\ \hline
c7552 & 
0.88 & 0.74 & 
& 
{1.00}& 0.69 & 
& 
{1.00}& 0.69 & 
& 
0.98 & 0.70 & 
& 
0.95 & 0.66 & 
& 
& 
0.88 & 0.58 & 
& 
{1.00}& 0.65 & 
& 
{1.00}& 0.66 & 
& 
0.94 & 0.61 & 
& 
0.96 & 0.67 
\\ \hline
\textbf{Average} & 
\textbf{0.82} & \textbf{0.64} & 
& 
\textbf{1.00} & \textbf{0.67} & 
& 
\textbf{0.99} & \textbf{0.65} & 
& 
\textbf{0.96} & \textbf{0.63} & 
& 
\textbf{0.95} & \textbf{0.67} & 
& 
& 
\textbf{0.82} & \textbf{0.62} & 
& 
\textbf{0.99} & \textbf{0.62} & 
& 
\textbf{1.00} & \textbf{0.63} & 
& 
\textbf{0.94} & \textbf{0.62} & 
& 
\textbf{0.95} & \textbf{0.67} 
\\ \hline
& 
\multicolumn{14}{c}{{ \textbf{K=256}}} & 
& 
& 
\multicolumn{14}{c}{{ \textbf{K=512}}} 
\\ \hline
b14\_C & 
0.84 & 0.63 & 
& 
{1.00}& 0.65 & 
& 
{1.00}& 0.65 & 
& 
0.93 & 0.62 & 
& 
0.95 & 0.66 & 
& 
& 
0.83 & 0.65 & 
& 
{1.00}& 0.60 & 
& 
{1.00}& 0.61 & 
& 
0.96 & 0.61 & 
& 
0.95 & 0.63 
\\ \hline
b15\_C & 
0.77 & 0.59 & 
& 
{1.00}& 0.59 & 
& 
{1.00}& 0.58 & 
& 
0.94 & 0.56 & 
& 
0.95 & 0.59 & 
& 
& 
0.75 & 0.60 & 
& 
{1.00}& 0.55 & 
& 
{1.00}& 0.57 & 
& 
0.95 & 0.55 & 
& 
0.95 & 0.59 
\\ \hline
b20\_C & 
0.86 & 0.65 & 
& 
{1.00}& 0.62 & 
& 
0.99 & 0.64 & 
& 
0.96 & 0.66 & 
& 
0.95 & 0.67 & 
& 
& 
0.84 & 0.64 & 
& 
0.99 & 0.59 & 
& 
0.99 & 0.62 & 
& 
0.94 & 0.60 & 
& 
0.95 & 0.64 
\\ \hline
b21\_C & 
0.85 & 0.65 & 
& 
0.99 & 0.67 & 
& 
0.99 & 0.64 & 
& 
0.99 & 0.60 & 
& 
0.99 & 0.67 & 
& 
& 
0.86 & 0.63 & 
& 
{1.00}& 0.62 & 
& 
{1.00}& 0.63 & 
& 
0.94 & 0.60 & 
& 
0.96 & 0.64 
\\ \hline
b22\_C & 
0.85 & 0.64 & 
& 
0.99 & 0.65 & 
& 
0.99 & 0.64 & 
& 
0.98 & 0.64 & 
& 
0.95 & 0.69 & 
& 
& 
0.85 & 0.63 & 
& 
0.99 & 0.64 & 
& 
0.99 & 0.64 & 
& 
0.96 & 0.61 & 
& 
0.96 & 0.65 
\\ \hline
b17\_C & 
0.72 & 0.61 & 
& 
{1.00}& 0.59 & 
& 
{1.00}& 0.57 & 
& 
0.94 & 0.57 & 
& 
0.96 & 0.60 & 
& 
& 
0.75 & 0.58 & 
& 
{1.00}& 0.56 & 
& 
{1.00}& 0.57 & 
& 
0.92 & 0.57 & 
& 
0.92 & 0.59 
\\ \hline
\textbf{Average} & 
\textbf{0.82} & \textbf{0.63} & 
& 
\textbf{0.99} & \textbf{0.63} & 
& 
\textbf{0.99} & \textbf{0.62} & 
& 
\textbf{0.96} & \textbf{0.61} & 
& 
\textbf{0.96} & \textbf{0.65} & 
& 
& 
\textbf{0.81} & \textbf{0.62} & 
& 
\textbf{0.99} & \textbf{0.59} & 
& 
\textbf{0.99} & \textbf{0.61} & 
& 
\textbf{0.95} & \textbf{0.59} & 
& 
\textbf{0.95} & \textbf{0.62} 
\\ \hline
\end{tabular}%
}
\end{table*}

\begin{figure}[tb]
\centering
\includegraphics[width=\textwidth]{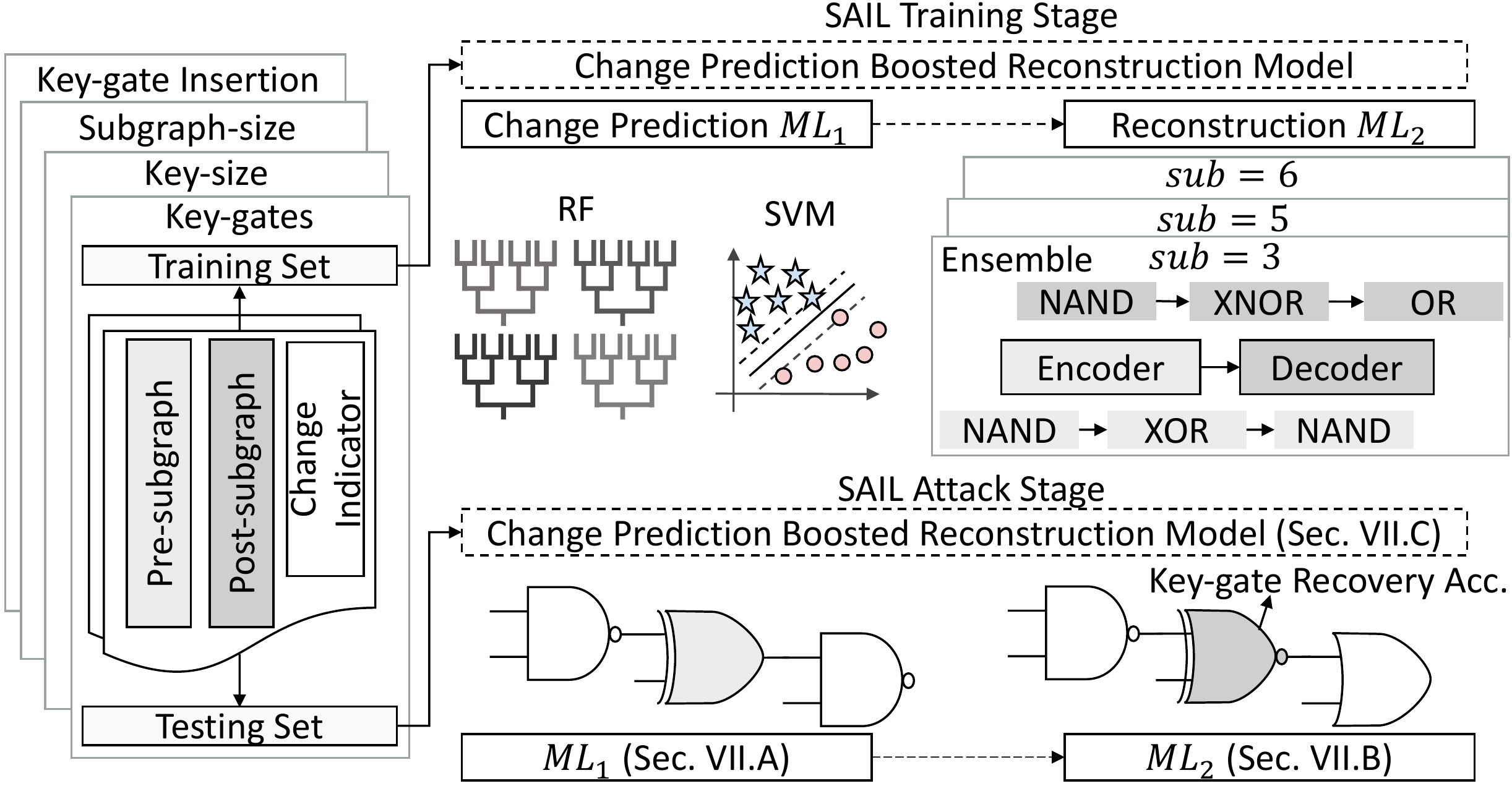}
\caption{Process for assessing the security of \textit{UNSAIL} against the SAIL attack.}
\label{fig:sail_exp}
\end{figure}

\section{Experimental Investigation}
\label{sec:security_analysis}

In this section, we first perform a detailed and thorough security analysis of our proposed \textit{UNSAIL} scheme, starting with the SAIL attack~\cite{chakraborty2018sail}. 
Fig.~\ref{fig:sail_exp} summarizes the evaluation process used for \textit{UNSAIL} against the SAIL attack. We consider the role of (i)~key-size, (ii)~key-gate type, (iii)~initial key-gate insertion algorithm, and (iv)~subgraph size.
When detailing the attack results, we initially report the classification accuracy of the change prediction model ${ML}_{1}$ using RF and SVM classifier models (Sec.~\ref{sec:accuracy}). 
Next, we report the key-gate recovery accuracy using the reconstruction model ${ML}_{2}$ implemented as a Seq2Seq ensemble model (Sec.~\ref{sec:reconstruction}).
Finally, we report the overall key-gate recovery accuracy when combining both ${ML}_{1}$ and ${ML}_{2}$ (Sec.~\ref{sec:sail_attack_accuracy}).

Our evaluation is expanded beyond the SAIL attack; the resilience of \textit{UNSAIL} is further demonstrated against the SWEEP attack and the redundancy attack in Sec.~\ref{Sec:Sweep_results} and Sec.~\ref{Sec:Redundancy_result}, respectively. 
Furthermore, the results of the output corruption evaluation for \textit{UNSAIL} are discussed in Sec.~\ref{sec:HD_analysis}, while the effect of \textit{UNSAIL} on structural testing is investigated in Sec.~\ref{sec:test}. 
The overheads of our defense are presented in Sec.~\ref{sec:methodology_overheads}. 
Finally, the experimental results on the DARPA OpenCores benchmark are discussed in Sec.~\ref{DARPA}.

\subsection{Change-Prediction Model Accuracy on \textit{UNSAIL} Vs.\ Traditional Logic Locking}
\label{sec:accuracy}

Initially, we investigate the classification accuracy to evaluate the performance of $\texttt{ML}_{\texttt{1}}$. 
The accuracy is defined as the number of correct predictions divided by the total number of predictions (i.e., key-size \texttt{K}). 
We test the model for different key-sizes, different key-gate structures, several sub-sizes, and two classification algorithms. 
Recall that the goal of \textit{UNSAIL} is to insert key-gates such that the complexity of the classification problem increases and the classification accuracy drops.
Indeed, for all considered cases, the model performs with much lower accuracy on \textit{UNSAIL}-locked instances.

\textbf{Varying the Key-Gate Type.} Pre- and post-subgraphs of size \texttt{sub=3} are extracted from all the locked instances. 
The model $\texttt{ML}_{\texttt{1}}$ is initially implemented as an RF model and trained separately for each benchmark, with the data extracted from \texttt{19} (in total \texttt{20}) instances for each locking variation.
Then, the classifier is tested on locked ISCAS-85 and ITC-99 benchmarks with \texttt{K=64} and \texttt{K=256}, respectively; results from both experiments are shown in Table~\ref{tab:RF_Classifier_sub3}.
Studying the classification accuracy of SAIL on RLL X(N)OR vs.\ RLL CL increases the performance for the latter case.
For example, the average accuracy on RLL X(N)OR is \texttt{82\%}, while on RLL CL\_v1, it is \texttt{99.5\%}.
Further investigation reveals most of the extracted post-subgraphs are affected by re-synthesis for the evaluated CL scheme.
This leads to an imbalanced data set where most of the subgraphs belong to one class, namely ``\textit{Changed},'' rendering classification less difficult.

Next, we study the effect of \textit{UNSAIL}. 
We note that our defense is capable of reducing the classification accuracy in all the considered test cases. 
On an average, \textit{UNSAIL} reduces the classification accuracy when using X(N)OR, CL\_v2, and CL\_v4 structures by \texttt{18pp}, \texttt{34pp}, and \texttt{28pp}, respectively, for \texttt{K=64}. 
\ul{Hence, \textit{UNSAIL} has an even more significant effect on reducing the classification accuracy when using CL techniques as compared to \textit{UNSAIL} X(N)OR locking.}

\textbf{Varying the Key-Size.} Next, we repeat the previous experiment for \texttt{K=128} and \texttt{K=512}, to study the effect of increasing the key-size on the classification accuracy.
Results from both the experiments are shown in Table~\ref{tab:RF_Classifier_sub3}.
We note that, on an average, the attack (classification accuracy of SAIL) performs slightly better for smaller key-sizes. 
While the average accuracy on RLL ISCAS-85 benchmarks is \texttt{94.4\%} and \texttt{94\%} for \texttt{K=64} and \texttt{K=128}, respectively, the average accuracy on RLL ITC-99 benchmarks is \texttt{94.4\%} and \texttt{93.8\%} for \texttt{K=256} and \texttt{K=512}, respectively.

\textit{UNSAIL} reduces the average classification accuracy when using X(N)OR, CL\_v2, and CL\_v4 structures by \texttt{20pp}, \texttt{37pp}, and \texttt{28pp}, respectively, for ISCAS-85 instances with \texttt{K=128}.
For ITC-99 instances with \texttt{K=512}, \textit{UNSAIL} reduces the average classification accuracy by \texttt{19pp}, \texttt{38pp}, and \texttt{33pp}, respectively, for the same structures/locking techniques.
We note that \textit{UNSAIL} has a larger impact on the instances locked with a larger key-size: the average reduction for classification accuracy is \texttt{29.2pp} for \texttt{K=64} and \texttt{30.8pp} for \texttt{K=128}.
\ul{These findings illustrate that \textit{UNSAIL} is effective for a varied range of benchmarks with varying key-sizes.}

\textbf{Varying the Sub-size.} In this set of experiments, we examine the effect of varying the sub-size on the classification accuracy for \textit{UNSAIL}.
Toward this end, we train and test $\texttt{ML}_{\texttt{1}}$ using \texttt{sub=5} and \texttt{sub=6}; see Table~\ref{tab:RF_Classifier_subs} for the results.

We observe that the classification accuracy for SAIL increases with an increase of sub-size (which is in agreement with the findings reported in~\cite{chakraborty2018sail}).
For example, the average classification accuracy for RLL X(N)OR increases from \texttt{82\%} for \texttt{sub=3} to \texttt{93\%} for \texttt{sub=6}. 
Increasing the sub-size leads to an imbalanced data set as most subgraphs are affected by re-synthesis, which results in higher classification accuracy. 
This is intuitive as a large subgraph has a higher probability of being affected by re-synthesis than a smaller subgraph. 

\ul{Even with such an increase in classification accuracy for larger sub-size, \textit{UNSAIL} remains successful in reducing accuracy.}
Comparing \textit{UNSAIL} vs. RLL, for CL\_v1 with \texttt{sub=3}, \texttt{sub=5}, and \texttt{sub=6}, the average classification accuracy is
reduced by \texttt{37pp}, \texttt{24pp}, and \texttt{21pp}, respectively, for ISCAS-85 benchmarks locked with \texttt{K=128}. 
We note that the classifier trained with \texttt{sub=3} is affected most by \textit{UNSAIL};
this is expected as \textit{UNSAIL} structures of size \texttt{sub=3} were added.

\textbf{Varying the Classifier Model.} To further investigate the efficacy of \textit{UNSAIL} against other classification algorithms, the SAIL model was implemented using SVM, trained, and tested using \texttt{sub=3} and different key-sizes; the related results are presented in Table~\ref{tab:SVM_Classifier_sub3}.
These experiments are in agreement with our earlier findings, namely that the SAIL classifier achieves slightly better accuracy on locked instances with smaller key-size. That is, the average classification accuracy on ISCAS-85 benchmarks locked using RLL with \texttt{K=64} is
\texttt{94.4\%} and reduces marginally to \texttt{93.8\%} with \texttt{K=128}.

The results on \textit{UNSAIL}-locked instances support our claim that
\ul{\textit{UNSAIL} incurs a more complex classification problem for different classifiers compared to RLL.}
The average classification accuracy for CL\_v1 was reduced by \texttt{37pp} and \texttt{36pp} using the RF model and SVM model, respectively, for ISCAS-85 benchmarks locked with \texttt{K=128} and \texttt{sub=3}.

\begin{table*}[tb]
\centering
\caption{Accuracy for SAIL Random Forest (RF) Classifier for Selected ISCAS-85 and ITC-99 Benchmarks Using \texttt{sub=5} and \texttt{sub=6}}
\label{tab:RF_Classifier_subs}
\resizebox{\textwidth}{!}{%
\begin{tabular}{ccccccccccccccccccccccccccccccc}
\hline
\multicolumn{31}{c}{\textbf{K=128}} 
\\ \hline
\textbf{Sub-size} & 
\multicolumn{14}{c}{{ \textbf{sub=5}}} & 
& 
& 
\multicolumn{14}{c}{\textbf{sub=6}} 
\\ \hline
\textbf{Key-gates} & \multicolumn{2}{c}{\textbf{X(N)OR}} & 
& 
\multicolumn{2}{c}{\textbf{CL\_v1}} & 
& 
\multicolumn{2}{c}{\textbf{CL\_v2}} & 
& 
\multicolumn{2}{c}{\textbf{CL\_v3}} & 
& 
\multicolumn{2}{c}{\textbf{CL\_v4}} & 
& 
& 
\multicolumn{2}{c}{\textbf{X(N)OR}} & 
& 
\multicolumn{2}{c}{\textbf{CL\_v1}} & 
& 
\multicolumn{2}{c}{\textbf{CL\_v2}} & 
& 
\multicolumn{2}{c}{\textbf{CL\_v3}} & 
& 
\multicolumn{2}{c}{\textbf{CL\_v4}} 
\\ \hline
\textbf{Insertion} & 
\textbf{RLL} & \textbf{UNSAIL} & 
& 
\textbf{RLL} & \textbf{UNSAIL} & 
& 
\textbf{RLL} & \textbf{UNSAIL} & 
& 
\textbf{RLL} & \textbf{UNSAIL} & 
& 
\textbf{RLL} & \textbf{UNSAIL} & 
& 
& 
\textbf{RLL} & \textbf{UNSAIL} & 
& 
\textbf{RLL} & \textbf{UNSAIL} & 
& 
\textbf{RLL} & \textbf{UNSAIL} & 
& 
\textbf{RLL} & \textbf{UNSAIL} & 
& 
\textbf{RLL} & \textbf{UNSAIL} 
\\ \hline
c880 & 
0.86 & 0.70 & 
& 
1.00 & 0.65 & 
& 
1.00 & 0.63 & 
& 
0.98 & 0.67 & 
& 
0.98 & 0.70 & 
& & 
0.88 & 0.72 & 
& 
1.00 & 0.70 & 
& 
1.00 & 0.71 & 
& 
0.98 & 0.76 & 
& 
0.98 & 0.71 
\\ \hline
c1355 & 
0.98 & 0.84 & 
& 
1.00 & 0.83 & 
& 
1.00 & 0.84 & 
& 
1.00 & 0.72 & 
& 
1.00 & 0.83 & 
& & 
0.98 & 0.89 & 
& 
1.00 & 0.90 & 
& 
1.00 & 0.88 & 
& 
1.00 & 0.85 & 
& 
0.99 & 0.85 
\\ \hline
c1908 & 
0.97 & 0.79 & 
& 
1.00 & 0.77 & 
& 
1.00 & 0.74 & 
& 
0.98 & 0.69 & 
& 
0.98 & 0.69 & 
& & 
0.97 & 0.76 & 
& 
1.00 & 0.83 & 
& 
1.00 & 0.80 & 
& 
0.98 & 0.84 & 
& 
0.98 & 0.83 
\\ \hline
c2670 & 0.86 & 0.66 & & 1.00 & 0.74 & & 1.00 & 0.83 & & 0.96 & 0.74 & & 0.97 & 0.80 & & & 0.89 & 0.67 & & 1.00 & 0.76 & & 1.00 & 0.84 & & 0.96 & 0.80 & & 0.97 & 0.80 \\ \hline
c3540 & 0.89 & 0.66 & & 1.00 & 0.79 & & 1.00 & 0.80 & & 0.97 & 0.72 & & 0.98 & 0.78 & & & 0.90 & 0.65 & & 1.00 & 0.80 & & 1.00 & 0.80 & & 0.98 & 0.80 & & 0.98 & 0.80 \\ \hline
c5315 & 0.89 & 0.66 & & 1.00 & 0.79 & & 1.00 & 0.66 & & 0.94 & 0.71 & & 0.97 & 0.77 & & & 0.88 & 0.70 & & 1.00 & 0.82 & & 1.00 & 0.70 & & 0.94 & 0.78 & & 0.97 & 0.77 \\ \hline
c6288 & 0.96 & 0.70 & & 1.00 & 0.77 & & 1.00 & 0.77 & & 0.98 & 0.75 & & 0.98 & 0.81 & & & 0.98 & 0.76 & & 1.00 & 0.80 & & 1.00 & 0.85 & & 0.99 & 0.81 & & 0.99 & 0.83 \\ \hline
c7552 & 0.95 & 0.73 & & 1.00 & 0.66 & & 1.00 & 0.77 & & 0.98 & 0.68 & & 0.98 & 0.69 & & & 0.95 & 0.70 & & 1.00 & 0.71 & & 1.00 & 0.82 & & 0.98 & 0.77 & & 0.98 & 0.75 \\ \hline
\textbf{Average} & \textbf{0.92} & \textbf{0.72} & & \textbf{0.99} & \textbf{0.75} & & \textbf{0.99} & \textbf{0.76} & & \textbf{0.97} & \textbf{0.71} & & \textbf{0.98} & \textbf{0.76} & & & \textbf{0.93} & \textbf{0.73} & & \textbf{1.00} & \textbf{0.79} & & \textbf{1.00} & \textbf{0.80} & & \textbf{0.98} & \textbf{0.80} & & \textbf{0.98} & \textbf{0.79} \\ \hline
\multicolumn{31}{c}{\textbf{K=512}} \\\hline
{\textbf{Sub-size}} & \multicolumn{14}{c}{{ \textbf{sub=5}}} & & & \multicolumn{14}{c}{\textbf{sub=6}} \\ \hline
b14\_C & 0.92 & 0.74 & & 1.00 & 0.73 & & 1.00 & 0.76 & & 0.98 & 0.68 & & 0.98 & 0.73 & & & 0.93 & 0.73 & & 1.00 & 0.77 & & 1.00 & 0.77 & & 0.98 & 0.71 & & 0.99 & 0.75 \\\hline
b15\_C & 0.87 & 0.70 & & 1.00 & 0.68 & & 1.00 & 0.71 & & 0.98 & 0.72 & & 0.98 & 0.73 & & & 0.91 & 0.71 & & 1.00 & 0.74 & & 1.00 & 0.76 & & 0.98 & 0.75 & & 0.98 & 0.77 \\\hline
b20\_C & 0.92 & 0.72 & & 0.99 & 0.71 & & 0.99 & 0.75 & & 0.98 & 0.72 & & 0.98 & 0.72 & & & 0.92 & 0.75 & & 0.99 & 0.79 & & 0.99 & 0.76 & & 0.98 & 0.76 & & 0.98 & 0.77 \\\hline
b21\_C & 0.94 & 0.71 & & 0.99 & 0.72 & & 0.99 & 0.76 & & 0.97 & 0.72 & & 0.98 & 0.72 & & & 0.96 & 0.71 & & 0.99 & 0.77 & & 0.99 & 0.81 & & 0.98 & 0.76 & & 0.98 & 0.74 \\\hline
b22\_C & 0.96 & 0.74 & & 1.00 & 0.72 & & 1.00 & 0.74 & & 0.96 & 0.72 & & 0.99 & 0.72 & & & 0.96 & 0.75 & & 1.00 & 0.75 & & 1.00 & 0.77 & & 0.97 & 0.76 & & 0.99 & 0.78 \\\hline
b17\_C & 0.83 & 0.67 & & 1.00 & 0.72 & & 1.00 & 0.68 & & 0.95 & 0.72 & & 0.98 & 0.74 & & & 0.86 & 0.67 & & 1.00 & 0.74 & & 1.00 & 0.72 & & 0.95 & 0.75 & & 0.98 & 0.76 \\\hline
\textbf{Average} & \textbf{0.91} & \textbf{0.71} & & \textbf{0.99} & \textbf{0.71} & & \textbf{0.99} & \textbf{0.73} & & \textbf{0.97} & \textbf{0.71} & & \textbf{0.98} & \textbf{0.73} & & & \textbf{0.92} & \textbf{0.72} & & \textbf{0.99} & \textbf{0.76} & & \textbf{0.99} & \textbf{0.76} & & \textbf{0.97} & \textbf{0.75} & & \textbf{0.98} & \textbf{0.76}\\\hline
\end{tabular}%
}
\end{table*}

\begin{table*}[tb]
\centering
\caption{Accuracy for SAIL Support Vector Machine (SVM) Classifier for Selected ISCAS-85 and ITC-99 Benchmarks Using \texttt{sub=3}}
\label{tab:SVM_Classifier_sub3}
\resizebox{\textwidth}{!}{%
\begin{tabular}{ccccccccccccccccccccccccccccccc}
\hline
& 
\multicolumn{14}{c}{\textbf{K=64}} & 
& 
& 
\multicolumn{14}{c}{\textbf{K=128}} 
\\ \hline
\textbf{Key-gates} & \multicolumn{2}{c}{\textbf{X(N)OR}} & 
& 
\multicolumn{2}{c}{\textbf{CL\_v1}} & 
& 
\multicolumn{2}{c}{\textbf{CL\_v2}} & 
& 
\multicolumn{2}{c}{\textbf{CL\_v3}} & 
& 
\multicolumn{2}{c}{\textbf{CL\_v4}} & 
& 
& 
\multicolumn{2}{c}{\textbf{X(N)OR}} & 
& 
\multicolumn{2}{c}{\textbf{CL\_v1}} & 
& 
\multicolumn{2}{c}{\textbf{CL\_v2}} & 
& 
\multicolumn{2}{c}{\textbf{CL\_v3}} & 
& 
\multicolumn{2}{c}{\textbf{CL\_v4}} 
\\ \hline

\textbf{Insertion} & 
\textbf{RLL} & \textit{\textbf{UNSAIL}} & 
& 
\textbf{RLL} & \textit{\textbf{UNSAIL}} & 
& 
\textbf{RLL} & \textit{\textbf{UNSAIL}} & 
& 
\textbf{RLL} & \textit{\textbf{UNSAIL}} & 
& 
\textbf{RLL} & \textit{\textbf{UNSAIL}} & 
& 
& 
\textbf{RLL} & \textit{\textbf{UNSAIL}} & 
& 
\textbf{RLL} & \textit{\textbf{UNSAIL}} & 
& 
\textbf{RLL} & \textit{\textbf{UNSAIL}} & 
& 
\textbf{RLL} & \textit{\textbf{UNSAIL}} & 
& 
\textbf{RLL} & \textit{\textbf{UNSAIL}} 
\\ \hline

c880 & 
0.71 & 0.58 & 
& 
{1.00}& 0.64 & 
& 
{1.00}& 0.62 & 
& 
0.97 & 0.56 & 
& 
0.94 & 0.61 & 
& 
& 
0.71 & 0.56 & 
& 
{1.00}& 0.58 & 
& 
{1.00}& 0.62 & 
& 
0.93 & 0.62 & 
& 
0.96 & 0.62 
\\ \hline

c1355 & 
0.92 & 0.67 & 
& 
{1.00}& 0.72 & 
& 
{1.00}& 0.75 & 
& 
0.97 & 0.60 & 
& 
0.98 & 0.74 & 
& 
& 
0.94 & 0.73 & 
& 
{1.00}& 0.78 & 
& 
{1.00}& 0.77 & 
& 
0.98 & 0.70 & 
& 
0.98 & 0.76 
\\ \hline

c1908 & 
0.82 & 0.74 & 
& 
{1.00}& 0.73 & 
& 
{1.00}& 0.70 & 
& 
0.97 & 0.64 & 
& 
0.97 & 0.74 & 
& 
& 
0.94 & 0.69 & 
& 
{1.00}& 0.68 & 
& 
{1.00}& 0.65 & 
& 
0.96 & 0.65 & 
& 
0.95 & 0.74 
\\ \hline

c2670 & 
0.69 & 0.56 & 
& 
{1.00}& 0.64 & 
& 
{1.00}& 0.59 & 
& 
0.94 & 0.59 & 
& 
0.95 & 0.64 & 
& 
& 
0.67 & 0.59 & 
& 
{1.00}& 0.54 & 
& 
{1.00}& 0.58 & 
& 
0.91 & 0.57 & 
& 
0.91 & 0.61 
\\ \hline

c3540 & 
0.80 & 0.56 & 
& 
{1.00}& 0.63 & 
& 
{1.00}& 0.64 & 
& 
0.89 & 0.59 & 
& 
0.95 & 0.59 & 
& 
& 
0.78 & 0.54 & 
& 
{1.00}& 0.58 & 
& 
{1.00}& 0.57 & 
& 
0.93 & 0.58 & 
& 
0.93 & 0.68 
\\ \hline

c5315 & 
0.83 & 0.62 & 
& 
{1.00}& 0.61 & 
& 
{1.00}& 0.62 & 
& 
0.97 & 0.55 & 
& 
0.89 & 0.66 & 
& 
& 
0.80 & 0.58 & 
& 
{1.00}& 0.59 & 
& 
{1.00}& 0.60 & 
& 
0.89 & 0.57 & 
& 
0.95 & 0.62 
\\ \hline

c6288 & 
0.88 & 0.61 & 
& 
{1.00}& 0.71 & 
& 
0.98 & 0.64 & 
& 
0.98 & 0.72 & 
& 
0.97 & 0.66 & 
& 
& 
0.87 & 0.67 & 
& 
0.99 & 0.66 & 
& 
0.99 & 0.66 & 
& 
0.98 & 0.65 & 
& 
0.95 & 0.72 
\\ \hline

c7552 & 
0.88 & 0.70 & 
& 
{1.00}& 0.67 & 
& 
{1.00}& 0.69 & 
& 
0.98 & 0.69 & 
& 
0.95 & 0.66 & 
& 
& 
0.88 & 0.63 & 
& 
{1.00}& 0.63 & 
& 
{1.00}& 0.66 & 
& 
0.94 & 0.61 & 
& 
0.96 & 0.68 
\\ \hline

\textbf{Average} & 
\textbf{0.82} & \textbf{0.63} & 
& 
\textbf{1.00} & \textbf{0.67} & 
& 
\textbf{0.99} & \textbf{0.66} & 
& 
\textbf{0.96} & \textbf{0.62} & 
& 
\textbf{0.95} & \textbf{0.66} & 
& 
& 
\textbf{0.82} & \textbf{0.62} & 
& 
\textbf{0.99} & \textbf{0.63} & 
& 
\textbf{0.99} & \textbf{0.64} & 
& 
\textbf{0.94} & \textbf{0.62} & 
& 
\textbf{0.95} & \textbf{0.68} 
\\ \hline
& 
\multicolumn{14}{c}{\textbf{K=256}} & 
& & 
\multicolumn{14}{c}{\textbf{K=512}} 
\\ \hline

b14\_C & 
0.84 & 0.64 & 
& 
{1.00}& 0.65 & 
& 
{1.00}& 0.66 & 
& 
0.93 & 0.63 & 
& 
0.95 & 0.64 & 
& 
& 
0.83 & 0.64 & 
& 
{1.00}& 0.59 & 
& 
{1.00}& 0.61 & 
& 
0.96 & 0.60 & 
& 
0.95 & 0.63 
\\ \hline

b15\_C & 
0.77 & 0.59 & 
& 
{1.00}& 0.58 & 
& 
{1.00}& 0.58 & 
& 
0.94 & 0.54 & 
& 
0.95 & 0.59 & 
& 
& 
0.75 & 0.59 & 
& 
{1.00}& 0.55 & 
& 
{1.00}& 0.57 & 
& 
0.95 & 0.55 & 
& 
0.95 & 0.59 
\\ \hline

b20\_C & 
0.86 & 0.64 & 
& 
0.99 & 0.63 & 
& 
0.99 & 0.64 & 
& 
0.99 & 0.66 & 
& 
0.99 & 0.68 & 
& 
& 
0.84 & 0.64 & 
& 
0.99 & 0.59 & 
& 
0.99 & 0.62 & 
& 
0.94 & 0.61 & 
& 
0.95 & 0.64 
\\ \hline

b21\_C & 
0.85 & 0.66 & 
& 
0.99 & 0.67 & 
& 
0.99 & 0.63 & 
& 0.98 & 0.60 & 
& 
0.95 & 0.66 & 
& 
& 
0.86 & 0.63 & 
& 
{1.00}& 0.62 & 
& 
{1.00}& 0.63 & 
& 
0.94 & 0.61 & 
& 
0.96 & 0.65 
\\ \hline

b22\_C & 
0.88 & 0.62 & 
& 
{1.00}& 0.64 & 
& 
0.99 & 0.63 & 
& 
0.96 & 0.65 & 
& 
0.95 & 0.69 & 
& 
& 
0.85 & 0.62 & 
& 
0.99 & 0.64 & 
& 
0.99 & 0.64 & 
& 
0.96 & 0.61 & 
& 
0.96 & 0.65 
\\ \hline

b17\_C & 
0.69 & 0.60 & 
& 
{1.00}& 0.59 & 
& 
{1.00}& 0.57 & 
& 
0.94 & 0.57 & 
& 
0.96 & 0.60 & 
& 
& 
0.75 & 0.58 & 
& 
{1.00}& 0.57 & 
& 
{1.00}& 0.58 & 
& 
0.92 & 0.56 & 
& 
0.96 & 0.59 
\\ \hline

\textbf{Average} & 
\textbf{0.82} & \textbf{0.62} & 
& 
\textbf{0.99} & \textbf{0.63} & 
& 
\textbf{0.99} & \textbf{0.62} & 
& 
\textbf{0.96} & \textbf{0.61} & 
& 
\textbf{0.96} & \textbf{0.64} & 
& 
& 
\textbf{0.81} & \textbf{0.62} & 
& 
\textbf{0.99} & \textbf{0.59} & 
& 
\textbf{0.99} & \textbf{0.61} & 
& 
\textbf{0.95} & \textbf{0.59} & 
& 
\textbf{0.95} & \textbf{0.62} 
\\ \hline
\end{tabular}
}
\end{table*}

\textbf{Integrating \textit{UNSAIL} with FLL and SLL.} We locked ISCAS-85 benchmarks using SLL and FLL with \texttt{K=128}. Each benchmark was locked once using SLL and once using FLL. 
Reusing the \texttt{20} X(N)OR-based RLL instances for each of those benchmarks, $\texttt{ML}_{\texttt{1}}$ (implemented as RF) was trained for
\texttt{sub=6} and then launched on the SLL and FLL instances. 
The average accuracy of the model on SLL and FLL benchmarks is \texttt{97\%} and \texttt{93\%}, respectively.
Comparing the performance of $\texttt{ML}_{\texttt{1}}$ on RLL (Table~\ref{tab:RF_Classifier_subs}) vs.\ $\texttt{ML}_{\texttt{1}}$ on FLL, the
results are largely consistent. 
In the case of SLL, we observe an increase in the accuracy of $\texttt{ML}_{\texttt{1}}$ when compared to RLL, namely by an average of \texttt{4pp}.

\begin{figure}[tb]
\captionsetup[subfigure]{labelformat=empty}
\centering
\subfloat[]{\includegraphics[width=.7\textwidth]{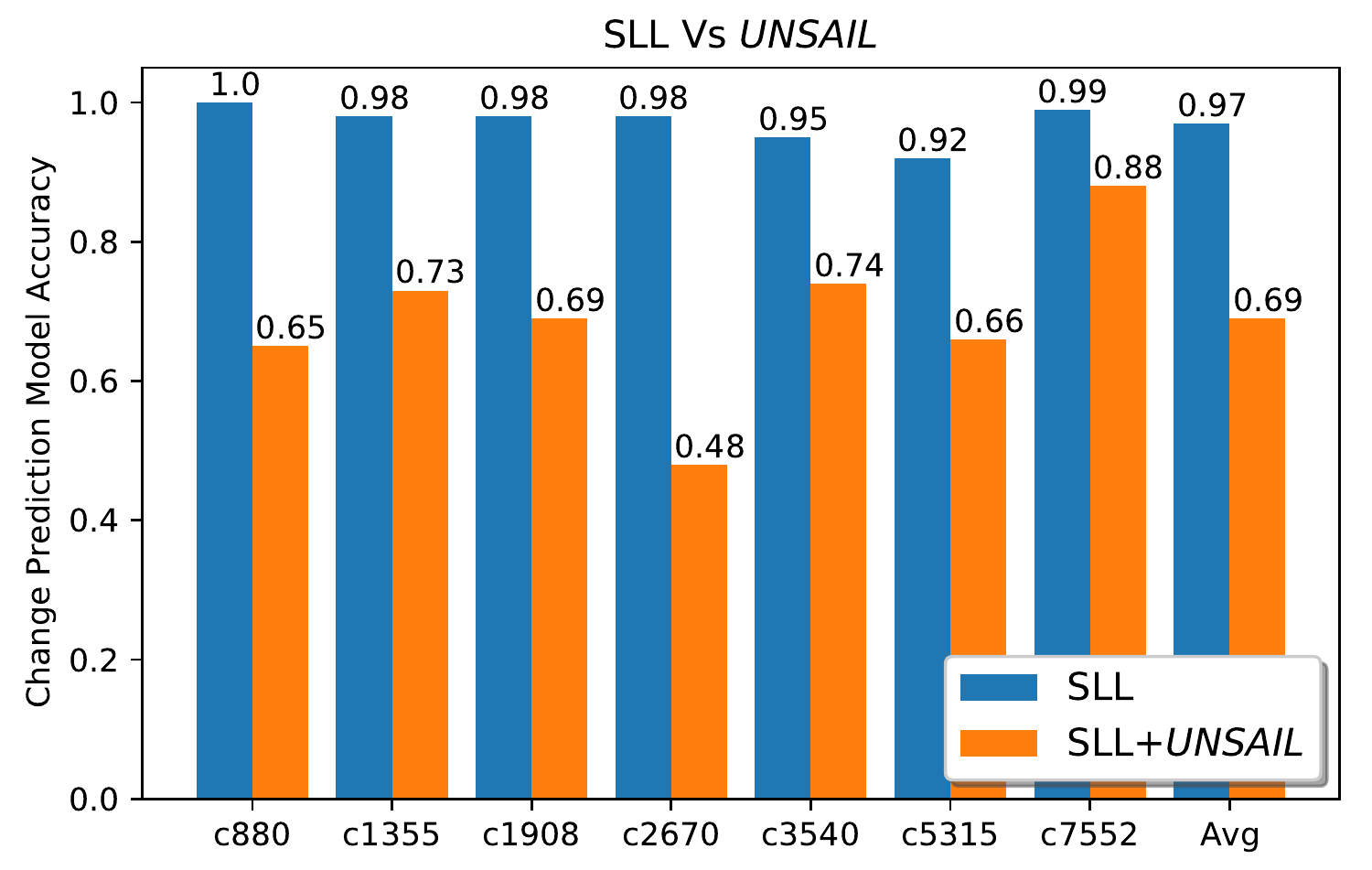}}\\
		\vspace{-4ex}
\subfloat[]{\includegraphics[width=.7\textwidth]{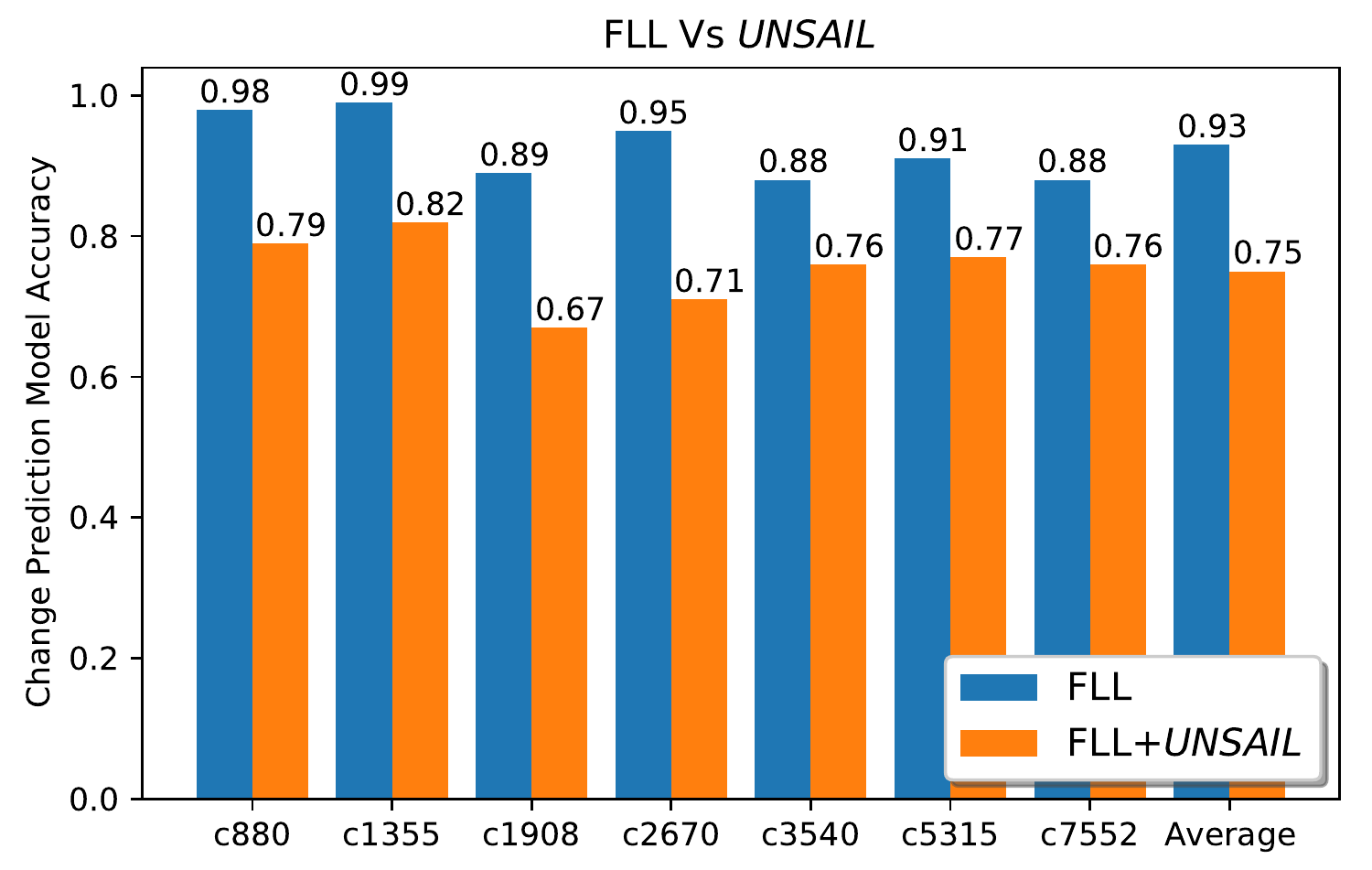}}
\caption{SAIL change-prediction model $\texttt{ML}_{\texttt{1}}$ accuracy on \textit{UNSAIL} vs.\ SLL and FLL when \texttt{K=128} and \texttt{sub=6} on selected ISCAS-85 benchmarks.}
\label{fig:UNSAIL_SLL_FLL}
\end{figure}

To evaluate the performance of \textit{UNSAIL} in protecting the FLL and SLL instances, we next integrate \textit{UNSAIL} with SLL and FLL. 
As usual, \texttt{64} key-gates are employed during initial locking, whereupon the remaining \texttt{64} key-gates are injected by \textit{UNSAIL}, thereby achieving \texttt{K=128}.
We train the model using RLL-based \textit{UNSAIL} instances and launch it on the SLL-based and the FLL-based \textit{UNSAIL} instances; this is fair since SLL and FLL are also using X(N)OR key-gate structures.
The results are reported in Fig.~\ref{fig:UNSAIL_SLL_FLL}. 
The average accuracy on the SLL-based \textit{UNSAIL} instances is \texttt{69\%}, implying a reduction of \texttt{28pp}; the average accuracy on the FLL-based \textit{UNSAIL} instances is \texttt{75\%}, indicating a reduction of \texttt{18pp}. 
These findings support our claim that \ul{\textit{UNSAIL} is capable of protecting any combinational logic locking technique from SAIL.}

\textbf{Summary.} $\texttt{ML}_{\texttt{1}}$ of SAIL was thoroughly tested on RLL and \textit{UNSAIL}-locked instances of selected ISCAS-85 and ITC-99 benchmarks.
We considered different types of key-gate, key-sizes, sub-sizes, and other classification models.
The classifier was additionally studied on SLL/FLL vs.\ SLL/FLL-based \textit{UNSAIL}.
Results show that the classification accuracy of $\texttt{ML}_{\texttt{1}}$ increases with an increase in sub-size, which is also reported in~\cite{chakraborty2018sail}.
It was also observed that $\texttt{ML}_{\texttt{1}}$ achieves better accuracy on benchmarks locked with smaller key-sizes.
Besides, $\texttt{ML}_{\texttt{1}}$ achieves a higher accuracy on (1) the CL-based RLL vs.\ X(N)OR-based RLL and (2) SLL instances when compared to RLL and FLL. 

Analysis of the results for \textit{UNSAIL}-locked instances shows that our defense is capable of decreasing the classification accuracy of $\texttt{ML}_{\texttt{1}}$ for all of the tested cases, achieving the best performance when \texttt{sub=3} is used. 
Our defense technique does not require a specific classification model or setup; it succeeds in increasing the complexity for classifying the key-gates and related subgraphs under all scenarios. 
Moreover, our defense does not require a specific locking technique and can protect any traditionally locked design.

\subsection{Reconstruction Model Accuracy on \textit{UNSAIL} Vs.\ Traditional Logic Locking}
\label{sec:reconstruction}
Here, we study $\texttt{ML}_{\texttt{2}}$ for RLL and \textit{UNSAIL}-locked instances. 
The accuracy of recovering the pre-synthesis key-gate is shown for ISCAS-85 benchmarks and ITC-99 benchmarks with \texttt{K=128} and
\texttt{K=512}, respectively, in Table~\ref{tab:attack_reconstruction}. 
Two important observations can be inferred from the results, which are discussed next.

\textbf{Varying the Key-Gate Type.} Recall that in this work, we also study the effect of different types of key-gates (Table~\ref{tab:variations}).
We believe that by varying the locking structures, the model will need to understand (have learned on) a larger variation of synthesis-induced changes, which tends to affect the underlying accuracy. 
Moreover, by randomizing the selection of key-gate types used, we can expect to limit the ``deterministic footprint'' for obfuscation inferred by the synthesis tools.

We observe that the average recovery accuracy reduces once we introduce more variations to the key-gate structures.
The average accuracy on RLL ISCAS-85 instances with \texttt{K=128} for X(N)OR, CL\_v1, CL\_v2, CL\_v3, and CL\_v4 key-gate structures is \texttt{64\%}, \texttt{51\%}, \texttt{35\%}, \texttt{28\%}, and \texttt{40\%}, respectively (Table~\ref{tab:attack_reconstruction}).
For the large ITC-99 benchmarks with \texttt{K=512}, the average key-gate recovery accuracy is \texttt{72\%}, \texttt{46\%}, \texttt{48\%}, \texttt{32\%}, and \texttt{43\%}, respectively.

\textbf{Effect of \textit{UNSAIL} Structures.} The accuracy of $\texttt{ML}_{\texttt{2}}$ is further reduced by the \textit{UNSAIL} structures, as shown in Table~\ref{tab:attack_reconstruction}. 
The average key-gate recovery accuracy for \textit{UNSAIL}-locked ISCAS-85 instances with \texttt{K=128}) for X(N)OR, CL\_v1, CL\_v2, CL\_v3, and
CL\_v4 techniques is \texttt{50\%}, \texttt{24\%}, \texttt{15\%}, \texttt{17\%}, and \texttt{23\%}, respectively, which represents an average
reduction of \texttt{17.8pp} when compared to RLL.
For the larger ITC-99 benchmarks with \texttt{K=512}, the average reduction is \texttt{18.6pp}.
Similarly, a consistent effect induced by \textit{UNSAIL} for reducing the accuracy is also observed when comparing \textit{UNSAIL} to SLL and
FLL, as shown in Table~\ref{tab:reconstruction_sll_fll}.

\textbf{Summary.} More variations in key-gate structures hinder the reconstruction model in general, and the \textit{UNSAIL} structures strengthen this effect further, for any logic locking technique.

\begin{table}[tb]
\centering
\caption{Key-Gate Detection Accuracy Using SAIL Reconstruction Model on \textit{UNSAIL} Vs.\ RLL}
\label{tab:attack_reconstruction}
\resizebox{\textwidth}{!}{%
\begin{tabular}{ccccccccccc}
\hline
& \multicolumn{10}{c}{\textbf{K=128}} 
\\ \hline
\textbf{Key-gates} & \multicolumn{2}{c}{\textbf{X(N)OR}} & 
 
\multicolumn{2}{c}{\textbf{CL\_v1}} & 
 
\multicolumn{2}{c}{\textbf{CL\_v2}} &
 
\multicolumn{2}{c}{\textbf{CL\_v3}} & 
 
\multicolumn{2}{c}{\textbf{CL\_v4}} 
\\ \hline
\textbf{Insertion} 
& \textbf{RLL} & \textit{\textbf{UNSAIL}} 

& \textbf{RLL} & \textit{\textbf{UNSAIL}} 

& \textbf{RLL} & \textit{\textbf{UNSAIL}} 

& \textbf{RLL} & \textit{\textbf{UNSAIL}} 

& \textbf{RLL} & \textit{\textbf{UNSAIL}} 
\\ \hline
c880 & 
0.72 & 0.46 & 
 
0.47 & 0.26 & 
 
0.31 & 0.16 & 
 
0.26 & 0.15 & 
 
0.38 & 0.17 
\\ \hline
c1355 & 
0.60 & 0.62 & 
 
0.56 & 0.26 & 
 
0.40 & 0.11 & 
 
0.33 & 0.24 & 
 
0.36 & 0.28 
\\ \hline
c1908 & 
0.64 & 0.56 & 
 
0.52 & 0.24 & 
 
0.34 & 0.16 & 
 
0.29 & 0.18 & 
 
0.45 & 0.27 
\\ \hline
c2670 & 
0.77 & 0.48 & 
 
0.57 & 0.26 & 
 
0.32 & 0.18 & 
 
0.24 & 0.14 & 
 
0.41 & 0.23 
\\ \hline
c3540 & 
0.62 & 0.46 & 
 
0.47 & 0.23 & 
 
0.30 & 0.17 & 
 
0.30 & 0.21 & 
 
0.34 & 0.26 
\\ \hline
c5315 & 
0.65 & 0.52 & 
 
0.45 & 0.22 & 
 
0.30 & 0.14 & 
 
0.30 & 0.10 & 
 
0.43 & 0.16 
\\ \hline
c6288 & 
0.57 & 0.45 & 
 
0.58 & 0.21 & 
 
0.42 & 0.13 & 
 
0.25 & 0.15 & 
 
0.41 & 0.33 
\\ \hline
c7552 & 
0.55 & 0.43 & 
 
0.47 & 0.23 & 
 
0.37 & 0.15 & 
 
0.24 & 0.16 & 
 
0.40 & 0.17 
\\ \hline
{ \textbf{Average}} & 
\textbf{0.64} & \textbf{0.50} & 
 
\textbf{0.51} & \textbf{0.24} & 
 
\textbf{0.35} & \textbf{0.15} & 
 
\textbf{0.28} & \textbf{0.17} & 
 
\textbf{0.40} & \textbf{0.23} 
\\ \hline
& \multicolumn{10}{c}{{ \textbf{K=512}}} 
\\ \hline
b14\_C & 
0.71 & 0.50 & 
 
0.45 & 0.24 & 
 
0.43 & 0.30 & 
 
0.31 & 0.22 & 
 
0.41 & 0.27 
\\ \hline
b15\_C & 
0.77 & 0.57 & 
 
0.48 & 0.23 & 
 
0.49 & 0.23 & 
 
0.32 & 0.19 & 
 
0.44 & 0.27 
\\ \hline
b20\_C & 
0.71 & 0.56 & 
 
0.47 & 0.23 & 
 
0.49 & 0.26 & 
 
0.32 & 0.18 & 
 
0.45 & 0.25 
\\ \hline
b21\_C & 
0.71 & 0.52 & 
 
0.46 & 0.25 & 
 
0.48 & 0.26 & 
 
0.29 & 0.18 & 
 
0.45 & 0.27 
\\ \hline
b22\_C & 
0.68 & 0.50 & 
 
0.46 & 0.30 & 
 
0.50 & 0.27 & 
 
0.34 & 0.17 & 
 
0.42 & 0.24 
\\ \hline
b17\_C & 
0.74 & 0.54 & 
 
0.46 & 0.22 & 
 
0.47 & 0.26 & 
 
0.34 & 0.2 & 
 
0.38 & 0.28 
\\ \hline
\textbf{Average} & 
\textbf{0.72} & \textbf{0.53} & 
 
\textbf{0.46} & \textbf{0.25} & 
 
\textbf{0.48} & \textbf{0.26} & 
 
\textbf{0.32} & \textbf{0.19} & 
 
\textbf{0.43} & \textbf{0.26} 
\\ \hline
\end{tabular}
}
\end{table}

\begin{table}[tb]
\centering
\scriptsize
\caption{Key-Gate Detection Accuracy Using SAIL Reconstruction Model on \textit{UNSAIL} Vs.\ SLL and FLL for \texttt{K=128}}
\label{tab:reconstruction_sll_fll}
\resizebox{\textwidth}{!}{
\begin{tabular}{ccccc}
\hline
\textbf{Insertion} & 
\textbf{SLL} & 
\textbf{SLL-based \textit{UNSAIL}} & 
\textbf{FLL} & 
\textbf{FLL-based \textit{UNSAIL}} 
\\ \hline
c880 & 0.75 & 0.35 & 0.44 & 0.41 
\\ \hline
c1355 & 0.46 & 0.46 & 0.37 & 0.42 
\\ \hline
c1908 & 0.44 & 0.44 & 0.47 & 0.32 
\\ \hline
c2670 & 0.61 & 0.33 & 0.34 & 0.43 
\\ \hline
c3540 & 0.58 & 0.42 & 0.41 & 0.49 
\\ \hline
c5315 & 0.38 & 0.24 & 0.51 & 0.4 
\\ \hline
c7552 & 0.44 & 0.44 & 0.48 & 0.37 
\\ \hline
\textbf{Average} & 
\textbf{0.52} & 
\textbf{0.38} & 
\textbf{0.43} & 
\textbf{0.41} 
\\ \hline
\end{tabular}
}
\end{table}

\subsection{Change-Prediction-Boosted Reconstruction Model Accuracy on \textit{UNSAIL} Vs.\ Traditional Logic Locking}
\label{sec:sail_attack_accuracy}

Finally, to evaluate the effectiveness of \textit{UNSAIL}, we launch the full SAIL attack where $\texttt{ML}_{\texttt{1}}$ is combined with $\texttt{ML}_{\texttt{2}}$ first to detect the subgraphs that went through changes due to synthesis and then revert those changes.
We report the accuracy of the full SAIL attack in Table~\ref{tab:full_attack_reconstruction}.
We use the RF classifier with \texttt{sub=6} since this configuration provided the highest classification accuracy, as previously shown in Sec.\ref{sec:accuracy}.

Analyzing the results obtained on RLL-based locking; first, we note that the effect of varying the key-gates, as observed in Sec.~\ref{sec:reconstruction} is still present. 
More specifically, we observe a reduction in the attack accuracy from an average of \texttt{68\%} (RLL-based X(N)OR) down to \texttt{27\%} (RLL-based CL\_v3) on ISCAS-85 benchmarks with \texttt{K=128}.
Similarly, a reduction in accuracy is also observed for the larger ITC-99 benchmarks with \texttt{K=512}.
Focusing on the results obtained on X(N)OR locking, it can be observed that the reconstruction accuracy is boosted when both the models of SAIL are merged. 
That is, comparing with the results in Table~\ref{tab:attack_reconstruction} for the case of RLL, the average accuracy on ISCAS-85 benchmarks is increased from \texttt{64\%} to \texttt{68\%}. 
Nevertheless, \ul{\textit{UNSAIL} is capable of reducing the accuracy by \texttt{11pp}, dropping it to \texttt{57\%}.}

This latter finding is further supported once the full SAIL attack is launched on SLL and FLL instances (Table~\ref{tab:boosted_sll_fll}). 
Comparing to the results in Table~\ref{tab:reconstruction_sll_fll}, the average accuracy increases by \texttt{3pp} for both SLL and FLL. 
On average, \textit{UNSAIL} reduces the accuracy by \texttt{9pp} for SLL and by \texttt{2pp} for FLL on ISCAS-85 benchmarks with \texttt{K=128}.
Although the reduction in accuracy is marginal for FLL-based \textit{UNSAIL}, \ul{the final accuracy of \texttt{44\%} is still below random-guessing (\texttt{50\%}).}

Analyzing the results of CL structures for RLL, we note that merging the two attack models did not boost the key-gate recovery accuracy for RLL to begin with.
This is because most of the key-gate structures added by RLL go through changes due to synthesis; recall that mainly one class of
subgraphs can be observed in Fig.~\ref{fig:tsne} for RLL-based CL.
Hence, using a classifier model to boost $\texttt{ML}_{\texttt{2}}$ will not provide a significant benefit.
For example, the average key-gate recovery accuracy for CL\_v4 was \texttt{40\%} using $\texttt{ML}_{\texttt{2}}$ on its own, which even (slightly) reduced to \texttt{39\%} for the combined and boosted attack setup.
In contrast, \textit{UNSAIL} ensures that two types of classes exist in training and, thus, the key-gate detection accuracy for
CL-based \textit{UNSAIL} is observed to increase.
Although the accuracy of the classifier was reduced by an average of \texttt{19.5pp} when using CL-based \textit{UNSAIL}, the classifier was still able to improve the overall accuracy of the attack.
Even then, \ul{the average key-gate recovery accuracy for CL-based \textit{UNSAIL} instances is \texttt{53\%}, which is just a shade better than random-guessing, rendering the full SAIL attack futile.}

\begin{table}[tb]
\centering
\caption{Key-Gate Detection Accuracy Using SAIL Change-Prediction-Boosted Reconstruction Model on \textit{UNSAIL} Vs.\ RLL using \texttt{sub=6} for the Change-Prediction Model}
\label{tab:full_attack_reconstruction}
\resizebox{\textwidth}{!}{%
\begin{tabular}{ccccccccccc}
\hline
& \multicolumn{10}{c}{\textbf{K=128}} 
\\ \hline
\textbf{Key-gates} & \multicolumn{2}{c}{\textbf{X(N)OR}} & 
 
\multicolumn{2}{c}{\textbf{CL\_v1}} & 
 
\multicolumn{2}{c}{\textbf{CL\_v2}} & 
 
\multicolumn{2}{c}{\textbf{CL\_v3}} & 
 
\multicolumn{2}{c}{\textbf{CL\_v4}} 
\\ \hline
\textbf{Insertion} & 
\textbf{RLL} & \textit{\textbf{UNSAIL}} & 
 
\textbf{RLL} & \textit{\textbf{UNSAIL}} & 
 
\textbf{RLL} & \textit{\textbf{UNSAIL}} & 
 
\textbf{RLL} & \textit{\textbf{UNSAIL}} & 
 
\textbf{RLL} & \textit{\textbf{UNSAIL}} 
\\ \hline
c880 & 
0.77 & 0.47 & 
 
0.52 & 0.59 & 
 
0.31 & 0.60 & 
 
0.26 & 0.41 & 
 
0.38 & 0.49 
\\ \hline
c1355 & 
0.66 & 0.63 & 
 
0.54 & 0.48 & 
 
0.38 & 0.40 & 
 
0.30 & 0.49 & 
 
0.36 & 0.50 
\\ \hline
c1908 & 
0.64 & 0.65 & 
 
0.51 & 0.54 & 
 
0.51 & 0.54 & 
 
0.29 & 0.50 & 
 
0.41 & 0.50 
\\ \hline
c2670 & 
0.81 & 0.63 & 
 
0.55 & 0.59 & 
 
0.32 & 0.55 & 
 
0.20 & 0.53 & 
 
0.41 & 0.52 
\\ \hline
c3540 & 
0.66 & 0.59 & 
 
0.47 & 0.58 & 
 
0.30 & 0.59 & 
 
0.30 & 0.58 & 
 
0.34 & 0.57 
\\ \hline
c5315 & 
0.69 & 0.57 & 
 
0.49 & 0.59 & 
 
0.30 & 0.54 & 
 
0.30 & 0.43 & 
 
0.43 & 0.52 
\\ \hline
c6288 & 
0.57 & 0.53 & 
 
0.54 & 0.57 & 
 
0.33 & 0.53 & 
 
0.25 & 0.52 & 
 
0.38 & 0.55 
\\ \hline
c7552 & 
0.62 & 0.45 & 
 
0.30 & 0.60 & 
 
0.28 & 0.59 & 
 
0.29 & 0.50 & 
 
0.40 & 0.55 
\\ \hline
\textbf{Average} & 
\textbf{0.68} & \textbf{0.57} & 
 
\textbf{0.49} & \textbf{0.57} & 
 
\textbf{0.34} & \textbf{0.54} & 
 
\textbf{0.27} & \textbf{0.50} & 
 
\textbf{0.39} & \textbf{0.53} 
\\ \hline
& \multicolumn{10}{c}{\textbf{K=512}} 
\\ \hline
b14\_C & 
0.71 & 0.59 & 
 
0.46 & 0.56 & 
 
0.43 & 0.6 & 
 
0.30 & 0.47 & 

0.42 & 0.48
\\ \hline
b15\_C & 
0.77 & 0.66 & 
 
0.49 & 0.58 & 
 
0.49 & 0.61 & 
 
0.32 & 0.49 & 

0.46 & 0.52
\\ \hline
b20\_C & 
0.71 & 0.64 & 
 
0.47 & 0.56 & 
 
0.50 & 0.59 & 
 
0.35 & 0.44 & 

0.45 & 0.49
\\ \hline
b21\_C &
0.67 & 0.59 & 
 
0.46 & 0.55 & 
 
0.48 & 0.56 & 
 
0.29 & 0.46 & 

0.45 & 0.51
\\ \hline
b22\_C &
0.68 & 0.58 & 
 
0.46 & 0.59 & 
 
0.50 & 0.57 & 
 
0.34 & 0.50 & 

0.42 & 0.50 
\\ \hline
b17\_C &
0.74 & 0.63 & 
 
0.48 & 0.49 & 
 
0.49 & 0.58 & 
 
0.32 & 0.48 & 
 
0.38 & 0.48 
\\ \hline
\textbf{Average} & 
\textbf{0.71} & \textbf{0.62} & 
 
\textbf{0.47} & \textbf{0.56} & 
 
\textbf{0.48} & \textbf{0.59} & 

\textbf{0.32} & \textbf{0.47} & 
 
\textbf{0.43} & \textbf{0.50} 
\\ \hline 
\end{tabular}
}
\end{table}

\begin{table}[tb]
\centering
\scriptsize
\caption{Key-Gate Detection Accuracy Using SAIL Change-Prediction-Boosted Reconstruction Mode on \textit{UNSAIL} Vs.\ SLL and FLL when \texttt{K=128} and Using \texttt{sub=6} for the Change-Prediction Model}
\label{tab:boosted_sll_fll}
\resizebox{\textwidth}{!}{
\begin{tabular}{ccccc}
\hline
\textbf{Insertion} & \textbf{SLL} & \textbf{SLL-based \textit{UNSAIL}} & \textbf{FLL} & 
\textbf{FLL-based \textit{UNSAIL}} 
\\ \hline
c880 & 0.73 & 0.49 & 0.47 & 0.51 
\\ \hline
c1355 & 0.51 & 0.51 & 0.48 & 0.5 
\\ \hline
c1908 & 0.45 & 0.52 & 0.46 & 0.45 
\\ \hline
c2670 & 0.63 & 0.33 & 0.36 & 0.39 
\\ \hline
c3540 & 0.67 & 0.57 & 0.45 & 0.57 
\\ \hline
c5315 & 0.37 & 0.38 & 0.49 & 0.39 
\\ \hline
c7552 & 0.52 & 0.45 & 0.54 & 0.28 
\\ \hline
\textbf{Average} & 
\textbf{0.55} & 
\textbf{0.46} & 
\textbf{0.46} & 
\textbf{0.44} 
\\ \hline
\end{tabular}
}
\end{table}

\subsection{SWEEP Attack~\cite{alaql2019sweep} on \textit{UNSAIL} Vs.\ RLL}
\label{Sec:Sweep_results}
Here, we launch the SWEEP attack on locked ISCAS-85 and ITC-99 benchmarks. 
The attack model is trained using the variations of key-gate types shown in Table~\ref{tab:variations}. 
The accuracy metric is used as suggested in~\cite{alaql2019sweep};
it denotes the percentage of correctly extracted key-bits out of the entire key-size.
The attack was launched on RLL and \textit{UNSAIL}-locked instances; results are documented in Table~\ref{tab:SWEEP_results}.

\begin{figure}[tb]
\centering
\includegraphics[width=0.8\textwidth]{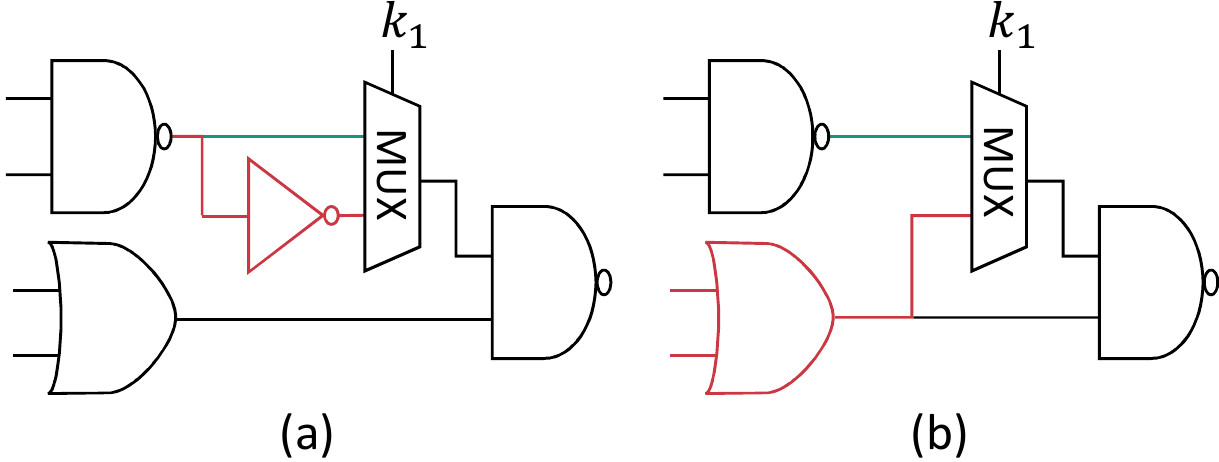}
\caption{Example of logic locking using MUXes. The true path is denoted by green, while the false path is denoted by red. 
(a) MUX key-gate inserted by \textit{UNSAIL}. 
The false wire is the negation of the true wire.
(b) MUX key-gate inserted by traditional logic locking. The false wire is taken from the design.}
\label{fig:MUX_locking_example}
\end{figure}

First, SWEEP does not cope well with X(N)OR locking, as indicated in~\cite{alaql2019sweep}. 
We experimentally verify this through the low accuracy of \texttt{33\%} observed on the RLL X(N)OR benchmarks.
Second, although SWEEP was explicitly developed to attack MUX-based locking, \ul{the accuracy of our MUX-based CL techniques for RLL is relatively low}, with an average of \texttt{41.63\%}.
For the related CL approach, we replace an X(N)OR key-gate by a MUX, with one input driven by the true wire/signal as is and the other input driven by the false signal, which is simply the true signal inverted (Fig.~\ref{fig:MUX_locking_example}(a)). 
Hence, depending on the key-bit assignment (to the MUX select line), the MUX key-gate will either be replaced by a buffer or an inverter by the synthesis tool run by SWEEP. 
Accordingly, few structural changes are induced, which can be extracted for the training of the SWEEP model. 
This is in contrast to other techniques broken by SWEEP, e.g., we observe that SWEEP was able to handle FLL (Fig.~\ref{fig:MUX_locking_example}(b)) significantly better, with an average accuracy of \texttt{76.3\%}.\footnote{%
For FLL, there is a specific algorithm underlying to select the true and false wires connected to the MUX key-gates~\cite{JV-Tcomp-2013}.
Depending on the key-bit, different fan-in cones will be fed to the MUX inputs (Fig.~\ref{fig:MUX_locking_example}(b)), resulting in various synthesis-induced changes for those fan-in cones, which enable SWEEP to learn the correlation between the extracted features and the correct key-bit. 
Even if the selection of wires is randomized, the wrong key-bit assignments still result in larger fan-in cones/logic structures on average when compared to the correct key-bit assignment, as indicated in~\cite{alaql2019sweep}.}

Third, SWEEP was launched on RLL vs.\ \textit{UNSAIL}-locked instances, to study the effect of adding \textit{UNSAIL} key-gate structures on the performance of SWEEP.
On average, \ul{\textit{UNSAIL} degrades the performance of the attack by \texttt{15pp}, which demonstrates that \textit{UNSAIL} hardens locking against another ML-based attack, not only SAIL.}

\begin{table}[tb]
\centering
\caption{Accuracy of SWEEP Attack~\cite{alaql2019sweep} on RLL Vs. \textit{UNSAIL}}
\label{tab:SWEEP_results}
\resizebox{\textwidth}{!}{%
\begin{tabular}{ccccccccccc}
\hline
& \multicolumn{10}{c}{\textbf{K=128}} 
\\ \hline
\textbf{Key-gates} & 
\multicolumn{2}{c}{\textbf{X(N)OR}} & 
 
\multicolumn{2}{c}{\textbf{CL\_v1}} & 
 
\multicolumn{2}{c}{\textbf{CL\_v2}} & 
 
\multicolumn{2}{c}{\textbf{CL\_v3}} & 
 
\multicolumn{2}{c}{\textbf{CL\_v4}} 
\\ \hline
\textbf{Insertion} & 
\textbf{RLL} & \textit{\textbf{UNSAIL}} & 
 
\textbf{RLL} & \textit{\textbf{UNSAIL}} & 
 
\textbf{RLL} & \textit{\textbf{UNSAIL}} & 
 
\textbf{RLL} & \textit{\textbf{UNSAIL}} & 
 
\textbf{RLL} & \textit{\textbf{UNSAIL}} 
\\ \hline
c880 & 
0.35 & 0.18 & 
 
0.24 & 0.06 & 
 
0.41 & 0.26 & 
 
0.36 & 0.28 & 
 
0.34 & 0.31 
\\ \hline
c1355 & 
0.04 & 0.06 & 
 
0.20 & 0.14 & 
 
0.16 & 0.07 & 
 
0.18 & 0.09 & 
 
0.36 & 0.21 
\\ \hline
c1908 & 
0.13 & 0.14 & 
 
0.16 & 0.12 & 
 
0.20 & 0.12 & 
 
0.20 & 0.15 & 
 
0.39 & 0.17 
\\ \hline
c2670 & 
0.41 & 0.23 & 
 
0.38 & 0.24 & 
 
0.29 & 0.17 & 
 
0.34 & 0.17 & 
 
0.44 & 0.23 
\\ \hline
c3540 & 
0.33 & 0.28 & 
 
0.37 & 0.25 & 
 
0.38 & 0.29 & 
 
0.31 & 0.27 & 
 
0.52 & 0.30 
\\ \hline
c5315 & 
0.41 & 0.19 & 
 
0.48 & 0.27 & 
 
0.50 & 0.27 & 
 
0.48 & 0.22 & 
 
0.62 & 0.27 
\\ \hline
c6288 & 
0.25 & 0.20 & 
 
0.30 & 0.27 & 
 
0.27 & 0.20 & 
 
0.36 & 0.22 & 
 
0.42 & 0.27 
\\ \hline
c7552 & 
0.27 & 0.12 & 
 
0.44 & 0.17 & 
 
0.37 & 0.20 & 
 
0.41 & 0.23 & 
 
0.43 & 0.29 
\\ \hline
\textbf{Average} & 
\textbf{0.27} & \textbf{0.17} & 
 
\textbf{0.32} & \textbf{0.19} & 
 
\textbf{0.32} & \textbf{0.20} & 
 
\textbf{0.33} & \textbf{0.21} & 
 
\textbf{0.44} & \textbf{0.25} 
\\ \hline
& \multicolumn{10}{c}{\textbf{K=512}} 
\\ \hline
b14\_C & 
0.39 & 0.28 & 
 
0.50 & 0.30 & 
 
0.46 & 0.27 & 
 
0.47 & 0.24 & 
 
0.53 & 0.35 
\\ \hline
b15\_C & 
0.41 & 0.33 & 
 
0.45 & 0.29 & 
 
0.44 & 0.25 & 
 
0.46 & 0.23 & 
 
0.52 & 0.39 
\\ \hline
b20\_C & 
0.37 & 0.38 & 
 
0.50 & 0.30 & 
 
0.44 & 0.28 & 
 
0.44 & 0.32 & 
 
0.50 & 0.40 
\\ \hline
b21\_C & 
0.38 & 0.33 & 
 
0.44 & 0.25 & 
 
0.44 & 0.25 & 
 
0.47 & 0.25 & 
 
0.53 & 0.37 
\\ \hline
b22\_C & 
0.38 & 0.29 & 
 
0.48 & 0.27 & 
 
0.47 & 0.30 & 
 
0.44 & 0.22 & 
 
0.53 & 0.40 
\\ \hline
b17\_C & 
0.43 & 0.34 & 
 
0.49 & 0.24 & 
 
0.47 & 0.25 & 
 
0.47 & 0.29 & 
 
0.57 & 0.38 
\\ \hline
\textbf{Average} & 
\textbf{0.39} & \textbf{0.33} & 
 
\textbf{0.48} & \textbf{0.27} & 
 
\textbf{0.45} & \textbf{0.27} & 
 
\textbf{0.46} & \textbf{0.26} & 
 
\textbf{0.53} & \textbf{0.38} 
\\ \hline
\end{tabular}
}
\end{table}

\subsection{Redundancy Attack~\cite{li2019piercing} on \textit{UNSAIL} Vs. RLL}
\label{Sec:Redundancy_result}

We launched the redundancy attack on RLL and \textit{UNSAIL}-locked ISCAS-85 and ITC-99 benchmarks for \texttt{K=128}.
We demonstrate the percentages of deciphered key-bits in Table~\ref{tab:redundancy_results}. 
In our analysis, we study the effect of different key-gates as well. 
The average attack accuracy on X(N)OR-based RLL-locked benchmarks is \texttt{42\%}. 
We observe that the attack is more successful on CL\_v4-based RLL, with an average accuracy of \texttt{52\%}. 
Note that CL\_v4 contains AND/OR key-gates where incorrect key-bits result in stuck-at-faults; such key-gates are more vulnerable to the redundancy attack.

The redundancy attack's accuracy is reported as high as \texttt{79.59\%} on X(N)OR-based RLL-locked ISCAS-85 benchmarks in~\cite{li2019piercing}, in contrast to the average accuracy of \texttt{25\%}, observed in our study for the same benchmarks. 
The degradation in accuracy observed here is contingent upon the re-synthesis step followed in our work. 
Note that industry-grade synthesis tools invoke redundancy checking and removal as an integral step of logic optimization~\cite{jiang2009logic,li2019piercing}. 
Thus, when the locked RLL benchmarks are re-synthesized, the synthesis tool removes redundancies in the netlist that could have been generated by incorrect key-bits. 
Consequently, we observe a low attack accuracy on both RLL and \textit{UNSAIL} schemes.\footnote{In fact, we have launched the redundancy attack on X(N)OR-based RLL-locked benchmarks without re-synthesis (in \textit{BENCH} format) and observed similar accuracy values as reported in~\cite{li2019piercing}.}
When comparing the resilience of baseline RLL with \textit{UNSAIL}, we note that \textit{UNSAIL} reduces the accuracy of the attack further by an average of \texttt{3.34pp} across all locking variations.

\begin{table}[tb]
\centering
\caption{Accuracy of Redundancy attack~\cite{li2019piercing} on \textit{UNSAIL} Vs. RLL}
\label{tab:redundancy_results}
\resizebox{\textwidth}{!}{%
\begin{tabular}{ccccccccccc}
\hline
& \multicolumn{10}{c}{\textbf{K=128}} \\ \hline
\textbf{Key-gates} & \multicolumn{2}{c}{\textbf{X(N)OR}} & \multicolumn{2}{c}{\textbf{CL\_v1}} & \multicolumn{2}{c}{\textbf{CL\_v2}} & \multicolumn{2}{c}{\textbf{CL\_v3}} & \multicolumn{2}{c}{\textbf{CL\_v4}} \\ \hline
\textbf{Insertion} & \textbf{RLL} & \textbf{\textit{UNSAIL}} & \textbf{RLL} & \textbf{\textit{UNSAIL}} & \textbf{RLL} & \textbf{\textit{UNSAIL}} & \textbf{RLL} & \textbf{\textit{UNSAIL}} & \textbf{RLL} & \textbf{\textit{UNSAIL}} \\ \hline
c880 & 0.10 & 0.06 & 0.08 & 0.10 & 0.12 & 0.12 & 0.11 & 0.12 & 0.34 & 0.25 \\ \hline
c1355 & 0.02 & 0.02 & 0.90 & 0.01 & 0.11 & 0.01 & 0.45 & 0.04 & 0.55 & 0.17 \\ \hline
c1908 & 0.08 & 0.08 & 0.01 & 0.11 & 0.05 & 0.06 & 0.05 & 0.07 & 0.29 & 0.14 \\ \hline
c2670 & 0.29 & 0.33 & 0.16 & 0.35 & 0.27 & 0.27 & 0.19 & 0.33 & 0.40 & 0.39 \\ \hline
c3540 & 0.52 & 0.54 & 0.42 & 0.56 & 0.43 & 0.36 & 0.38 & 0.41 & 0.80 & 0.47 \\ \hline
c5315 & 0.43 & 0.39 & 0.32 & 0.43 & 0.48 & 0.40 & 0.45 & 0.39 & 0.55 & 0.52 \\ \hline
c6288 & 0.31 & 0.26 & 0.20 & 0.23 & 0.31 & 0.26 & 0.27 & 0.26 & 0.52 & 0.31 \\ \hline
c7552 & 0.30 & 0.32 & 0.27 & 0.22 & 0.26 & 0.22 & 0.27 & 0.26 & 0.31 & 0.35 \\ \hline
\textbf{Average} & \textbf{0.26} & \textbf{0.25} & \textbf{0.29} & \textbf{0.25} & \textbf{0.25} & \textbf{0.21} & \textbf{0.27} & \textbf{0.23} & \textbf{0.47} & \textbf{0.32} \\ \hline
& \multicolumn{10}{c}{\textbf{K=128}} \\ \hline
b14 & 0.50 & 0.42 & 0.44 & 0.47 & 0.45 & 0.40 & 0.48 & 0.44 & 0.49 & 0.55 \\ \hline
b15 & 0.63 & 0.59 & 0.60 & 0.57 & 0.52 & 0.61 & 0.50 & 0.55 & 0.64 & 0.58 \\ \hline
b20 & 0.54 & 0.32 & 0.42 & 0.44 & 0.33 & 0.53 & 0.39 & 0.39 & 0.64 & 0.54 \\ \hline
b21 & 0.52 & 0.42 & 0.42 & 0.37 & 0.43 & 0.39 & 0.41 & 0.48 & 0.51 & 0.52 \\ \hline
b22 & 0.63 & 0.38 & 0.46 & 0.43 & 0.43 & 0.47 & 0.48 & 0.43 & 0.53 & 0.49 \\ \hline
b17 & 0.73 & 0.75 & 0.53 & 0.59 & 0.54 & 0.62 & 0.60 & 0.65 & 0.60 & 0.65 \\ \hline
\textbf{Average} & \textbf{0.59} & \textbf{0.48} & \textbf{0.48} & \textbf{0.48} & \textbf{0.45} & \textbf{0.50} & \textbf{0.48} & \textbf{0.49} & \textbf{0.57} & \textbf{0.56} \\ \hline
\end{tabular}%
}
\end{table}
\subsection{Hamming Distance (HD) and Output Error Rate (OER) Analysis on \textit{UNSAIL} Vs.\ RLL}
\label{sec:HD_analysis}

Next, we calculate the HD and OER between the original benchmark outputs and the outputs of the RLL and \textit{UNSAIL}-locked instances by applying random keys. 
This is done to quantify the level of functional obfuscation for the logic locking techniques in general, independent of an actual attack.
For each benchmark, \texttt{100} random keys are chosen, and the locked instances' outputs are compared with the golden outputs by applying \texttt{10,000} random input patterns.
The results are documented in Table~\ref{tab:HD}.

It is observed that, with an increase in key-size, HD increases for ISCAS-85 and ITC-99 benchmarks.
This is intuitive, as an increase in key-size/number of key-inputs allows for a more widespread propagation of potentially false key-bit assignments (albeit that remains subject to the netlist structure) and, in turn, the possibility for more output corruption. 

Comparing \textit{UNSAIL} vs. RLL, RLL achieves a higher HD, by an average of \texttt{3.25pp}. 
This is because \textit{UNSAIL} inserts key-gates in specific locations that would affect the performance of the ML models used by SAIL but might
not be effective in propagating the effect of faults to the outputs when incorrect key-bits are applied.
Next, we also calculate OER for \textit{UNSAIL}-locked benchmarks; ideally, OER should be \texttt{100\%}. 
The average OER obtained for \textit{UNSAIL} on ISCAS-85 benchmarks (\texttt{K=64} and \texttt{K=128}) is \texttt{99.95\%}, whereas
the OER obtained for ITC-99 benchmarks with \texttt{K=256} and \texttt{K=512} is \texttt{100\%} for all the test cases.

\begin{table*}[tb]
\centering
\caption{HD Results for RLL and \textit{UNSAIL} Schemes on Selected ISCAS-85 Benchmarks Upon Applying \texttt{100} Random Keys and \texttt{10,000} Random Input Patterns for each Key Assignment}
\label{tab:HD}
\resizebox{\textwidth}{!}{%
\begin{tabular}{ccccccccccccccccccccccccccccccc}
\hline
& 
\multicolumn{14}{c}{\textbf{RLL}} & 
& 
& \multicolumn{14}{c}{\textit{\textbf{UNSAIL}}} 
\\ \hline
\textbf{Key-gates} & 
\multicolumn{2}{c}{\textbf{X(N)OR}} & 
& 
\multicolumn{2}{c}{\textbf{CL\_v1}} & 
& 
\multicolumn{2}{c}{\textbf{CL\_v2}} & 
& 
\multicolumn{2}{c}{\textbf{CL\_v3}} & 
& 
\multicolumn{2}{c}{\textbf{CL\_v4}} & 
& 
& 
\multicolumn{2}{c}{\textbf{X(N)OR}} & 
& 
\multicolumn{2}{c}{\textbf{CL\_v1}} & 
& 
\multicolumn{2}{c}{\textbf{CL\_v2}} & 
& 
\multicolumn{2}{c}{\textbf{CL\_v3}} & 
& 
\multicolumn{2}{c}{\textbf{CL\_v4}} 
\\ \hline
\textbf{Key-size} & 
\textbf{64} & \textbf{128} & 
& 
\textbf{64} & \textbf{128} & 
& 
\textbf{64} & \textbf{128} & 
& 
\textbf{64} & \textbf{128} & 
& 
\textbf{64} & \textbf{128} & 
& 
& 
\textbf{64} & \textbf{128} & 
& 
\textbf{64} & \textbf{128} & 
& 
\textbf{64} & \textbf{128} & 
& 
\textbf{64} & \textbf{128} & 
& 
\textbf{64} & \textbf{128} 
\\ \hline
c880 & 
0.28 & 0.34 & 
& 
0.29 & 0.38 & 
& 
0.26 & 0.37 & 
& 
0.25 & 0.37 & 
& 
0.29 & 0.29 & 
& 
& 
0.18 & 0.39 & 
& 
0.28 & 0.40 & 
& 
0.24 & 0.36 & 
& 
0.30 & 0.34 & 
& 
0.29 & 0.37 
\\ \hline
c1355 & 
0.33 & 0.46 & 
& 
0.22 & 0.46 & 
& 
0.32 & 0.43 & 
& 
0.33 & 0.43 & 
& 
0.29 & 0.44 & 
& 
& 
0.24 & 0.38 & 
& 
0.16 & 0.32 & 
& 
0.19 & 0.36 & 
& 
0.25 & 0.39 & 
& 
0.28 & 0.34 
\\ \hline
c1908 & 
0.33 & 0.43 & 
& 
0.32 & 0.43 & 
& 
0.30 & 0.44 & 
& 
0.29 & 0.40 & 
& 
0.30 & 0.33 & 
& 
& 
0.23 & 0.35 & 
& 
0.28 & 0.35 & 
& 
0.27 & 0.39 & 
& 
0.33 & 0.32 & 
& 
0.23 & 0.35 
\\ \hline
c2670 & 
0.09 & 0.11 & 
& 
0.09 & 0.12 & 
& 
0.11 & 0.14 & 
& 
0.06 & 0.13 & 
& 
0.11 & 0.12 & 
& 
& 
0.10 & 0.11 & 
& 
0.09 & 0.11 & 
& 
0.09 & 0.14 & 
& 
0.08 & 0.12 & 
& 
0.08 & 0.13 
\\ \hline
c3540 & 
0.35 & 0.44 & 
& 
0.30 & 0.36 & 
& 
0.30 & 0.40 & 
& 
0.34 & 0.39 & 
& 
0.31 & 0.40 & 
& 
& 
0.27 & 0.43 & 
& 
0.31 & 0.34 & 
& 
0.27 & 0.33 & 
& 
0.29 & 0.35 & 
& 
0.29 & 0.37 
\\ \hline
c5315 & 
0.17 & 0.21 & 
& 
0.15 & 0.22 & 
& 
0.15 & 0.20 & 
& 
0.14 & 0.22 & 
& 
0.15 & 0.19 & 
& 
& 
0.11 & 0.18 & 
& 
0.11 & 0.15 & 
& 
0.14 & 0.17 & 
& 
0.12 & 0.16 & 
& 
0.10 & 0.15 
\\ \hline
c6288 & 
0.35 & 0.42 & 
& 
0.36 & 0.41 & 
& 
0.35 & 0.45 & 
& 
0.37 & 0.45 & 
& 
0.33 & 0.43 & 
& 
& 
0.36 & 0.37 & 
& 
0.33 & 0.39 & 
& 
0.36 & 0.35 & 
& 
0.32 & 0.40 & 
& 
0.33 & 0.36 
\\ \hline
c7552 & 
0.14 & 0.21 & 
& 
0.16 & 0.19 & 
& 
0.15 & 0.18 & 
& 
0.13 & 0.16 & 
& 
0.13 & 0.19 & 
& 
& 
0.10 & 0.13 & 
& 
0.12 & 0.16 & 
& 
0.10 & 0.17 & 
& 
0.17 & 0.16 & 
& 
0.14 & 0.15 
\\ \hline
\textbf{Average} & 
\textbf{0.25} & \textbf{0.33} & 
& 
\textbf{0.24} & \textbf{0.32} & 
& 
\textbf{0.24} & \textbf{0.33} & 
& 
\textbf{0.24} & \textbf{0.32} & 
& 
\textbf{0.24} & \textbf{0.30} & 
& 
& 
\textbf{0.20} & \textbf{0.29} & 
& 
\textbf{0.21} & \textbf{0.28} & 
& 
\textbf{0.21} & \textbf{0.28} & 
& 
\textbf{0.23} & \textbf{0.28} & 
& 
\textbf{0.22} & \textbf{0.28} 
\\ \hline
\\ \hline
\textbf{Key-size} & 
\textbf{256} & \textbf{512} & 
& 
\textbf{256} & \textbf{512} & 
& 
\textbf{256} & \textbf{512} & 
& 
\textbf{256} & \textbf{512} & 
& 
\textbf{256} & \textbf{512} & 
& 
& 
\textbf{256} & \textbf{512} & 
& 
\textbf{256} & \textbf{512} & 
& 
\textbf{256} & \textbf{512} & 
& 
\textbf{256} & \textbf{512} & 
& 
\textbf{256} & \textbf{512} 
\\ \hline
b14\_C & 
0.17 & 0.25 & 
& 
0.16 & 0.29 & 
& 
0.16 & 0.30 & 
& 
0.18 & 0.30 & 
& 
0.16 & 0.26 & 
& 
& 
0.17 & 0.22 & 
& 
0.17 & 0.20 & 
& 
0.12 & 0.21 & 
& 
0.13 & 0.20 & 
& 
0.15 & 0.27 
\\ \hline
b15\_C & 
0.15 & 0.23 & 
& 
0.14 & 0.24 & 
& 
0.14 & 0.22 & 
& 
0.13 & 0.24 & 
& 
0.12 & 0.18 & 
& 
& 
0.09 & 0.18 & 
& 
0.12 & 0.15 & 
& 
0.11 & 0.19 & 
& 
0.12 & 0.15 & 
& 
0.08 & 0.18 
\\ \hline
b20\_C & 
0.10 & 0.20 & 
& 
0.13 & 0.18 & 
& 
0.12 & 0.18 & 
& 
0.09 & 0.17 & 
& 
0.10 & 0.15 & 
& 
& 
0.10 & 0.15 & 
& 
0.07 & 0.17 & 
& 
0.08 & 0.15 & 
& 
0.07 & 0.12 & 
& 
0.06 & 0.15 
\\ \hline
b21\_C & 
0.12 & 0.17 & 
& 
0.12 & 0.20 & 
& 
0.10 & 0.19 & 
& 
0.12 & 0.17 & 
& 
0.10 & 0.16 & 
& 
& 
0.08 & 0.15 & 
& 
0.07 & 0.16 & 
& 
0.11 & 0.14 & 
& 
0.08 & 0.15 & 
& 
0.07 & 0.14 
\\ \hline
b22\_C & 
0.07 & 0.16 & 
& 
0.07 & 0.16 & 
& 
0.08 & 0.13 & 
& 
0.08 & 0.16 & 
& 
0.06 & 0.12 & 
& 
& 
0.07 & 0.15 & 
& 
0.09 & 0.10 & 
& 
0.07 & 0.10 & 
& 
0.06 & 0.11 & 
& 
0.03 & 0.12 
\\ \hline
b17\_C & 
0.05 & 0.10 & 
& 
0.05 & 0.09 & 
& 
0.05 & 0.08 & 
& 
0.05 & 0.10 & 
& 
0.05 & 0.09 & 
& 
& 
0.04 & 0.09 & 
& 
0.03 & 0.08 & 
& 
0.02 & 0.07 & 
& 
0.03 & 0.07 & 
& 
0.03 & 0.06 
\\ \hline
\textbf{Average} & 
\textbf{0.11} & \textbf{0.19} & 
& 
\textbf{0.11} & \textbf{0.19} & 
& 
\textbf{0.11} & \textbf{0.18} & 
& 
\textbf{0.11} & \textbf{0.19} & 
& 
\textbf{0.10} & \textbf{0.16} & 
& 
& 
\textbf{0.09} & \textbf{0.16} & 
& 
\textbf{0.09} & \textbf{0.14} & 
& 
\textbf{0.08} & \textbf{0.14} & 
& 
\textbf{0.08} & \textbf{0.13} & 
& 
\textbf{0.07} & \textbf{0.15} 
\\ \hline
\end{tabular}
}
\end{table*}

\subsection{\textit{UNSAIL} Test and Fault Coverage}
\label{sec:test}

Here, we study the impact of \textit{UNSAIL} on the testability of the overall design.
We report the fault coverage and test coverage for locked benchmarks without key
constraints, as recommended in~\cite{yasin_DATE_2016}.
For context, we also obtain the coverage values for the original designs.
Fault coverage represents the percentage of detected faults out of the total faults in the design while test coverage represents the percentage
of the detected faults out of the detectable faults in the design~\cite{bushnell2004essentials}.

The test coverage for the original ISCAS-85 and ITC-99 benchmarks is \texttt{100\%} for all benchmarks. 
The average fault coverage is \texttt{99.98\%} and \texttt{99.92\%} for ISCAS-85 and ITC-99 benchmarks, respectively. 
For \textit{UNSAIL}-locked benchmarks, the test coverage remains at \texttt{100\%} for all locked benchmarks. 
Moreover, the average fault coverage is \texttt{99.71\%} and \texttt{99.96\%} for the locked ISCAS-85 and ITC-99 benchmarks, respectively. 
These experiments illustrate that \textit{UNSAIL} does not negatively impact the testability of the underlying designs.

\subsection{Implementation Overheads}
\label{sec:methodology_overheads}

\textbf{Obfuscation Time.} To obfuscate a design using \textit{UNSAIL}, the design must be initially locked using a traditional logic locking algorithm (e.g., RLL, FLL, etc.) and then synthesized using a synthesis tool of choice. 
The pre- and post-subgraphs must be extracted and compared, and the additional \textit{UNSAIL} key-gates structures must be inserted.
In our experiments, the extraction of subgraphs took less than a second for ISCAS-85 benchmarks for both \texttt{K=64} and \texttt{K=128}, respectively. 
For ITC-99 benchmarks, the extraction of subgraphs took between \texttt{70s-138s} for \texttt{K=512}.
In case of locking the GPS module in the ORPSoC design, the extraction took on average \texttt{6.5h}. 
Insertion of \textit{UNSAIL} key-gates took less than a second for ISCAS-85 benchmarks, around \texttt{105s} for the ITC-99 benchmarks, and on
average \texttt{6h} for the GPS module, which demonstrates the scalability of our approach also for large designs. 

\begin{figure}[tb]
\centering
\subfloat[]{\includegraphics[width=.48\textwidth]{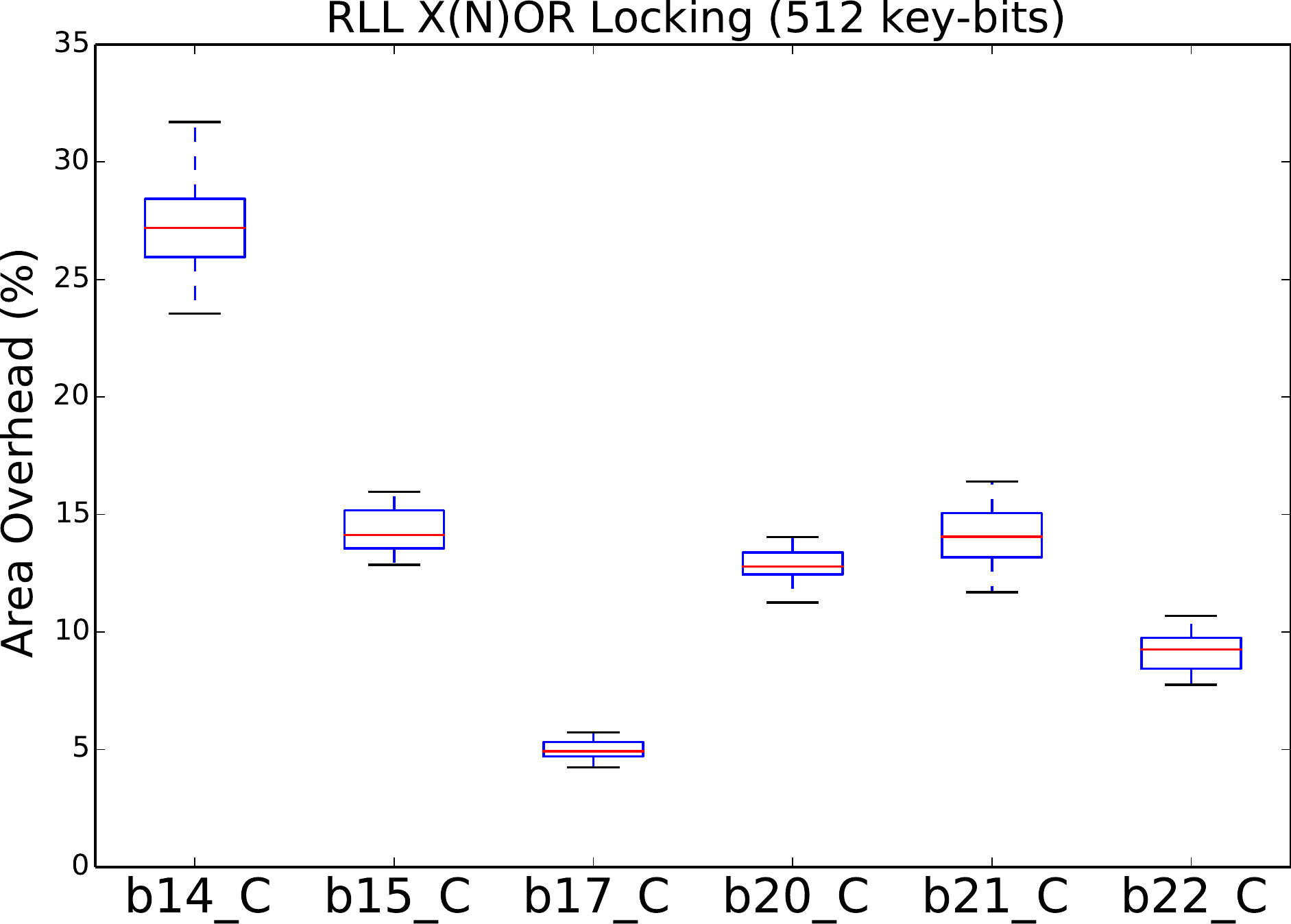}}\hfill
\subfloat[]{\includegraphics[width=.48\textwidth]{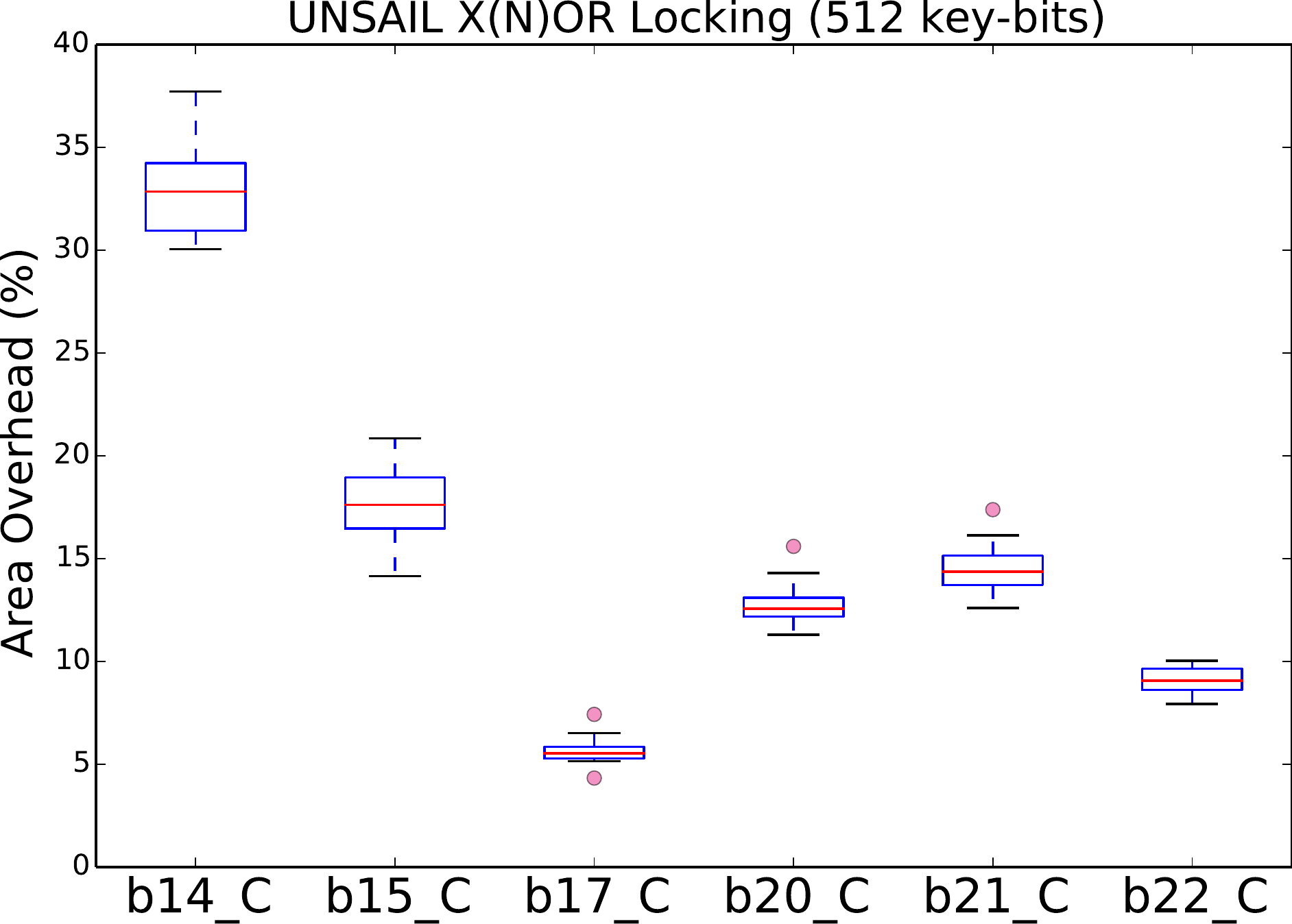}}\\
\caption{Area overheads for selected ITC-99 benchmarks locked with \texttt{K=512} at iso-performance of \texttt{200} MHz. 
(a)~RLL with X(N)OR key-gates.
(b)~\textit{UNSAIL} integrated with RLL using X(N)OR key-gates.
Each box consists of \texttt{20} trials, the boxes span from the 5th to the 95th percentile, the whiskers indicate the minimum and maximum values, the red bars indicate the median, and the red dots represent outliers, respectively.}
\label{fig:area_overheads_RLL_UNSAIL_X(N)OR}
\end{figure}

\begin{figure}[tb]
\centering
\subfloat[]{\includegraphics[width=.485\textwidth]{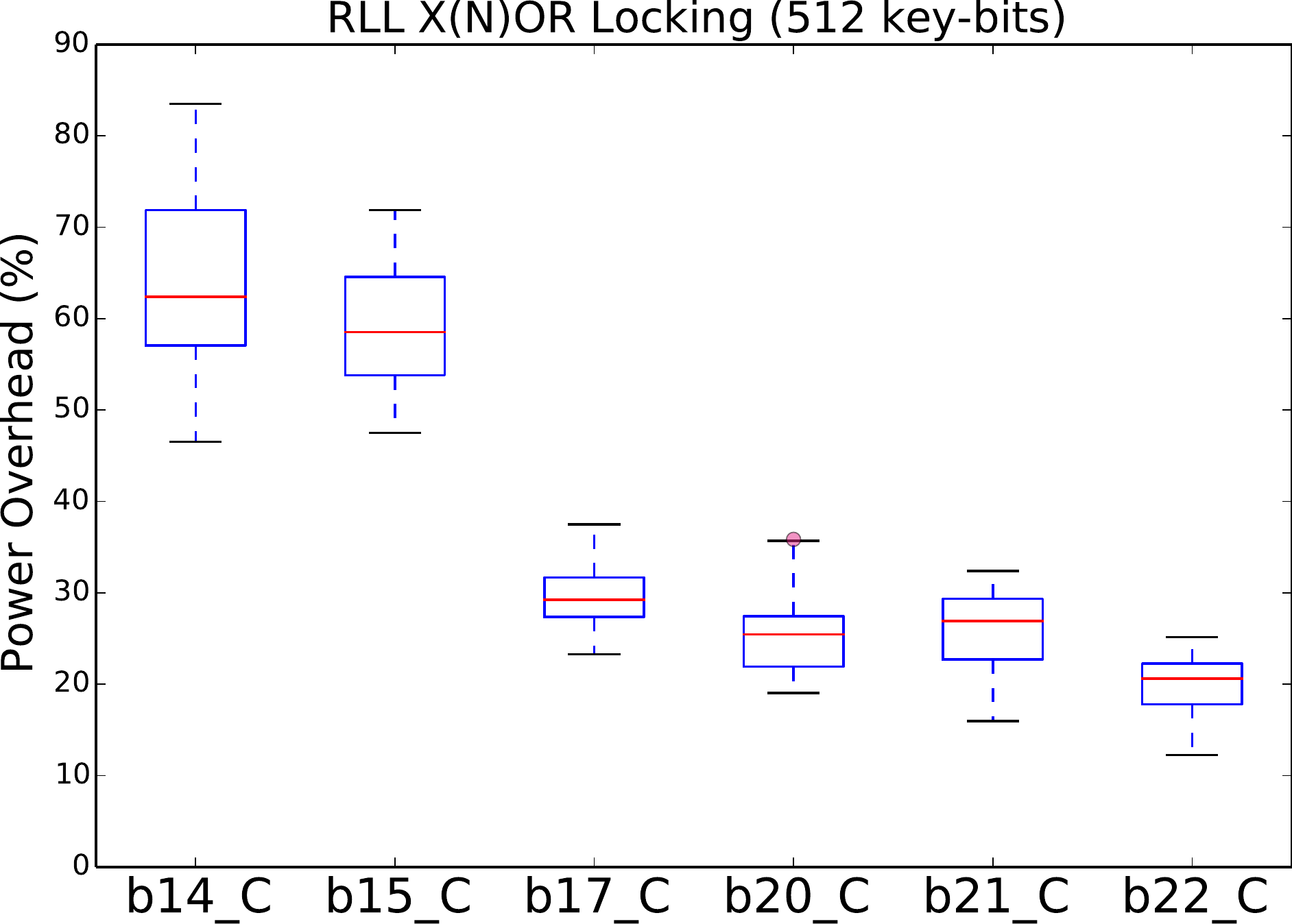}}\hfill
\subfloat[]{\includegraphics[width=.485\textwidth]{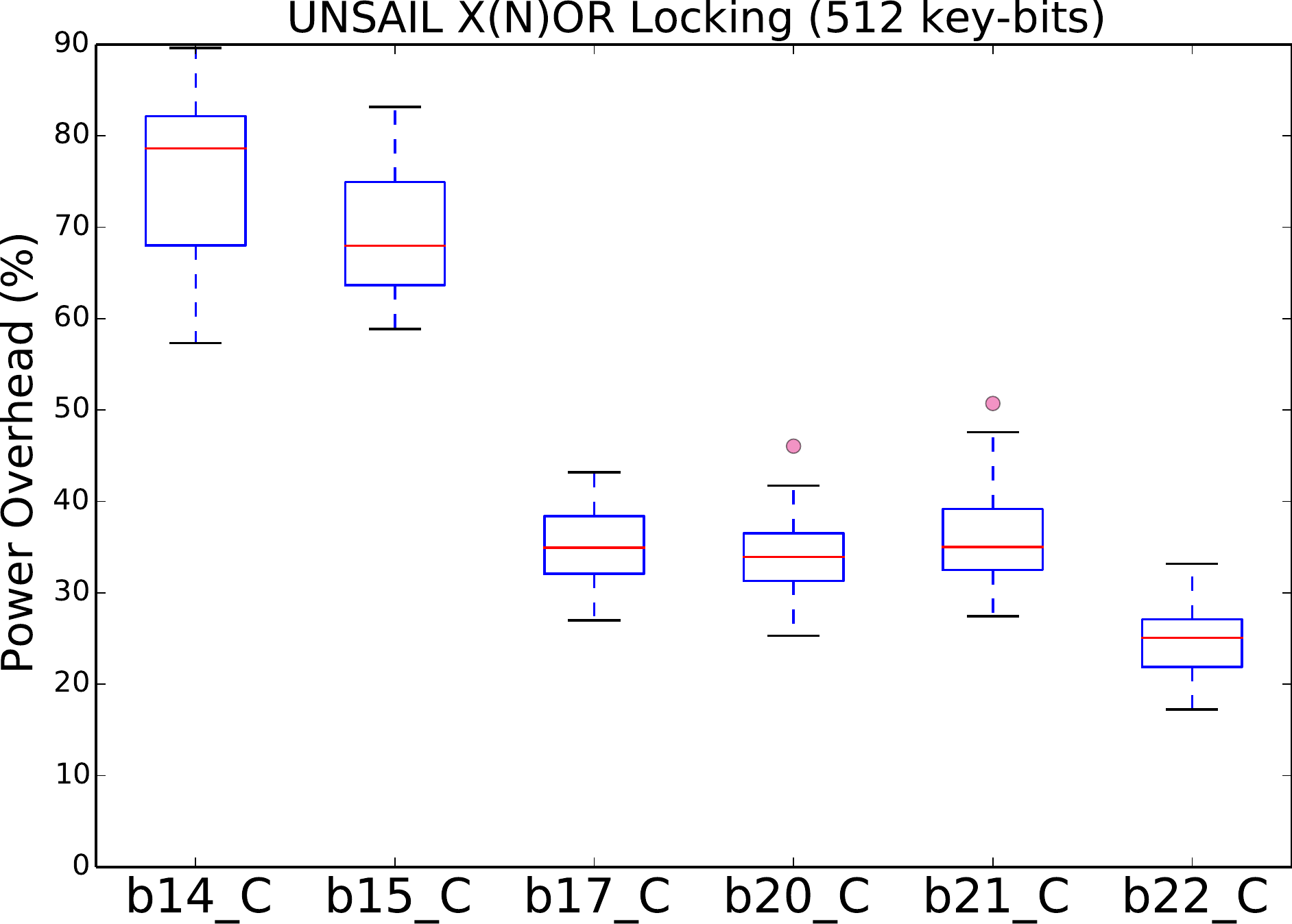}}\\
\caption{Power overheads for selected ITC-99 benchmarks locked with \texttt{K=512} at iso-performance of \texttt{200} MHz.
(a)~RLL with X(N)OR key-gates.
(b)~\textit{UNSAIL} integrated with RLL using X(N)OR key-gates.
The details regarding boxes are the same as in Fig.~\ref{fig:area_overheads_RLL_UNSAIL_X(N)OR}.}
\label{fig:power_overheads_RLL_UNSAIL_X(N)OR}
\end{figure}

\textbf{Layout Cost Incurred by \textit{UNSAIL}.} The area and power overheads for RLL and \textit{UNSAIL}-locked instances of ITC-99 benchmarks are reported for the largest key-size of \texttt{K=512}.
Area and power overheads for all the benchmarks have been obtained for an iso-performance layout implementation considering \texttt{5} ns timing constraint.
Area overheads for RLL and \textit{UNSAIL}-locked instances using X(N)OR key-gates are shown in Fig.~\ref{fig:area_overheads_RLL_UNSAIL_X(N)OR}(a) and Fig.~\ref{fig:area_overheads_RLL_UNSAIL_X(N)OR}(b), respectively; the related power overheads are illustrated in Fig.~\ref{fig:power_overheads_RLL_UNSAIL_X(N)OR}(a) and Fig.~\ref{fig:power_overheads_RLL_UNSAIL_X(N)OR}(b).

We note that \textit{UNSAIL} increases the area overheads (by \texttt{0.37\%}--\texttt{5.79\%}) and power overheads (by \texttt{4.88\%}--\texttt{14.17\%}) compared to the baseline RLL, i.e., at iso-performance.
This is because of two reasons playing out at the synthesis and physical-layout level. First, the synthesis tool is unconstrained after the
insertion of \textit{UNSAIL} key-gate structures. Thus, while more secure than RLL, the \textit{UNSAIL} netlists are less optimized with regards to area and power. 
Second, as our physical-layout flow is optimized for timing closure, algorithms internally invoked by \textit{Cadence Innovus}, like the insertion of buffers and/or upsizing of gates, as well as further re-routing of nets cause an increase in both area and power overheads.
However, the cost for \textit{UNSAIL} is amortized for large, million-gate-based designs, as illustrated next.

\subsection{Results on DARPA OpenCores Benchmark~\cite{CEP_github}}
\label{DARPA}
For the \textit{DARPA CEP} benchmark~\cite{CEP_github}, also known as ORPSoC, we lock the sensitive GPS module using X(N)OR locking.
The SAIL RF classifier was trained and tested on both RLL and \textit{UNSAIL} with \texttt{K=512}. 
The classification accuracy for the case of RLL with \texttt{sub=3} is \texttt{77\%}, whereas, for the case of \textit{UNSAIL}, it is \texttt{55\%}, i.e., only slightly better than random-guessing.
For \texttt{sub=6}, \textit{UNSAIL} still succeeds in lowering the classification accuracy, namely from \texttt{91\%} to \texttt{79\%}.
The model $\texttt{ML}_{\texttt{2}}$ is tested as well, and the key-gate recovery accuracy for RLL and \textit{UNSAIL} is \texttt{73\%} and \texttt{53\%}, respectively. 
Once both models are combined and the full SAIL attack is launched, the key-gate detection accuracy for RLL vs.\ \textit{UNSAIL} is reduced from \texttt{73\%} to \texttt{66\%}.
The area and power overheads for \textit{UNSAIL} using \texttt{K=512} are \texttt{0.26\%} and \texttt{0.61\%}, respectively, for iso-performance
at \texttt{100} MHz.

{In summary, we demonstrated that \textit{UNSAIL} is both effective and cost-efficient when protecting large designs against SAIL.}
\section{Discussion}
\label{sec:discussion}

\subsection{Impact of Re-Synthesizing \textit{UNSAIL}-locked Designs}

Essentially, for logic locking using \textit{UNSAIL}, several key-gate structures are added to confuse ML-based attacks.
A defender would like to have those structures injected by \textit{UNSAIL} ideally untouched. 
One might argue that an attacker could re-synthesize the \textit{UNSAIL}-locked designs to remove the subterfuge added by our defense. 
However, doing so will only increase the complexity of SAIL, as explained next.

The goal of SAIL is to obtain the locked netlist before re-synthesis, let us call it netlist \texttt{A}. 
This locked netlist is re-synthesized for obfuscation, providing netlist \texttt{B}. 
Next, \textit{UNSAIL} additionally locks the re-synthesized netlist, providing netlist \texttt{C}. 
If an attacker re-synthesizes the final locked design one more time, he/she will end up with netlist \texttt{D}. 
In such case, an even more powerful attack must be developed to revert all the changes, i.e., to go back from netlist $\texttt{D} \longrightarrow
\texttt{C} \longrightarrow \texttt{B} \longrightarrow \texttt{A}$.

\subsection{Impact of Layout Optimization on \textit{UNSAIL}}

We also compared the post-layout netlists to the pre-layout (i.e., post-synthesis) netlists,
to investigate whether the \textit{UNSAIL} key-gate structures are carried over or resolved by layout-level optimization techniques.
On average, we note that \texttt{10\%} of all key-gate structures are affected, i.e., they go through some optimization.
Still, we argue that such a transformation will not affect the overall resilience offered by \textit{UNSAIL}.
This is because an enhanced, yet-to-be-demonstrated two-step SAIL attack, capable of working on post-layout designs, would have to infer the
additional changes incurred due to layout-level optimizations.
Even when assuming such a powerful attack exists,
the attacker would still be left only with the post-synthesis netlist, which remains protected by \textit{UNSAIL}.

\subsection{Extended Threat Model}
\label{sec:Discussion_Threat}

As discussed, various attacks have challenged logic locking while leveraging an oracle~\cite{chakraborty2019keynote}; this
fact has resulted in broad efforts to protect against such \textit{oracle-guided} attacks.
However, recent ML-based,
\textit{oracle-less} attacks are considered more powerful, as they have shown to undermine the security promises of logic locking
already during the early stages of the IC supply chain, without requiring an oracle.

In this work, we have proposed and demonstrated \textit{UNSAIL} to protect logic locking against such potent \textit{oracle-less} attacks.
Recall that \textit{UNSAIL} is compatible with any traditional locking scheme of choice.
Given that traditional locking techniques are often incorporated with locking solutions resilient against SAT-based attacks,
we argue that by integrating such a resilient scheme also with \textit{UNSAIL}, the design could be protected from both
\textit{oracle-guided} and \textit{oracle-less} attacks at once.
Related efforts shall, however, remain scope for future work.

Note that pairing \textit{UNSAIL} with a SAT-resilient technique would not compromise the resilience offered by \textit{UNSAIL} against
SAIL. This is because any SAT-resilient technique is independent and separate from the \textit{UNSAIL} structures. Furthermore, SAT-resilient
techniques differ significantly from traditional locking in terms of logic and structural properties. Thus, in its current form, one cannot readily
apply SAIL to those resilient techniques, and it remains to be seen if SAIL could be tailored for such SAT-resilient techniques at all.
\section{Conclusion}
\label{sec:conclusion}

In this work, we initially implemented a reference platform for the SAIL attack, to thoroughly investigate the security of logic locking against
such an \textit{oracle-less} machine learning (ML)-based attack.
For the first time, our study considers various key-sizes, key-gate structures, and key-gate insertion heuristics. 
Among others, we find that compound logic locking, i.e., where various key-gate structures are randomly selected and used at once, tend to be more resilient.
Second, we presented a defense mechanism called \textit{UNSAIL}, which targets specifically at the training stage of such an ML-based attack.
The presented defense can be integrated with any combinational logic locking scheme, and we have considered various traditional logic locking schemes toward that end.

\textit{UNSAIL} serves to confuse the SAIL models by introducing additional structural transformations that these models cannot distinguish from regular ones (i.e., those introduced by synthesis tools for the sake of obfuscation, as is common practice with logic locking).
We have initially motivated the notion of \textit{UNSAIL} using Fisher's discriminant ratio, which demonstrated more complex classification
problems for SAIL in particular and classification-based attacks in general.
Besides SAIL, we show that our defense can hinder another potent oracle-less, ML-based attack, called SWEEP.
For both SAIL and SWEEP, we have performed a thorough evaluation
when different attack models and configurations are utilized.
Reflection of the results argues that \textit{UNSAIL} degrades the accuracy of all the stages/models of SAIL, achieving an overall reduction of
attack accuracy of \texttt{11} percentage points (\texttt{pp}); \textit{UNSAIL} also degrades the performance of the SWEEP attack by an average of \texttt{15pp}; all while
inducing only marginal layout overheads. 
We have demonstrated that \textit{UNSAIL} is further capable of thwarting non-ML-based
\textit{oracle-less} attacks, i.e., the redundancy attack specifically, which can recover the key-bits of \textit{UNSAIL}-locked
designs only with a low accuracy of \texttt{38\%} on average. Finally, \textit{UNSAIL}-locked designs can be activated post-testing, ensuring high
fault coverage and test quality all while additionally offering protection from an untrusted test facility.

\section*{Acknowledgments}

The authors are grateful to Leon Li and Prof.\ Alex Orailoglu (University of California, San Diego) for providing the redundancy attack~\cite{li2019piercing} for the experimental analysis.

\def\bibfont{\footnotesize}
\bibliography{main}
\bibliographystyle{IEEEtran}

\begin{IEEEbiography}[{\includegraphics[width=1in,height=1.25in,clip,keepaspectratio]{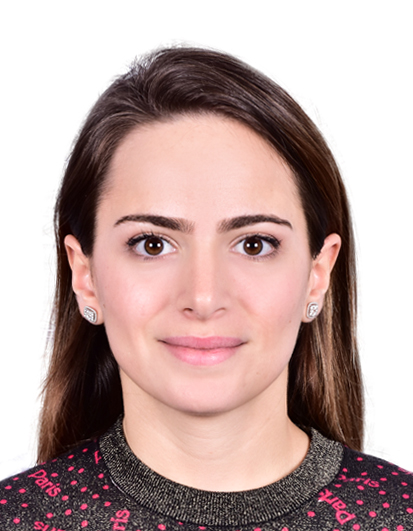}}]{Lilas Alrahis} received the M.Sc.\ degree and the Ph.D.\ degree in Electrical and Computer Engineering from Khalifa University, UAE, in 2016 and 2020, respectively. Her current research interests include hardware security, design-for-trust and applied machine learning. Lilas won the MWSCAS Myril B.\ Reed Best Paper Award in 2016 and the Best Paper Award at the Applied Research Competition held in conjunction with Cyber Security Awareness Week, in 2019. 
\end{IEEEbiography}

\begin{IEEEbiography}[{\includegraphics[width=1in,height=1.25in,clip,keepaspectratio]
{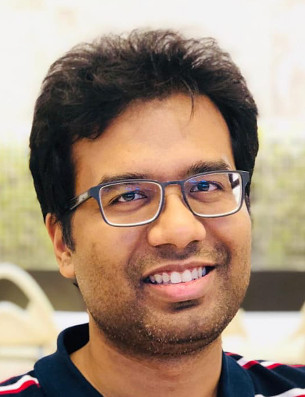}}]{Satwik Patnaik} received the B.E.\ degree in electronics and telecommunications from the University of Pune, India, the M.Tech.\ degree in computer science and engineering with a specialization in VLSI design from the Indian Institute of Information Technology and Management, Gwalior, India, and the Ph.D.\ degree in Electrical engineering from Tandon School of Engineering, New York University, Brooklyn, NY, USA in September 2020.

He is currently a Post-Doctoral Researcher with the Department of Electrical and Computer Engineering, Texas A\&M University, College Station, TX, USA.
His current research interests include hardware security, trust and reliability issues for CMOS and emerging devices with particular focus on low-power VLSI Design.
Dr.\ Patnaik received the Bronze Medal in the Graduate Category at the ACM/SIGDA Student Research Competition held at ICCAD 2018, and the Best Paper Award at the Applied Research Competition held in conjunction with Cyber Security Awareness Week, in 2017.
\end{IEEEbiography}

\begin{IEEEbiography}[{\includegraphics[width=1in,height=1.25in,clip,keepaspectratio]{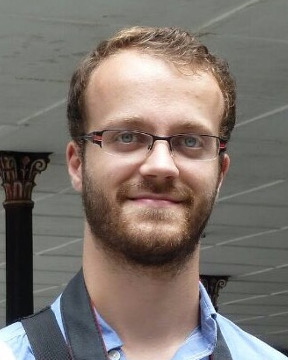}}]{Johann Knechtel}
received the M.Sc.\ degree in Information Systems Engineering (Dipl.-Ing.) and the Ph.D.\ degree in Computer Engineering (Dr.-Ing., summa cum laude) from TU Dresden, Germany, in 2010 and 2014, respectively.

 He is a Research Scientist with New York University Abu Dhabi, United Arab Emirates.
 From 2015 to 2016, he was a Postdoctoral Researcher with the Masdar Institute of Science and Technology, Abu Dhabi;
 from 2010 to 2014, he was a Ph.D.\ Scholar with the DFG Graduate School ``Nano- and Biotechnologies for Packaging of Electronic
 Systems'' hosted at TU Dresden;
 in 2012, he was a Research Assistant with the
 Chinese University of Hong Kong;
 and in 2010, he was a Visiting Research Student with the
 University of Michigan at Ann Arbor, MI, USA.
 His research interests cover VLSI physical design automation, with particular focus on emerging technologies and hardware security.
 He has (co-)authored around 50 publications.
\end{IEEEbiography}

\begin{IEEEbiography}[{\includegraphics[width=1in,height=1.25in,clip,keepaspectratio]{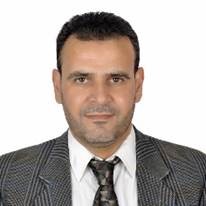}}]{Hani Saleh}
is an associate professor of electronic engineering at Khalifa University since
2012. He is a co-founder/active researcher in the
KSRC (Khalifa University Research Center) and
the System on Chip Research Center (SOCC).
Hani has a total of 19 years of industrial experience in ASIC chip design.
Prior to academia, Hani worked for many leading chip design companies including Apple, Intel,
AMD, Qualcomm, Synopsys, Fujitsu and Motorola.
Hani received a Bachelor of Science degree in
Electrical Engineering from the University of Jordan, a Master of Science
degree in Electrical Engineering from the University of Texas at San
Antonio, and a Ph.D.\ degree in Electrical and Computer Engineering
from the University of Texas at Austin. Hani's research interest includes
IoT Devices design, deep learning, DSP algorithms design, computer.
\end{IEEEbiography}

\begin{IEEEbiography}[{\includegraphics[width=1in,height=1.25in,clip,keepaspectratio]{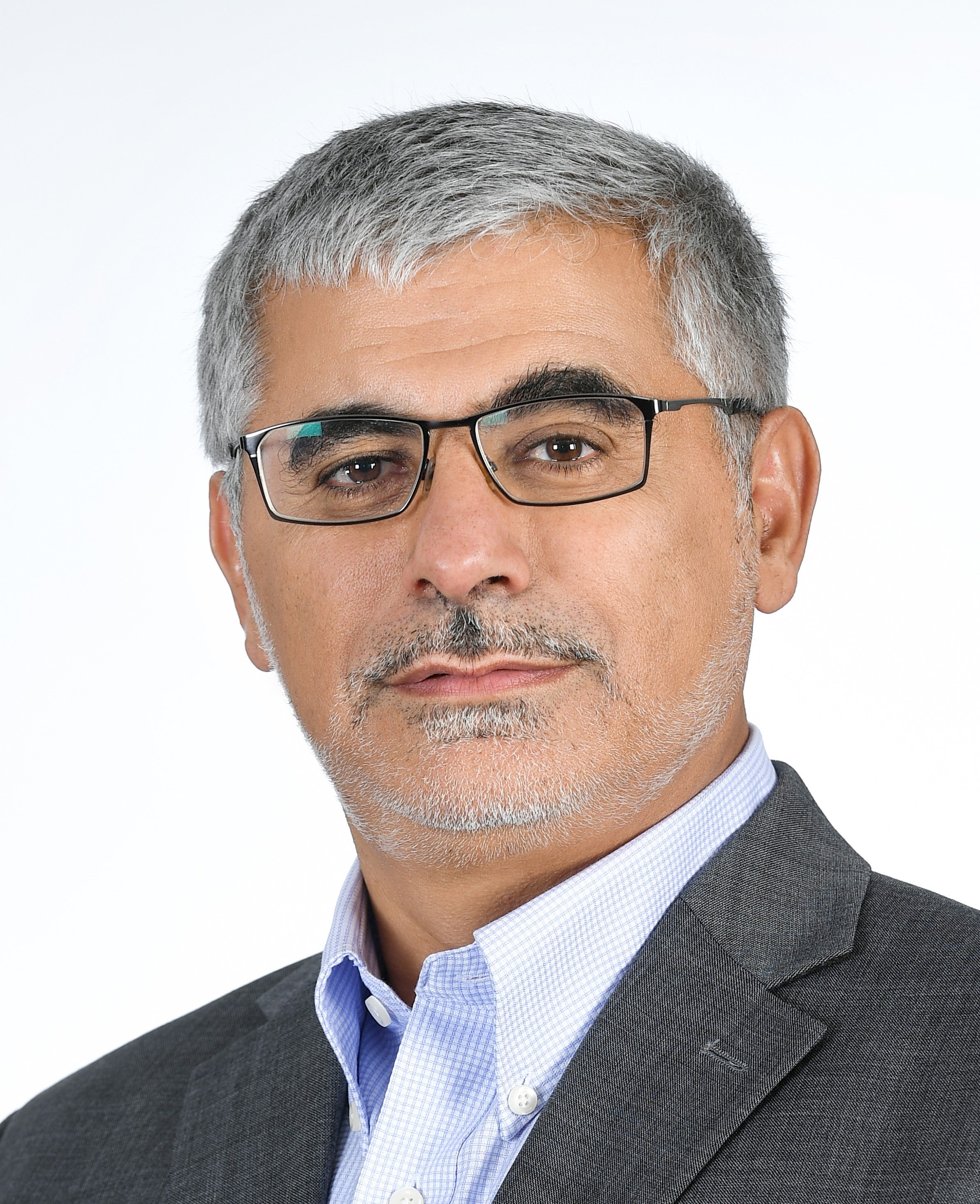}}]{Baker Mohammad} received his Ph.D.\ from University of Texas at Austin, his M.S.\ degree from Arizona State University, Tempe, and BS degree from the University of New Mexico, Albuquerque, all in ECE. 
Baker is the director of System on Chip Center and associate professor of ECE at Khalifa University. 
Prior to joining Khalifa University Baker was a Senior staff Engineer/Manager at Qualcomm, Austin, USA for 6-years, where he was engaged in designing high performance and low power DSP processor used for communication and multi-media application. 
Before joining Qualcomm he worked for 10 years at Intel Corporation on a wide range of micro-processors design from high performance, server chips $>$ 100 Watt (IA-64), to mobile embedded processor low power sub 1 watt (xscale). 
His research interest includes VLSI, power efficient computing, Design with emerging technology such as Memristor, STTRAM, In Memory-Computing, Efficient hardware accelerator for search engine, image processing, and Artificial Intelligent hardware.
\end{IEEEbiography}

\begin{IEEEbiography}[{\includegraphics[width=1in,height=1.25in,clip,keepaspectratio]{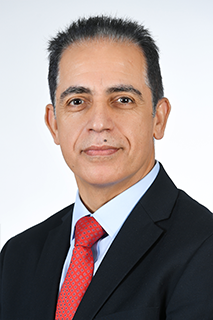}}]{Mahmoud Al-Qutayri} received the B.Eng.\ degree from Concordia University, Canada, 1984,
the M.Sc.\ degree from the University of Manchester, U.K., 1987, and the Ph.D.\ degree from
the University of Bath, U.K., 1992 all in electrical
and electronic engineering. He is currently a Full
Professor with the Department of Electrical and
Computer Engineering and the Associate Dean
for Graduate Studies, College of Engineering at
Khalifa University, UAE. Prior to joining Khalifa
University, he worked at De Montfort University, UK and University of Bath, UK. Al-Qutayri current research interests include wireless sensor networks, embedded systems design, in-memory computing, mixed-signal integrated circuits design and test, and hardware
security.
\end{IEEEbiography}

\begin{IEEEbiography}[{\includegraphics[width=1in,height=1.25in,clip,keepaspectratio]{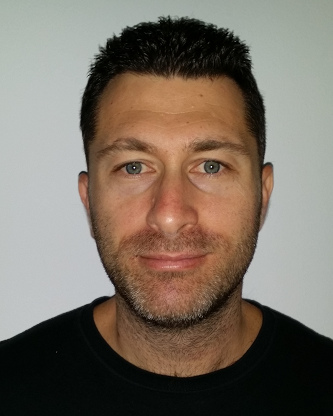}}]{Ozgur Sinanoglu} is a professor of electrical and computer engineering at New York University Abu Dhabi. 
He obtained his Ph.D.\ in Computer Science and Engineering from University of California San Diego. 
He has industry experience at TI, IBM and Qualcomm, and has been with NYU Abu Dhabi since 2010. 
During his Ph.D.\ he won the IBM Ph.D.\ fellowship award twice. 
He is also the recipient of the best paper awards at IEEE VLSI Test Symposium 2011 and ACM Conference on Computer and Communication Security 2013. 
Prof.\ Sinanoglu’s research interests include design-for-test, design-for-security and design-for-trust for VLSI circuits, where he has more than 200 conference and journal papers, and 20 issued and pending US Patents. 
Prof.\ Sinanoglu is the director of the Center for CyberSecurity at NYU Abu Dhabi. 
His recent research in hardware security and trust is being funded by US National Science Foundation, US Department of Defense, Semiconductor Research Corporation, Intel Corp, and Mubadala Technology.
\end{IEEEbiography}

\end{document}